\documentclass{article}
\usepackage{amssymb}
\usepackage{amsfonts}
\usepackage{amsmath}

\catcode`\@=11
\def\numberbysection{\@addtoreset{equation}{section}
        \def\theequation{\thesection.\arabic{equation}}}
\numberbysection

\begin{document}

\begin{titlepage}     
\vspace{0.5cm}
\begin{center}
{\Large\bf On the algebraic Bethe ansatz: Periodic boundary conditions}\\
\vspace{1cm}
{\large A. Lima-Santos\footnote{e-mail: dals@df.ufscar.br} \hspace{.5cm} } \\
\vspace{1cm}
{\large \em Universidade Federal de S\~ao Carlos, Departamento de F\'{\i}sica \\
Caixa Postal 676, CEP 13569-905~~S\~ao Carlos, Brasil}\\
\end{center}
\vspace{1.2cm}

\begin{abstract}
In this paper, the algebraic Bethe ansatz with periodic boundary conditions is used
to investigate trigonometric vertex models associated with the fundamental representations of the non-exceptional
Lie algebras. This formulation allow us to present explicit expressions for the eigenvectors and eigenvalues of the respective
transfer matrices.
\end{abstract}

\vspace{2cm}
\begin{center}
\end{center}
\vfill
\begin{center}
\small{\today}
\end{center}
\end{titlepage}

\section{Introduction}

One of the main branches of theoretical and mathematical physics is the
theory of exactly solvable models. The most successful approach to construct
integrable two-dimensional lattice models of statistical mechanics and
integrable ($1+1$)-dimensional quantum field theory is through the solution
of the Yang-Baxter equation \cite{Yang, Baxter}. Given a solution of this
equation, depending on a spectral parameter $\lambda $, one can define the
local Boltzmann weights of a commuting family of transfer matrix $T(\lambda
) $ \cite{Baxter1, KBI} and the factorizable $S$-matrix in a two-dimensional
quantum field theory \cite{Abdalla}.

The structure of the solutions of the Yang-Baxter equation based on simple
Lie algebras is by now fairly well understood \cite{Drinfeld}. In
particular, explicit expressions for the $\mathcal{R}$--matrices related to
non-exceptional affine Lie algebras were exhibited in \cite{Bazhanov} and 
\cite{Jimbo}. Since then, many other $\mathcal{R}$-matrices associated to
higher dimensional representations of these algebras have also been
determined \cite{Delius}.

A complete understanding of the vertex models living in a planar lattice
include the exact diagonalization of the row-to-row transfer matrices which
can provide informations about the on-shell physical properties such as
free-energy thermodynamics and quasi-particle excitation behavior. The most
efficient method for achieving this is the algebraic Bethe ansatz \cite%
{Fadeev}, though its coordinate version is usually more efficient for
finding the energy spectrum of concrete models \cite{Bethe}. A long-standing
open problem is the diagonalization of transfer matrices of vertex models
associated with solutions of the Yang-Baxter equation.

One possible method of finding the eigenvalues of a given transfer matrix is
the so-called analytical Bethe ansatz \cite{Reshe1}. This technique relies
on the unitarity, crossing and analyticity properties of the transfer matrix
and, in some cases, an extra amount of phenomenological input is also
required. This method has been applied to some of the models which we are
going to consider in this paper, more precisely for the systems B$_{n}^{(1)}$%
, C$_{n}^{(1)}$, D$_{n}^{(1)}$ \cite{Reshe2, Kuniba}. Unfortunately, the
explicit construction of eigenvectors of the transfer matrix is beyond the
scope of the analytical Bethe ansatz. The construction of exact eigenvectors
is certainly an important step in the program of solving integrable models 
\cite{IK1}

The importance of the algebraic Bethe ansatz does not rely on the
calculation of the energy spectrum of a given model, but to also to supply
information on the nature of the eigenfunctions. Thus is crucial in the
investigation of off-shell properties such as correlators of physical
operators \cite{KBI} as well as for the calculation of form-factors \cite%
{Babujian}.

A unified formulation of the quantum inverse scattering method for lattice
vertex models associated to non-exceptional Lie algebras has been developed
in the last years by Martins and collaborators \cite{Martins1, Martins2,
Martins3}. In their works the Yang-Baxter algebra is recast in terms of
novel commutation relations among creation, annihilation and diagonal
fields. In particular, the solution of the twisted D$_{n+1}^{(2)}$ vertex
models is accommodated in their unification by the solution of a
sixteen-vertex model \cite{Martins4}.

In this work, we will describe a detailed account of this method which
complements results on the literature \cite{Martins1}, by the investigation
of the trigonometric vertex models associated with the affine Lie algebras B$%
_{n}^{(1)}$, C$_{n}^{(1)}$, D$_{n}^{(1)}$, A$_{2n}^{(2)}$ and A$%
_{2n-1}^{(2)} $. The D$_{n+1}^{(2)}$ vertex models are out of the scope of
this paper.

The outline of the paper is as follows:

In section $2$ we present the models through \ their $\mathcal{R}$ matrices.
In section $3$ the eigenvalue problem of the transfer matrix is formulated.
In section $4$ we perform a detailed construction of the intertwining
relation in order to derive the fundamental commutation relations. In
sections $5$ and $6$ the eigenvalue problem is executed in full detail for
one and two-particle Bethe states, respectively, emphasizing the subtleties
of each case and developing the language used in the text. In sections $7$
the multi-particle cases are solved. The section $8$ is reserved for our
conclusions. In the appendix\ the A$_{n}^{(1)}$ vertex models are considered
for sake of completeness.

\section{The vertex models}

The search for integrable models through solutions of the Yang-Baxter
equation has been performed by the quantum group approach in \cite{KR},
where the problem is reduced to a linear one. Indeed, $\mathcal{R}$-matrices
corresponding to vector representations of all non-exceptional affine Lie
algebras were determined in this way in \cite{Jimbo}.

Quantum $\mathcal{R}$-matrices for the vertex models associated to the B$%
_{n}^{(1)},$C$_{n}^{(1)},\ $D$_{n}^{(1)},$\ A$_{2n}^{(2)}$ and A$%
_{2n-1}^{(2)}$ affine Lie algebras as presented by Jimbo have the form \cite%
{Jimbo}: 
\begin{eqnarray}
\mathcal{R}^{(l)} &=&x_{1}^{(l)}\sum_{\alpha \neq \alpha ^{\prime
}}^{N_{l}}E_{\alpha \alpha }\otimes E_{\alpha \alpha
}+x_{2}^{(l)}\sum_{\alpha \neq \beta ,\ \beta ^{\prime }}^{N_{l}}E_{\alpha
\alpha }\otimes E_{\beta \beta }+x_{3}^{(l)}\sum_{\alpha <\beta ,\ \alpha
\neq \beta ^{\prime }}^{N_{l}}E_{\alpha \beta }\otimes E_{\beta \alpha } 
\notag \\
&&+x_{4}^{(l)}\sum_{\alpha >\beta ,\ \alpha \neq \beta ^{\prime
}}^{N_{l}}E_{\alpha \beta }\otimes E_{\beta \alpha }+\sum_{\alpha ,\beta
}^{N_{l}}y_{\alpha \beta }^{(l)}\ E_{\alpha \beta }\otimes E_{\alpha
^{\prime }\beta ^{\prime }}  \label{mod.1}
\end{eqnarray}%
\ where we have introduced a label $l$, $\ l=0,1,...,n-1$ in order to work
with the nesting structure presents in the nested Bethe Ansatz construction.
\ For a given value of $n$, \ the label $l$ identifies a particular vertex
model among those models with $n-l\leq n$. \ We can name the label $l=0$ as
the \emph{ground} and the remained ones as the \emph{layers }in the nest
build. Here, $E_{ij}$ denotes the elementary $N_{l}$ by $N_{l}$ matrices ($%
(E_{\alpha \beta })_{ab}=\delta _{\alpha a}\delta _{\beta b}$), where $%
N_{l}=2(n-l)$ \ for C$_{n-l}^{(1)}$, D$_{n-l}^{(1)}$ \ and A$%
_{2(n-l)-1}^{(2)}$ and $N_{l}=2(n-l)+1$ for B$_{n-l}^{(1)}$ \ and A$%
_{2(n-l)}^{(2)}$.

The Boltzmann weights with functional dependence on the spectral parameter $%
\lambda $ are given by 
\begin{eqnarray}
x_{1}^{(l)}(\lambda ) &=&(\mathrm{e}^{\lambda }-q^{2})(\mathrm{e}^{\lambda
}-\xi _{l}),\qquad \ \ x_{2}^{(l)}(\lambda )=q(\mathrm{e}^{\lambda }-1)(%
\mathrm{e}^{\lambda }-\xi _{l}),  \notag \\
x_{3}^{(l)}(\lambda ) &=&-(q^{2}-1)(\mathrm{e}^{\lambda }-\xi _{l}),\qquad \
x_{4}^{(l)}(\lambda )=\mathrm{e}^{\lambda }x_{3}^{(l)}(\lambda )
\label{mod.2}
\end{eqnarray}%
and%
\begin{equation}
y_{\alpha \beta }^{(l)}(\lambda )=\left\{ 
\begin{array}{cc}
(q^{2}\mathrm{e}^{\lambda }-\xi _{l})(\mathrm{e}^{\lambda }-1) & (\alpha
=\beta ,\ \alpha \neq \alpha ^{\prime }) \\ 
q(\mathrm{e}^{\lambda }-\xi _{l})(\mathrm{e}^{\lambda }-1)+(\xi
_{l}-1)(q^{2}-1)\mathrm{e}^{\lambda } & (\alpha =\beta ,\ \alpha =\alpha
^{\prime }) \\ 
(q^{2}-1)\ \left( \varepsilon _{\alpha }\varepsilon _{\beta }\ \xi _{l}q^{%
\overset{\_}{\alpha }-\overset{\_}{\beta }}(\mathrm{e}^{\lambda }-1)-\delta
_{\alpha \beta ^{\prime }}(\mathrm{e}^{\lambda }-\xi _{l})\right) & (\alpha
<\beta ) \\ 
(q^{2}-1)\mathrm{e}^{\lambda }\left( \varepsilon _{\alpha }\varepsilon
_{\beta }\ q^{\overset{\_}{\alpha }-\overset{\_}{\beta }}(\mathrm{e}%
^{\lambda }-1)-\delta _{\alpha \beta ^{\prime }}(\mathrm{e}^{\lambda }-\xi
_{l})\right) & (\alpha >\beta )%
\end{array}%
\right.  \label{mod.3}
\end{equation}%
where $q=e^{-2\eta }$ denotes an arbitrary parameter and $\ \alpha ^{\prime
}=N_{l}+1-\alpha $. The sign functions $\varepsilon _{\alpha }=1\ (1\leq
\alpha \leq n-l)$, $\ \epsilon _{\alpha }=-1\ (n-l+1\leq \alpha \leq 2(n-l))$
for C$_{n-l}^{(1)}$ \ and $\varepsilon _{\alpha }=1$ for the remaining cases.

Here $\xi $ and $\overset{\_}{\alpha }$ are given respectively by 
\begin{equation}
\xi _{l}=q^{2(n-l)-1},\ q^{2(n-l)+2},\ q^{2(n-l)-2},\ -q^{2(n-l)+1},\
-q^{2(n-l)}  \label{mod.4}
\end{equation}%
for B$_{n-l}^{(1)}$, C$_{n-l}^{(1)}$, D$_{n-l}^{(1)}$, A$_{2(n-l)}^{(2)}$, A$%
_{2(n-l)-1}^{(2)}$; 
\begin{equation}
\overset{\_}{\alpha }=\left\{ 
\begin{array}{c}
\ \alpha -1/2\ \qquad \quad \quad \quad \qquad \quad \ \ \ (1\leq \alpha
\leq n-l) \\ 
\\ 
\alpha +1/2\ \qquad \quad \ \ \ (n-l+1\leq \alpha \leq 2(n-l))%
\end{array}%
\right.  \label{mod.5}
\end{equation}%
for C$_{n-l}^{(1)}$, and%
\begin{equation}
\overset{\_}{\alpha }=\left\{ 
\begin{array}{c}
\!\!\alpha +1/2\ \qquad \quad \qquad \ \ \ (1\leq \alpha <\frac{N_{l}+1}{2})
\\ 
\\ 
\alpha \ \qquad \qquad \qquad \ \ \qquad \ (\quad \alpha =\frac{N_{l}+1}{2}%
\quad ) \\ 
\\ 
\ \alpha -1/2\ \qquad \quad \ \ \ \qquad \ (\frac{N_{l}+1}{2}<\alpha \leq
N_{l})%
\end{array}%
\right.  \label{mod.6}
\end{equation}%
in the remaining cases.

These $\mathcal{R}$-matrices are regular satisfying {\small PT}-symmetry and
unitarity: 
\begin{equation}
\mathcal{R}^{(l)}(0)=x_{1}^{(l)}(0)\mathcal{P}^{(l)},\qquad \mathcal{R}%
_{21}^{(l)}(\lambda )=\mathcal{P}_{12}^{(l)}\mathcal{R}_{12}^{(l)}(\lambda )%
\mathcal{P}_{12}^{(l)},\qquad \mathcal{R}_{12}^{(l)}(\lambda )\mathcal{R}%
_{21}^{(l)}(-\lambda )=x_{1}^{(l)}(\lambda )x_{1}^{(l)}(-\lambda )\mathbf{,}
\label{mod.7}
\end{equation}%
where $\mathcal{P}$ is the permutation matrix: $\mathcal{P}^{(l)}\left\vert
\alpha \right\rangle $ $\otimes \left\vert \beta \right\rangle =\left\vert
\beta \right\rangle $ $\otimes \left\vert \alpha \right\rangle $.\ 

\section{The eigenvalue problem}

In the context of two-dimensional classical statistical systems, each
Yang-Baxter solution $\mathcal{R}(\lambda )$ is interpreted as local
Boltzmann weights of an integrable vertex model on a square lattice of size $%
L\times L$. \ A physical state on this lattice is defined by the assignment
of a \emph{state variable }to each lattice edge. If one takes the horizontal
direction as space and the vertical one as time, the transfer matrix $\tau
_{L}(\lambda )$ plays the role of a discrete evolution operator acting on
the Hilbert space $\mathcal{H}^{(L)}$ spanned by the \emph{row states} which
are defined by the set of vertical link variables on the same row. Thus, the
transfer matrix elements can be understood as the transition probability of
the one row state to project on \ the consecutive one after a unit of time.

The main problem is the diagonalization of the $\tau _{L}(\lambda )$ matrix
for these lattice systems. \ To do this we request the algebraic Bethe
ansatz where the row-to-row transfer matrix can be constructed from a local
vertex operator $\mathcal{L}_{ai}(\lambda )$, the Lax operator, which is
viewed as a matrix on the $N$-dimensional auxiliary space $V_{a}$,
corresponding in the vertex model to the space of states of the horizontal
degrees of freedom. Its matrix elements are operators on the $L$-product
Hilbert space $\mathcal{H}^{(L)}=V_{i}^{\otimes L}$, where $V_{i}$
corresponds to the space of vertical degrees of freedom and $i$ denotes the
sites of the one-dimensional lattice. An ordered product of Lax operators
defines the $N^{2L}$ by $N^{2L}$ monodromy matrix%
\begin{equation}
T(\lambda )=\mathcal{L}_{aL}(\lambda )\mathcal{L}_{aL-1}(\lambda )\cdots 
\mathcal{L}_{a1}(\lambda ).  \label{eig.1}
\end{equation}%
which can be written as an $N$ by $N$ matrix with entries%
\begin{equation}
T_{ij}(\lambda )=\sum_{k_{1},...,k_{L-1}=1}^{L}\mathcal{L}%
_{ik_{1}}^{(L)}(\lambda )\otimes \mathcal{L}_{k_{1}k_{2}}^{(L-1)}(\lambda
)\otimes \cdots \otimes \mathcal{L}_{k_{L-1}j}^{(1)}(\lambda )  \label{eig.2}
\end{equation}%
where $\mathcal{L}_{ij}^{(n)}(\lambda )$ are $N$ by $N$ matrices acting on
the quantum space $V_{n}$.

The transfer matrix of the vertex model with periodic boundary conditions
can be written as a trace of the monodromy matrix on the auxiliary space $%
V_{a}$%
\begin{equation}
\tau _{L}(\lambda )=\mathrm{Tr}_{a}[T(\lambda
)]=\sum_{i=1}^{N}T_{ii}(\lambda )  \label{eig.3}
\end{equation}%
and the eigenvalue problem is defined by%
\begin{equation}
\tau _{L}(\lambda )\Psi =\Lambda (\lambda )\Psi  \label{eig.4}
\end{equation}%
where the eigenfunction $\Psi $ is obtained from the action of the
non-diagonal matrix elements of $T(\lambda )$ on a reference state.

Here one uses the fact that the Yang-Baxter{\small \ }equation%
\begin{equation}
\mathcal{R}_{12}(\lambda -\mu )\mathcal{R}_{13}(\lambda )\mathcal{R}%
_{23}(\mu )=\mathcal{R}_{23}(\mu )\mathcal{R}_{13}(\lambda )\mathcal{R}%
_{12}(\lambda -\mu ).  \label{eig.5}
\end{equation}%
can be recast in the form of commutation relations for the matrix elements
of the monodromy matrix which play the role of creation and annihilation
operators. The commutation relations are derived from the global
intertwining relation%
\begin{equation}
S(\lambda -\mu )T(\lambda )\otimes T(\mu )=T(\mu )\otimes T(\lambda
)S(\lambda -\mu )  \label{eig.6}
\end{equation}%
where we have used that the intertwining matrix $S(\lambda -\mu )$ is
defined on the tensor product $V_{a}\otimes V_{a}$ and satisfy the relation $%
S(\lambda -\mu )=\mathcal{PR}(\lambda -\mu )$ when the auxiliary space $%
V_{a} $ and the quantum space $V_{i}$ are equivalent and \ the Lax operator
identified with the $\mathcal{R}$ matrix, i.e., $\mathcal{L(}\lambda 
\mathcal{)}\circeq \mathcal{R(}\lambda \mathcal{)}$. The indices in the
matrix $\mathcal{R}$ denote the spaces where its action is not trivial.

In this paper a sufficiently general recipe is supplied to derive the
fundamental commutation relations among the monodromy elements for the
trigonometric vertex models associated with the B$_{n}^{(1)}$, C$_{n}^{(1)}$%
, D$_{n}^{(1)}$, A$_{2n}^{(2)}$ and A$_{2n-1}^{(2)}$ affine Lie algebras. \ 

For the vertex models (\ref{mod.1}) the monodromy matrix (\ref{eig.3}) can
be written as an $N_{l}$ by $N_{l}$ matrix%
\begin{equation}
T^{(l)}=\left( 
\begin{array}{cccccc}
A_{1}^{(l)} & B_{1}^{(l)} & B_{2}^{(l)} & \cdots & B_{N_{l}-2}^{(l)} & 
B_{N_{l}-1}^{(l)} \\ 
C_{1}^{(l)} & D_{1,1}^{(l)} & D_{1,2}^{(l)} & \cdots & D_{1,N_{l}-2}^{(l)} & 
B_{N_{l}}^{(l)} \\ 
C_{2}^{(l)} & D_{2,1}^{(l)} & D_{2,2}^{(l)} & \cdots & D_{2,N_{l}-2}^{(l)} & 
B_{N_{l}+1}^{(l)} \\ 
\vdots & \vdots & \vdots & \ddots & \vdots & \vdots \\ 
C_{N_{l}-2}^{(l)} & D_{N_{l}-2,1}^{(l)} & D_{N_{l}-2,2}^{(l)} & \cdots & 
D_{N_{l}-2,N_{l}-2}^{(l)} & B_{2N_{l}-3}^{(l)} \\ 
C_{N_{l}-1}^{(l)} & C_{N_{l}}^{(l)} & C_{N_{l}+1}^{(l)} & \cdots & 
C_{2N_{l}-3}^{(l)} & A_{3}^{(l)}%
\end{array}%
\right) .  \label{eig.7}
\end{equation}%
The usual reference state 
\begin{equation}
\left\vert 0_{L}\right\rangle ^{(l)}=\prod\limits_{i=1}^{L}\otimes
\left\vert 0\right\rangle _{i},\qquad \left\vert 0\right\rangle _{i}=\left( 
\begin{array}{c}
1 \\ 
0 \\ 
\vdots \\ 
0%
\end{array}%
\right) _{N_{l}}  \label{eig.8}
\end{equation}%
where $N_{l}$ is the length of the vectors $\left\vert 0\right\rangle _{i}$,
is the highest vector of $T^{(l)}(\lambda )$ 
\begin{eqnarray}
A_{1}^{(l)}(\lambda )\left\vert 0_{L}\right\rangle ^{(l)}
&=&X_{1}^{(l)}(\lambda )\left\vert 0_{L}\right\rangle ^{(l)},\quad \
A_{3}^{(l)}(\lambda )\left\vert 0_{L}\right\rangle
^{(l)}=X_{3}^{(l)}(\lambda )\left\vert 0_{L}\right\rangle ^{(l)}  \notag \\
D_{\alpha \alpha }^{(l)}(\lambda )\left\vert 0_{L}\right\rangle ^{(l)}
&=&X_{2}^{(l)}(\lambda )\left\vert 0_{L}\right\rangle ^{(l)},\quad D_{\alpha
\beta }^{(l)}(\lambda )\left\vert 0_{L}\right\rangle ^{(l)}=0,\quad  \notag
\\
B_{\alpha }^{(l)}(\lambda )\left\vert 0_{L}\right\rangle ^{(l)} &\neq &\{0,\
\left\vert 0_{L}\right\rangle ^{(l)}\},\qquad \ \ \!C_{\alpha
}^{(l)}(\lambda )\left\vert 0_{L}\right\rangle ^{(l)}=0,\quad  \notag \\
\qquad \alpha &\neq &\beta =1,2,\ldots ,N_{l}-2  \label{eig.9}
\end{eqnarray}%
where 
\begin{equation}
X_{1}^{(l)}(\lambda )=[x_{1}^{(l)}(\lambda )]^{L},\qquad X_{2}^{(l)}(\lambda
)=[x_{2}^{(l)}(\lambda )]^{L},\qquad X_{3}^{(l)}(\lambda
)=[y_{N_{l}N_{l}}^{(l)}(\lambda )]^{L}  \label{eig.10}
\end{equation}

This triangular property suggests to write $T^{(l)}(\lambda )$ as a $3$ by $%
3 $ matrix: 
\begin{equation}
T^{(l)}(\lambda )=\left( 
\begin{array}{ccc}
A_{1}^{(l)}(\lambda ) & \mathcal{B}^{(l)}(\lambda ) & B_{N_{l}-1}(\lambda )
\\ 
\mathcal{C}^{(l)}(\lambda ) & \mathcal{D}^{(l)}(\lambda ) & \mathcal{B}%
^{\ast (l)}(\lambda ) \\ 
C_{N_{l}-1}(\lambda ) & \mathcal{C}^{\ast (l)}(\lambda ) & 
A_{3}^{(l)}(\lambda )%
\end{array}%
\right)  \label{eig.11}
\end{equation}%
where\textit{\ }one can identify four scalars%
\begin{equation}
A_{1}^{(l)}(\lambda ),\qquad B_{N_{l}-1}^{(l)}(\lambda ),\qquad
C_{N_{l}-1}^{(l)}(\lambda ),\qquad A_{3}^{(l)}(\lambda ),  \label{scalar}
\end{equation}%
as well as, four ($N_{l}-2)$-dimensional vectors 
\begin{equation}
\mathcal{B}^{(l)}(\lambda )=\left( B_{1}^{(l)}(\lambda ),B_{2}^{(l)}(\lambda
),\cdots ,B_{N_{l}-2}^{(l)}(\lambda )\right) ,\qquad \mathcal{B}^{\ast
(l)}(\lambda )=\left( 
\begin{array}{c}
B_{N_{l}}^{(l)}(\lambda ) \\ 
B_{N_{l}+1}^{(l)}(\lambda ) \\ 
\vdots \\ 
B_{2N_{l}-3}^{(l)}(\lambda )%
\end{array}%
\right) ,  \label{eig.13}
\end{equation}%
\begin{equation}
\mathcal{C}^{(l)}(\lambda )=\left( 
\begin{array}{c}
C_{1}^{(l)}(\lambda ) \\ 
C_{2}^{(l)}(\lambda ) \\ 
\vdots \\ 
C_{N_{l}-2}^{(l)}(\lambda )%
\end{array}%
\right) ,\qquad \mathcal{C}^{\ast (l)}(\lambda )=\left(
C_{N_{l}}^{(l)}(\lambda ),C_{N_{l}+1}^{(l)}(\lambda ),\cdots
,C_{2N_{l}-3}^{(l)}(\lambda )\right) ,  \label{eig.14}
\end{equation}%
besides an ($N_{l}-2$ ) by ($N_{l}-2$) matrix denoted by 
\begin{equation}
\mathcal{D}^{(l)}(\lambda )=\left( 
\begin{array}{cccc}
D_{11}^{(l)}(\lambda ) & D_{12}^{(l)}(\lambda ) & \cdots & 
D_{1,N_{l}-2}^{(l)}(\lambda ) \\ 
D_{21}^{(l)}(\lambda ) & D_{22}^{(l)}(\lambda ) & \cdots & 
D_{2,N_{l}-2}^{(l)}(\lambda ) \\ 
\vdots & \vdots & \ddots & \vdots \\ 
D_{N_{l}-2,1}^{(l)}(\lambda ) & D_{N_{l}-2,2}^{(l)}(\lambda ) & \cdots & 
D_{N_{l}-2,N_{l}-2}^{(l)}(\lambda )%
\end{array}%
\right) .  \label{eig.15}
\end{equation}

The problem of diagonalization of the transfer matrix becomes%
\begin{equation}
\tau _{L}^{(l)}(\lambda )\Psi _{m}^{(l)}=\left[ A_{1}^{(l)}(\lambda
)+\sum_{\alpha =1}^{N_{l}-2}D_{\alpha \alpha }^{(l)}(\lambda
)+A_{3}^{(l)}(\lambda )\right] \Psi _{m}^{(l)}=\Lambda _{L}^{(l)}(\lambda
|\{\lambda _{i}\})\Psi _{m}^{(l)}.  \label{eig.16}
\end{equation}%
where $\Psi _{m}^{(l)}$ $=\Psi _{m}^{(l)}(\lambda _{1},...,\lambda _{m})$ is
a scalar function named the $m$-particle state for the eigenvalue problem. \
In particular, the eigenvalue of the reference state (\ref{eig.8}) is%
\begin{equation}
\Lambda _{L}^{(l)}(\lambda |0)=[x_{1}^{(l)}(\lambda
)]^{L}+(N_{l}-2)[x_{2}^{(l)}(\lambda )]^{L}+[y_{N_{l}N_{l}}^{(l)}(\lambda
)]^{L}.  \label{eig.17}
\end{equation}%
In order to construct other eigenvalues, one has to find the commutation
relations among the elements of $T^{(l)}(\lambda )$.

\newpage

\section{Fundamental Commutation Relations}

In contrast to what happens to the six-vertex model and its multi-states
generalizations it is a rather complicated task to find the commutation
relations among the matrix elements of the monodromy matrix in a general
case. However, as we are going to see, this construction can be simplified
because all informations about the commutation relations are already encoded
in the simplest cases.

For $N_{l}\geq 3$ \ the corresponding $N_{l}$ by $N_{l}$ intertwining $%
S^{(l)}$ matrix can also be suitably written as a $9$ by $9$ matrix in the
form%
\begin{equation}
\lbrack S^{(l)}]=\left( 
\begin{array}{ccccccccc}
x_{1}^{(l)} & 0 & 0 & 0 & 0 & 0 & 0 & 0 & 0 \\ 
0 & x_{4}^{(l)} & 0 & x_{2}^{(l)} & 0 & 0 & 0 & 0 & 0 \\ 
0 & 0 & y_{N_{l}1}^{(l)} & 0 & \hat{Y}_{N_{l}2}^{(l)} & 0 & 
y_{N_{l}N_{l}}^{(l)} & 0 & 0 \\ 
0 & x_{2}^{(l)} & 0 & x_{3}^{(l)} & 0 & 0 & 0 & 0 & 0 \\ 
0 & 0 & \hat{Y}_{21}^{(l)} & 0 & \hat{Y}^{(l)} & 0 & \hat{Y}_{2N_{l}}^{(l)}
& 0 & 0 \\ 
0 & 0 & 0 & 0 & 0 & x_{4}^{(l)} & 0 & x_{2}^{(l)} & 0 \\ 
0 & 0 & y_{11}^{(l)} & 0 & \hat{Y}_{12}^{(l)} & 0 & y_{1N_{l}}^{(l)} & 0 & 0
\\ 
0 & 0 & 0 & 0 & 0 & x_{2}^{(l)} & 0 & x_{3}^{(l)} & 0 \\ 
0 & 0 & 0 & 0 & 0 & 0 & 0 & 0 & x_{1}^{(l)}%
\end{array}%
\right)  \label{fcr.1}
\end{equation}%
where one can identify four $N_{l+1}^{2}$-dimensional vectors%
\begin{equation}
\hat{Y}_{N_{l},2}^{(l)}(\lambda
)=\sum_{i=1}^{N_{l}-2}y_{N_{l},i+1}^{(l)}(\lambda )\ (E_{i}\otimes
E_{i^{\prime }})^{t},\quad \hat{Y}_{2N_{l}}^{(l)}(\lambda
)=\sum_{i=1}^{N_{l}-2}y_{i+1,N_{l}}^{(l)}(\lambda )\ E_{i^{\prime }}\otimes
E_{i}  \label{fcr.2}
\end{equation}%
\begin{equation}
\hat{Y}_{12}^{(l)}(\lambda )=\sum_{i=1}^{N_{l}-2}y_{1,i+1}^{(l)}(\lambda )\
(E_{i}\otimes E_{i^{\prime }})^{t},\quad \quad \hat{Y}_{21}^{(l)}(\lambda
)=\sum_{i=1}^{N_{l}-2}y_{i+1,1}^{(l)}(\lambda )\ E_{i^{\prime }}\otimes E_{i}
\label{fcr.3}
\end{equation}%
and a $N_{l+1}^{2}$ by $N_{l+1}^{2}$ matrix $\hat{Y}^{(l)}=\mathcal{P}%
^{(l+1)}Y^{(l)}$, \ with $Y^{(l)}$ obtained from the $\mathcal{R}^{(l)}$
matrix (\ref{mod.1}) by the reduction 
\begin{eqnarray}
Y^{(l)} &=&x_{1}^{(l)}\sum_{i\neq i^{\prime }}^{N_{l}-2}E_{ii}\otimes
E_{ii}+x_{2}^{(l)}\sum_{i\neq j,j^{\prime }}^{N_{l}-2}E_{ii}\otimes
E_{jj}+x_{3}^{(l)}\sum_{i<j,i\neq j^{\prime }}^{N_{l}-2}E_{ij}\otimes E_{ji}
\notag \\
&&+x_{4}^{(l)}\sum_{i>j,i\neq j^{\prime }}^{N_{l}-2}E_{ij}\otimes
E_{ji}+\sum_{i,j=1}^{N_{l}-2}y_{i+1j+1}^{(l)}\ E_{i,j}\otimes E_{i^{\prime
},j^{\prime }}  \label{fcr.4}
\end{eqnarray}%
The remaining entries of $[S^{(l)}]$ \ are five scalars \{$%
x_{1}^{(l)},y_{11}^{(l)},y_{1N_{l}}^{(l)},y_{N_{l}1}^{(l)},y_{N_{l}N_{l}}^{(l)} 
$\} and three $N_{l+1}$ by $N_{l+1}$ identity matrices \{$%
x_{2}^{(l)},x_{3}^{(l)},x_{4}^{(l)}$\}. The $E_{i}$ in the definition
relations of the $\hat{Y}_{i,j}^{(l)}$ are column vectors of length $N_{l+1}$
with only unitary element at $\ i$\textit{th} position and $t$ means
transposition.

Here we note that although these vectors and matrices act on the \emph{layer}
$l+1$ ($N_{l+1}=N_{l}-2$), their Boltzmann weights are written in term of
the $l$-\textit{th} layer Boltzmann weights. Therefore, our label $l$ in a
particular Boltzmann weight is indicating the model that it belongs.

Now, using (\ref{fcr.1}) \ and (\ref{eig.11}) we can write the fundamental
relation in the form%
\begin{equation}
\lbrack S^{(l)}](\lambda -\mu )T^{(l)}(\lambda )\otimes T^{(l)}(\mu
)=T^{(l)}(\mu )\otimes T^{(l)}(\lambda )[S^{(l)}](\lambda -\mu )
\label{fcr.5}
\end{equation}%
in order to get $81$ equations for the commutation relations among the
entries of the monodromy matrix.

Before we look at these equations we would like to show what happens in the
simplest cases. \ For $N_{l}=3$, all entries of $[S^{(l)}]$ are reduced to
the scalar status and we have%
\begin{equation}
\lbrack S]=\left( 
\begin{array}{ccccccccc}
x_{1} & 0 & 0 & 0 & 0 & 0 & 0 & 0 & 0 \\ 
0 & x_{4} & 0 & x_{2} & 0 & 0 & 0 & 0 & 0 \\ 
0 & 0 & y_{31} & 0 & y_{32} & 0 & y_{33} & 0 & 0 \\ 
0 & x_{2} & 0 & x_{3} & 0 & 0 & 0 & 0 & 0 \\ 
0 & 0 & y_{21} & 0 & y_{22} & 0 & y_{23} & 0 & 0 \\ 
0 & 0 & 0 & 0 & 0 & x_{4} & 0 & x_{2} & 0 \\ 
0 & 0 & y_{11} & 0 & y_{12} & 0 & y_{13} & 0 & 0 \\ 
0 & 0 & 0 & 0 & 0 & x_{2} & 0 & x_{3} & 0 \\ 
0 & 0 & 0 & 0 & 0 & 0 & 0 & 0 & x_{1}%
\end{array}%
\right)  \label{fcr.5a}
\end{equation}%
This is the intertwining $S$ matrix for the B$_{1}^{(1)}$ and A$_{2}^{(2)}$
vertex models whose $\mathcal{R}$-matrices are given by (\ref{mod.1}) with $%
\ N=3$. Moreover, the matrix elements of (\ref{eig.11}) are scalars in the
auxiliary space 
\begin{equation}
T=\left( 
\begin{array}{ccc}
A_{1} & B_{1} & B_{2} \\ 
C_{1} & D_{11} & B_{3} \\ 
C_{2} & C_{3} & A_{3}%
\end{array}%
\right)  \label{fcr.5b}
\end{equation}

Substituting (\ref{fcr.5a}) and (\ref{fcr.5b}) into (\ref{fcr.5}) we get the
commutation rules for the entries of (\ref{fcr.5b}) proposed previously by
Tarasov in the context of the Izergin-Korepin vertex model \cite{Tarasov}.

It is also worth note that the case $N_{l}=2$ can be added to our
discussion. \ In this case all entries with tensor status are removed from $%
[S^{(l)}]$. The result is%
\begin{equation}
\lbrack S]=\left( 
\begin{array}{cccc}
x_{1} & 0 & 0 & 0 \\ 
0 & y_{21} & y_{22} & 0 \\ 
0 & y_{11} & y_{12} & 0 \\ 
0 & 0 & 0 & x_{1}%
\end{array}%
\right)  \label{fcr.5c}
\end{equation}%
which is the intertwining $S$ matrix for the C$_{1}^{(1)},$D$_{1}^{(1)}$ and
A$_{1}^{(2)}$ vertex models whose $\mathcal{R}$-matrices are also given by (%
\ref{mod.1}). Consequently, their monodromy matrices preserve only the
scalar entries (\ref{scalar}):%
\begin{equation}
T=\left( 
\begin{array}{cc}
A_{1} & B_{1} \\ 
C_{1} & A_{3}%
\end{array}%
\right)  \label{fcr.5d}
\end{equation}%
Of course this is not a simple limit from the general case because their
Bethe states are different by construction. However, we can derive the
commutation relations for the entries of (\ref{fcr.5d}) by vanishing all
entries which are vector or matrix in the general commutation relations
which we are going to derive. Note also that C$_{1}^{(1)}$ and A$_{1}^{(2)}$
are six-vertex models while D$_{1}^{(1)}$ is neither a six-vertex model nor
a regular model as we can see from they $\mathcal{R}$ matrices.

In order to derive these so important commutation relations we shall proceed
in the following way: we denote by $E[i,j]=0$ the $(i,j)$ component of the
matrix equation (\ref{fcr.5}) and collecting \ them in $25$ blocks $B[i,j]$, 
$(i=1,...,5,\ j=i,...,10-i)$, defined by%
\begin{equation}
B[i,j]=\left\{ 
\begin{array}{c}
E_{ij}=E[i,j],\qquad e_{ij}=E[j,i], \\ 
\overset{\_}{E}_{ij}=E[10-i,10-j,\qquad \overset{\_}{e}_{ij}=E[10-j,10-i]%
\end{array}%
\right.  \label{fcr.6}
\end{equation}%
For a given block $B[i,j]$, the equation $\overset{\_}{e}_{ij}$ can be read
from the equation $E_{ij}$ ( and the equation $\overset{\_}{E}_{ij}$ from
the equation $e_{ij}$ ) by the interchanging%
\begin{eqnarray}
A_{1}^{(l)}(\lambda ) &\leftrightarrow &A_{3}^{(l)}(\mu ),\quad
B_{N_{l}-1}(\lambda )\leftrightarrow B_{N_{l}-1}(\mu ),\quad \mathcal{B}%
^{(l)}(\lambda )\leftrightarrow \mathcal{B}^{\ast (l)}(\mu ),  \notag \\
\mathcal{C}^{(l)}(\lambda ) &\leftrightarrow &\mathcal{C}^{\ast (l)}(\mu
),\quad C_{N_{l}-1}(\lambda )\leftrightarrow C_{N_{l}-1}(\mu ),\quad 
\mathcal{D}^{(l)}(\lambda )\leftrightarrow \mathcal{D}^{(l)}(\mu ),  \notag
\\
&&  \notag \\
x_{4}^{(l)} &\leftrightarrow &x_{3}^{(l)},\quad
y_{1,N_{l}}^{(l)}\leftrightarrow y_{N_{l},1}^{(l)}\quad \hat{Y}%
_{21}^{(l)}\leftrightarrow \hat{Y}_{12}^{(l)},\quad \hat{Y}%
_{N_{l},2}^{(l)}\leftrightarrow \hat{Y}_{2,N_{l}}^{(l)}  \label{fcr.7}
\end{eqnarray}%
Here we note that the Boltzmann weights (\ref{mod.3}) satisfy the relation $%
y_{\alpha ,\beta }^{(l)}(\lambda )=y_{\beta ^{\prime },\alpha ^{\prime
}}^{(l)}(\lambda )$. Taking into account these identifications, the
computation to find commutation relations is considerably simplified. \ 

For instance, in the block $B[1,4]$ \ we can solve the equations $E_{14}=0$
and $\overset{\_}{e}_{14}=0$ \ in order to find%
\begin{equation}
A_{1}^{(l)}(\lambda )\mathcal{B}^{(l)}(\mu )=z^{(l)}(\mu -\lambda )\mathcal{B%
}^{(l)}(\mu )A_{1}^{(l)}(\lambda )-\frac{x_{3}^{(l)}(\mu -\lambda )}{%
x_{2}^{(l)}(\mu -\lambda )}\mathcal{B}^{(l)}(\lambda )A_{1}^{(l)}(\mu )
\label{fcr.8}
\end{equation}%
and%
\begin{equation}
A_{3}^{(l)}(\lambda )\mathcal{B}^{\ast (l)}(\mu )=z^{(l)}(\lambda -\mu )%
\mathcal{B}^{\ast (l)}(\mu )A_{3}^{(l)}(\lambda )-\frac{x_{4}^{(l)}(\lambda
-\mu )}{x_{2}^{(l)}(\lambda -\mu )}\mathcal{B}^{\ast (l)}(\lambda
)A_{3}^{(l)}(\mu ).  \label{fcr.9}
\end{equation}%
where we have introduced the function%
\begin{equation}
z^{(l)}(\lambda )=\frac{x_{1}^{(l)}(\lambda )}{x_{2}^{(l)}(\lambda )}
\label{fcr.9a}
\end{equation}%
since it will appears many times in the text.

Let us consider here two further blocks: \ the equations in the block $%
B[2,7] $ yield the following intertwining relations%
\begin{eqnarray}
A_{1}^{(l)}(\lambda )\mathcal{B}^{\ast (l)}(\mu ) &=&\frac{x_{2}^{(l)}(\mu
-\lambda )}{y_{N_{l}N_{l}}^{(l)}(\mu -\lambda )}\mathcal{B}^{\ast (l)}(\mu
)A_{1}^{(l)}(\lambda )-\mathcal{B}^{(l)}(\lambda )\otimes \mathcal{D}%
^{(l)}(\mu )\frac{\hat{Y}_{2N_{l}}^{(l)}(\mu -\lambda )}{%
y_{N_{l}N_{l}}^{(l)}(\mu -\lambda )}  \notag \\
&&+\frac{x_{4}^{(l)}(\mu -\lambda )}{y_{N_{l}N_{l}}^{(l)}(\mu -\lambda )}%
B_{N_{l}-1}(\mu )\mathcal{C}^{(l)}(\lambda )-\frac{y_{1N_{l}}^{(l)}(\mu
-\lambda )}{y_{N_{l}N_{l}}^{(l)}(\mu -\lambda )}B_{N_{l}-1}(\lambda )%
\mathcal{C}^{(l)}(\mu )  \label{fcr.10}
\end{eqnarray}%
and%
\begin{eqnarray}
A_{3}^{(l)}(\lambda )\mathcal{B}^{(l)}(\mu ) &=&\frac{x_{2}^{(l)}(\lambda
-\mu )}{y_{N_{l}N_{l}}^{(l)}(\lambda -\mu )}\mathcal{B}^{(l)}(\mu
)A_{3}^{(l)}(\lambda )-\frac{\hat{Y}_{N_{l}2}^{(l)}(\lambda -\mu )}{%
y_{N_{l}N_{l}}^{(l)}(\lambda -\mu )}\mathcal{B}^{\ast (l)}(\lambda )\otimes 
\mathcal{D}^{(l)}(\mu )  \notag \\
&&+\frac{x_{3}^{(l)}(\lambda -\mu )}{y_{N_{l}N_{l}}^{(l)}(\lambda -\mu )}%
B_{N_{l}-1}(\mu )\mathcal{C}^{\ast (l)}(\lambda )-\frac{y_{N_{l}1}^{(l)}(%
\lambda -\mu )}{y_{N_{l}N_{l}}^{(l)}(\lambda -\mu )}B_{N_{l}-1}(\lambda )%
\mathcal{C}^{\ast (l)}(\mu )  \label{fcr.11}
\end{eqnarray}%
where the tensor structure coming from (\ref{fcr.5}) with scalars, vectors
and matrices defined previously. \ It is also worth note the correspondence
between (\ref{fcr.10}) and (\ref{fcr.11}) via the exchange property (\ref%
{fcr.7}) and the matrices product order. \ 

A very important information is given by the commutation relations derived
from the block $B[2,5]$

\begin{eqnarray}
\mathcal{D}^{(l)}(\lambda )\otimes \mathcal{B}^{(l)}(\mu ) &=&\mathcal{B}%
^{(l)}(\mu )\otimes \mathcal{D}^{(l)}(\lambda )\frac{s^{(l)}(\lambda -\mu )}{%
x_{2}^{(l)}(\lambda -\mu )}-B_{N_{l}-1}(\mu )\mathcal{C}^{(l)}(\lambda )%
\frac{\hat{Y}_{N_{l}2}^{(l)}(\mu -\lambda )}{y_{N_{l}N_{l}}^{(l)}(\mu
-\lambda )}\frac{s^{(l)}(\lambda -\mu )}{x_{2}^{(l)}(\lambda -\mu )}  \notag
\\
&&-\frac{x_{4}^{(l)}(\lambda -\mu )}{x_{2}^{(l)}(\lambda -\mu )}\mathcal{B}%
^{(l)}(\lambda )\otimes \mathcal{D}^{(l)}(\mu )+\frac{x_{4}^{(l)}(\lambda
-\mu )}{x_{2}^{(l)}(\lambda -\mu )}B_{N_{l}-1}(\lambda )\mathcal{C}%
^{(l)}(\mu )\frac{\hat{Y}_{N_{l}2}^{(l)}(\lambda -\mu )}{%
y_{N_{l}N_{l}}^{(l)}(\lambda -\mu )}  \notag \\
&&+\mathcal{B}^{\ast (l)}(\lambda )A_{1}^{(l)}(\mu )\frac{\hat{Y}%
_{N_{l}2}^{(l)}(\lambda -\mu )}{y_{N_{l}N_{l}}^{(l)}(\lambda -\mu )}
\label{fcr.12}
\end{eqnarray}%
and

\begin{eqnarray}
\mathcal{D}^{(l)}(\lambda )\otimes \mathcal{B}^{\ast (l)}(\mu ) &=&\frac{%
s^{(l)}(\mu -\lambda )}{x_{2}^{(l)}(\mu -\lambda )}\mathcal{B}^{\ast
(l)}(\mu )\otimes \mathcal{D}^{(l)}(\lambda )-\frac{s^{(l)}(\mu -\lambda )}{%
x_{2}^{(l)}(\mu -\lambda )}\frac{\hat{Y}_{2N_{l}}^{(l)}(\lambda -\mu )}{%
y_{N_{l}N_{l}}^{(l)}(\lambda -\mu )}B_{N_{l}-1}(\mu )\mathcal{C}^{\ast
(l)}(\lambda )  \notag \\
&&-\frac{x_{3}^{(l)}(\mu -\lambda )}{x_{2}^{(l)}(\mu -\lambda )}\mathcal{B}%
^{\ast (l)}(\lambda )\otimes \mathcal{D}^{(l)}(\mu )+\frac{x_{3}^{(l)}(\mu
-\lambda )}{x_{2}^{(l)}(\mu -\lambda )}\frac{\hat{Y}_{2N_{l}}^{(l)}(\mu
-\lambda )}{y_{N_{l}N_{l}}^{(l)}(\mu -\lambda )}B_{N_{l}-1}(\lambda )%
\mathcal{C}^{\ast (l)}(\mu )  \notag \\
&&+\frac{\hat{Y}_{2N_{l}}^{(l)}(\mu -\lambda )}{y_{N_{l}N_{l}}^{(l)}(\mu
-\lambda )}\mathcal{B}^{(l)}(\lambda )A_{3}^{(l)}(\mu )  \label{fcr.13}
\end{eqnarray}%
where we have defined \ a $N_{l+1}^{2}$ by $N_{l+1}^{2}$ matrix 
\begin{eqnarray}
s^{(l)}(\lambda ) &=&\hat{Y}^{(l)}(\lambda )-\frac{1}{y_{N_{l}N_{l}}^{(l)}(%
\lambda )}\hat{Y}_{N_{l}2}^{(l)}(\lambda )\otimes \hat{Y}_{2N_{l}}^{(l)}(%
\lambda )  \notag \\
&=&\hat{Y}^{(l)}(\lambda )-\frac{1}{y_{N_{l}N_{l}}^{(l)}(\lambda )}\hat{Y}%
_{2N_{l}}^{(l)}(\lambda )\otimes \hat{Y}_{N_{l}2}^{(l)}(\lambda )
\label{fcr.14}
\end{eqnarray}%
which satisfies the permuted version of the Yang-Baxter equation%
\begin{equation}
s_{12}^{(l)}(\lambda )s_{23}^{(l)}(\lambda +\mu )s_{12}^{(l)}(\lambda
)=s_{23}^{(l)}(\mu )s_{12}^{(l)}(\lambda +\mu )s_{23}^{(l)}(\mu )
\label{fcr.15}
\end{equation}%
In the definition (\ref{fcr.14}) the entries of $s^{(l)}(\lambda )$ were
written in terms of the Boltzmann weight of the vertex model with label $l$.
However, due to (\ref{fcr.15}), its label must be $l+1$. To write the matrix 
$s(\lambda )$ with its Boltzmann weights labeled correctly we can use the
identities%
\begin{equation}
\frac{s^{(l)}(\lambda )}{x_{a}^{(l)}(\lambda )}=\frac{S^{(l+1)}(\lambda )}{%
x_{a}^{(l+1)}(\lambda )},\qquad a=1,2.  \label{fcr.16}
\end{equation}%
where $S^{(l+1)}=\mathcal{P}^{(l+1)}\mathcal{R}^{(l+1)}$ and $\mathcal{R}%
^{(l+1)}$ is given by (\ref{mod.1}) replacing $l$ by $l+1$. These relations
give the emphasis of the meaning of the label $l$. \ Moreover, in (\ref%
{fcr.12}) and (\ref{fcr.13}) we have used the following matrix identities%
\begin{eqnarray}
\frac{\hat{Y}_{N_{l}2}^{(l)}(\mu -\lambda )}{y_{N_{l}N_{l}}^{(l)}(\mu
-\lambda )}s^{(l)}(\lambda -\mu ) &=&-\hat{Y}_{12}^{(l)}(\lambda -\mu )+%
\frac{y_{1N_{l}}^{(l)}(\lambda -\mu )}{y_{N_{l}N_{l}}^{(l)}(\lambda -\mu )}%
\hat{Y}_{N_{l}2}^{(l)}(\lambda -\mu )  \notag \\
s^{(l)}(\lambda -\mu )\frac{\hat{Y}_{2N_{l}}^{(l)}(\mu -\lambda )}{%
y_{N_{l}N_{l}}^{(l)}(\mu -\lambda )} &=&-\hat{Y}_{21}^{(l)}(\lambda -\mu )+%
\frac{y_{N_{l}1}^{(l)}(\lambda -\mu )}{y_{N_{l}N_{l}}^{(l)}(\lambda -\mu )}%
\hat{Y}_{2N_{l}}^{(l)}(\lambda -\mu )  \label{fcr.17}
\end{eqnarray}%
and the scalar relation%
\begin{equation}
\frac{x_{4}^{(l)}(\lambda )}{x_{2}^{(l)}(\lambda )}+\frac{%
x_{3}^{(l)}(-\lambda )}{x_{2}^{(l)}(-\lambda )}=0  \label{fcr.18}
\end{equation}

Many other commutation relations will be used in this paper. In the right
time we will derive them recalling the block $B[i,j]$ once more.

\section{The one-particle Bethe state}

In the quantum inverse scattering method the eigenstates of the transfer
matrix are constructing by the action of the creators operators on the
reference state. Such procedure results in excitations with multi-particle
structure, characterized by a set of rapidities $\left\{ \lambda
_{i}\right\} $ which are determined by solving the Bethe equations. In the
general case the vector $\mathcal{B}^{(l)}(\mu )$\ has $N_{l}-2$ components
and is used to define the one-particle state which is a scalar function
obtained by the linear combination%
\begin{equation}
\Psi _{1}^{(l)}(\lambda _{1})=\mathcal{B}^{(l)}(\lambda _{1})\mathcal{F}%
_{1}^{(l)}\left\vert 0_{L}^{(l)}\right\rangle  \label{one.1}
\end{equation}%
where $\mathcal{F}_{1}^{(l)}$ is a vector with $N_{l+1}$ components $%
f^{(l)\alpha }$.

The action of the transfer matrix $\tau _{L}^{(l)}(\lambda )$ on this state
is determined by (\ref{eig.9}) and the intertwining relations (\ref{fcr.5}).
The components of (\ref{fcr.5}) needed for the construction of the nested
Bethe ansatz of the one-particle state are the commutation relations (\ref%
{fcr.8}), (\ref{fcr.11}) and (\ref{fcr.12}). In particular, the matrix
relation (\ref{fcr.12}) must be written in terms of its components in order
to get the commutation relation for the scalar $\sum_{\alpha }D_{\alpha
\alpha }^{(l)}(\lambda )$: 
\begin{eqnarray}
\sum_{\alpha =1}^{N_{l}-2}D_{\alpha \alpha }^{(l)}(\lambda )\mathcal{B}%
^{(l)}(\mu ) &=&\mathcal{B}^{(l)}(\mu )\mathrm{Tr}_{a}[\frac{\mathcal{L}%
_{a1}^{(l+1)}(\lambda -\mu )}{x_{2}^{(l+1)}(\lambda -\mu )}\mathcal{D}%
^{(l)}(\lambda )]-\frac{x_{4}^{(l)}(\lambda -\mu )}{x_{2}^{(l)}(\lambda -\mu
)}\mathcal{B}^{(l)}(\lambda )\mathcal{D}^{(l)}(\mu )  \notag \\
&&+\frac{\hat{Y}_{N_{l}2}^{(l)}(\lambda -\mu )}{y_{N_{l}N_{l}}^{(l)}(\lambda
-\mu )}[B^{\ast (l)}(\lambda )\otimes \mathbf{1}^{(l)}]A_{1}^{(l)}(\mu ) 
\notag \\
&&-B_{N_{l}-1}(\mu )\frac{\hat{Y}_{N_{l}2}^{(l)}(\mu -\lambda )}{%
y_{N_{l}N_{l}}^{(l)}(\mu -\lambda )}\frac{S^{(l+1)}(\lambda -\mu )}{%
x_{2}^{(l+1)}(\lambda -\mu )}[\mathcal{C}^{(l)}(\lambda )\otimes \mathbf{1}%
^{(l)}]  \notag \\
&&+B_{N_{l}-1}(\lambda )\frac{x_{4}^{(l)}(\lambda -\mu )}{%
x_{2}^{(l)}(\lambda -\mu )}\frac{\hat{Y}_{N_{l}2}^{(l)}(\lambda -\mu )}{%
y_{N_{l}N_{l}}^{(l)}(\lambda -\mu )}[\mathcal{C}^{(l)}(\mu )\otimes \mathbf{1%
}^{(l)}]  \label{one.2}
\end{eqnarray}%
where $1^{(l)}$ is the $N_{l+1}$ by $N_{l+1}$ matrix identity and $\mathrm{Tr%
}_{a}$ is the trace in the auxiliary space. In (\ref{one.2}), we also have
used (\ref{fcr.16}) to write the matrix $s^{(l)}$ as $S^{(l+1)}$ and
identified its permuted matrix $\mathcal{R}^{(l+1)}$ with the Lax operator $%
\mathcal{L}^{(l+1)}$.

In the nested Bethe ansatz procedure we always begin at the ground level $%
l=0 $ and from now we shall omitte such label in the expressions of the
first eigenvalue problem.

The eigenvalue problem is accomplished by the action of $\tau _{L}(\lambda )$
on $\Psi _{1}(\lambda _{1})$ 
\begin{eqnarray}
\tau _{L}(\lambda )\Psi _{1}(\lambda _{1}) &=&\left( A_{1}(\lambda
)+\sum_{\alpha =1}^{N-2}D_{\alpha \alpha }(\lambda )+A_{3}(\lambda )\right) 
\mathcal{B}(\lambda _{1})\mathcal{F}_{1}\left\vert 0_{L}\right\rangle  \notag
\\
&\circeq &\Lambda _{L}(\lambda |\{\lambda _{1}\})\Psi _{1}(\lambda _{1})
\label{one.3}
\end{eqnarray}%
Taking into account the commutation relations (\ref{fcr.8}), (\ref{fcr.11})
, (\ref{one.2}) \ and (\ref{eig.9}) we are able to turn the operators $%
A_{1}(\lambda )$, $D_{\alpha \alpha }(\lambda )$ and $A_{3}(\lambda )$ over
the creation operator $\mathcal{B}(\lambda _{1})$ and as result we have the
following expression for the first eigenvalue problem:%
\begin{eqnarray}
\tau _{L}(\lambda )\Psi _{1}(\lambda _{1}) &=&X_{1}(\lambda )z(\lambda
_{1}-\lambda )\Psi _{1}(\lambda _{1})+X_{3}(\lambda )\frac{x_{2}(\lambda
-\lambda _{1})}{y_{NN}(\lambda -\lambda _{1})}\Psi _{1}(\lambda _{1})  \notag
\\
&&+\mathcal{B}(\lambda _{1})\mathrm{Tr}_{a}\left[ \frac{\mathcal{L}%
_{a1}^{(1)}(\lambda -\lambda _{1})}{x_{2}^{(1)}(\lambda -\lambda _{1})}%
\mathcal{D}(\lambda )\right] \mathcal{F}_{1}\left\vert 0_{L}\right\rangle 
\notag \\
&&-\mathcal{B}(\lambda )[X_{1}(\lambda _{1})\frac{x_{3}(\lambda _{1}-\lambda
)}{x_{2}(\lambda _{1}-\lambda )}+X_{2}(\lambda _{1})\frac{x_{4}(\lambda
-\lambda _{1})}{x_{2}(\lambda -\lambda _{1})}]\mathcal{F}_{1}\left\vert
0_{L}\right\rangle  \notag \\
&&-\frac{\hat{Y}_{N2}(\lambda -\lambda _{1})}{y_{NN}(\lambda -\lambda _{1})}%
\mathcal{B}^{\ast }(\lambda )\otimes \mathbf{1[}X_{2}(\lambda
_{1})-X_{1}(\lambda _{1})]\mathcal{F}_{1}\left\vert 0_{L}\right\rangle
\label{one.4}
\end{eqnarray}%
Of course, the terms proportional to $\Psi _{1}(\lambda _{1})$ are the
wanted terms and contribute to the eigenvalue $\Lambda _{L}(\lambda
|\{\lambda _{1}\})$. The remaining ones are the so-called unwanted terms and
they have to be eliminated by imposing restrictions on the rapidity $\lambda
_{1}$.

In this case it is easy to see that we have two unwanted terms which are
directly eliminated by the condition $X_{1}(\lambda _{1})=X_{2}(\lambda
_{1}) $. However, the trace term on the right hand side of (\ref{one.4})
does not give its wanted part directly. \ 

By simple inspection of the $\mathcal{R}$-matrices (\ref{mod.1}) for the C$%
_{n}^{(1)},$D$_{n}^{(1)}$ and A$_{2n-1}^{(2)}$ vertex models, one can see
that they trace in the auxiliary space is proportional to the $N$ by $N$
identity matrix $\mathbf{I}$ 
\begin{equation}
\mathrm{Tr}_{a}[\mathcal{L}_{a1}^{(1)}(\lambda )]=\left( x_{1}^{(1)}(\lambda
)+(N_{1}-2)x_{2}^{(1)}(\lambda )+y_{N_{1}N_{1}}^{(1)}(\lambda )\right) 
\mathbf{I,\qquad }N_{1}=4,6,8,...  \label{one.5}
\end{equation}%
It means that $\Psi _{1}(\lambda _{1})$ is the eigenstate of $\tau
_{L}(\lambda )$ with eigenvalue%
\begin{eqnarray}
\Lambda _{L}(\lambda |\{\lambda _{1}\}) &=&X_{1}(\lambda )z(\lambda
_{1}-\lambda )+X_{3}(\lambda )\frac{x_{2}(\lambda -\lambda _{1})}{%
y_{NN}(\lambda -\lambda _{1})}  \notag \\
&&+X_{2}(\lambda )\left( \frac{X_{1}^{(1)}(\lambda )}{X_{2}^{(1)}(\lambda )}%
+(N_{1}-2)+\frac{X_{3}^{(1)}(\lambda )}{X_{2}^{(1)}(\lambda )}\right)
\label{one.6}
\end{eqnarray}%
provided that 
\begin{equation}
\frac{X_{1}(\lambda _{1})}{X_{2}(\lambda _{1})}=1  \label{one.7}
\end{equation}%
where we have used the notation%
\begin{equation}
X_{3}^{(1)}(\lambda )=y_{N_{1}N_{1}}^{(1)}(\lambda -\lambda _{1}),\qquad
X_{a}^{(1)}(\lambda )=x_{a}^{(1)}(\lambda -\lambda _{1}),\quad a=1,2
\label{one.7a}
\end{equation}%
Here we make some remarks in respect to (\ref{one.6}). Although its
computation is made on the ground ($l=0$), the result also contains
Boltzmann weights of the model in the first \emph{layer} ($l=1$). Of course,
we can recall (\ref{fcr.16}) to write (\ref{one.6}) with all Boltzmann
weights of the model on the ground. Nevertheless, as we will see later, this 
\emph{layer} formation is very important -even when there is no nest to
build. For instance, the eigenvalue (\ref{one.6}) is not valid for the D$%
_{2}^{(1)}$ vertex model because its first \emph{layer} $l=1$ (the D$%
_{1}^{(1)}$ model) is not regular. It is also curious to note that no
further constraint is necessary for the vector $\mathcal{F}_{1}$. \ 

However, when $N_{l}$ is an odd number, the situation is not so simple
because the \textrm{Tr}$_{a}[\mathcal{L}_{a1}^{(l)}]$ is not more
proportional to the identity matrix due to the weight $y_{\alpha \beta
}^{(l)}(\lambda )$ with $\alpha =\beta $ and$\ \alpha =\alpha ^{\prime }$ (%
\ref{mod.3}), the element common to the diagonals of $\mathcal{R}^{(l)}$.

The trace in (\ref{one.4}) represents the transfer matrix of $L+1$ site
chain with one inhomogeneous site coming from the Lax operator $\mathcal{L}%
_{a1}^{(1)}(\lambda -\lambda _{1})$. To solve the eigenvalue problem \ of $%
\tau _{L+1}^{(1)}(\lambda ,\{\lambda _{1}\})\mathcal{F}_{1}\left\vert
0_{L}\right\rangle $, we first note that $\tau _{L+1}^{(1)}\in \mathcal{H}%
^{(L)}\otimes \mathcal{H}^{(1)}$ \ and the part of $\tau _{L+1}^{(1)}$
involving $\mathcal{D}(\lambda )\in \mathcal{H}^{(L)}$ commutes with $%
\mathcal{F}_{1}\in \mathcal{H}^{(1)}$ and can hit directly the reference
state $\left\vert 0_{L}\right\rangle $ 
\begin{equation}
\tau _{L+1}^{(1)}(\lambda ,\{\lambda _{1}\})\mathcal{F}_{1}\left\vert
0_{L}\right\rangle =X_{2}(\lambda )\left( \tau _{1}^{(1)}(\lambda ,\{\lambda
_{1}\})\mathcal{F}_{1}\right) \left\vert 0_{L}\right\rangle .  \label{one.8}
\end{equation}%
\ It means that the eigenvalue condition for (\ref{one.4}) leads to the
requirement that $\mathcal{F}_{1}$ ought to be an eigenvavector of $\tau
_{1}^{(1)}(\lambda ,\{\lambda _{1}\})$. Therefore, if we suppose that%
\begin{equation}
\tau _{1}^{(1)}(\lambda ,\{\lambda _{1}\})\mathcal{F}_{1}=\Lambda
_{1}^{(1)}(\lambda ,\{\lambda _{1}\}|\cdots )\mathcal{F}_{1},  \label{one.9}
\end{equation}%
the last wanted term in (\ref{one.4}) has the form%
\begin{equation}
X_{2}(\lambda )\frac{\Lambda _{1}^{(1)}(\lambda ,\{\lambda _{1}\}|\cdots )}{%
X_{2}^{(1)}(\lambda )}\mathcal{B}(\lambda _{1})\mathcal{F}_{1}\left\vert
0_{L}\right\rangle  \label{one.10}
\end{equation}%
and we conclude for B$_{n}^{(1)}$ and A$_{2n}^{(2)}$ vertex models that $%
\Psi _{1}(\lambda _{1})$ is an eigenfunction of $\tau _{L}(\lambda )$ with
eigenvalue%
\begin{equation}
\Lambda _{L}(\lambda |\{\lambda _{1}\})=X_{1}(\lambda )z(\lambda
_{1}-\lambda )+X_{2}(\lambda )\frac{\Lambda _{1}^{(1)}(\lambda ,\{\lambda
_{1}\}|\cdots )}{X_{2}^{(1)}(\lambda )}+X_{3}(\lambda )\frac{x_{2}(\lambda
-\lambda _{1})}{y_{NN}(\lambda -\lambda _{1})}  \label{one.11}
\end{equation}%
provided that%
\begin{equation}
\frac{X_{1}(\lambda _{1})}{X_{2}(\lambda _{1})}=1  \label{one.12}
\end{equation}%
The eigenvalue (\ref{one.11}) is partial because we still have to solve the
eigenvalue problem (\ref{one.9}) in order to know the value of $\Lambda
_{1}^{(1)}(\lambda ,\{\lambda _{1}\}|\cdots )$. Here we have reached a point
which is typical of nested Bethe ansatz problems. \ It means that we have to
solve an another eigenvalue problem for the transfer matrix $\tau _{1}^{(1)}$
with its $\mathcal{R}$-matrix given by (\ref{mod.1}) but with $l=1$, i.e.,
the first \emph{layer} of the nest.

The row-to-row $1$-site inhomogeneous transfer matrix $\tau
_{1}^{(1)}(\lambda ,\{\lambda _{1}\})$ is given by%
\begin{equation}
\tau _{1}^{(1)}(\lambda )=A_{1}^{(1)}(\lambda )+\sum_{\alpha
=1}^{N_{1}-2}D_{\alpha \alpha }^{(1)}(\lambda )+A_{1}^{(1)}(\lambda )
\label{one.13}
\end{equation}%
The notation used in (\ref{one.13}) is to be considered as a shorthand as
these terms depend also on the inhomogeneous parameter $\lambda _{1}$. The
reference state $\left\vert 0_{1}\right\rangle ^{(1)}$ is given by (\ref%
{one.8}) and it is a highest vector 
\begin{eqnarray}
A_{1}^{(1)}(\lambda )\left\vert 0_{1}\right\rangle ^{(1)}
&=&X_{1}^{(1)}(\lambda )|\left\vert 0_{1}\right\rangle ^{(1)},\quad
D_{\alpha \alpha }^{(1)}(\lambda )\left\vert 0_{1}\right\rangle
^{(1)}=X_{2}^{(1)}(\lambda )\left\vert 0_{1}\right\rangle ^{(1)},\quad 
\notag \\
A_{3}^{(1)}(\lambda )\left\vert 0_{1}\right\rangle ^{(1)}
&=&X_{3}^{(1)}(\lambda )\left\vert 0_{1}\right\rangle ^{(1)},\quad \
D_{\alpha \beta }^{(1)}(\lambda )\left\vert 0_{1}\right\rangle ^{(1)}=0,\quad
\notag \\
C_{\alpha }^{(1)}(\lambda )\left\vert 0_{1}\right\rangle ^{(1)} &=&0,\quad
B_{\alpha }^{(1)}(\lambda )\left\vert 0_{1}\right\rangle ^{(1)}\neq \left\{
0,\left\vert 0_{1}\right\rangle ^{(1)}\right\} ,\quad \alpha \neq \beta
=1,2,...,N_{1}-2.  \label{one.14}
\end{eqnarray}%
with eigenvalue%
\begin{equation}
\Lambda _{1}^{(1)}(\lambda ,\{\lambda _{1}\}|0)=X_{1}^{(1)}(\lambda
)+(N_{1}-2)X_{2}^{(1)}(\lambda )+X_{3}^{(1)}(\lambda )  \label{one.15}
\end{equation}%
where $X_{a}^{(1)}(\lambda ),\ a=1,2,3.$ are given by (\ref{one.7a}).

The condition that $\mathcal{F}_{1}$ ought to be an eigenvector of $\tau
_{1}^{(1)}(\lambda )$ requires the diagonalization of $\tau
_{1}^{(1)}(\lambda )$, which can be carried out by a second Bethe ansatz. \
There are two candidates for $\mathcal{F}_{1}\in \mathcal{H}^{(1)}$: the own
reference state $\left\vert 0_{1}\right\rangle ^{(1)}$ and the one-particle
excitation $\mathcal{B}^{(1)}(\lambda _{1}^{(1)})\left\vert
0_{1}\right\rangle ^{(1)}$.

The choice $\mathcal{F}_{1}=\left\vert 0_{1}\right\rangle ^{(1)}$ implies in
a particular linear combination for the one-particle state (\ref{one.1}) and
in this case $\Psi _{1}(\lambda _{1})$ is an eigenfunction of $\tau
_{L}(\lambda )$ for the B$_{n}^{(1)}$ and A$_{2n-1}^{(2)}$ with the
eigenvalue expression equal to (\ref{one.6}) and the Bethe equation equal to
(\ref{one.7}). However, the second choice seems to be the most general but
we will not consider it now. This case will be presented later in a more
general context where the multi-particle state is treated.

Notice that we have not yet given an explicit rule to eliminate the unwanted
terms, until now they were merely canceled out. In order to find such rule
we will consider the two-particle state with more details.

\section{The two-particle Bethe state}

In analogy with the scalar case \cite{Tarasov, ALS} we have two types of
linearly independent \ contributions $\mathcal{B}^{(l)}\otimes \mathcal{B}%
^{(l)}$ and $B_{N_{l}-1}$ for the vector $\Phi _{2}^{(l)}(\lambda
_{1},\lambda _{2})$. We seek vectors in the form%
\begin{equation}
\Phi _{2}^{(l)}(\lambda _{1},\lambda _{2})=\mathcal{B}^{(l)}(\lambda
_{1})\otimes \mathcal{B}^{(l)}(\lambda _{2})+B_{N_{l}-1}(\lambda _{1})\Gamma
(\lambda _{1},\lambda _{2})  \label{two.1}
\end{equation}%
where $\Gamma (\lambda _{1},\lambda _{2})$ is an operator-valued vector
which has to be fixed such that $\Phi _{2}^{(l)}(\lambda _{1},\lambda _{2})$
is unique.

It was demonstrated in \cite{Tarasov} that $\Phi _{2}^{(l)}(\lambda
_{1},\lambda _{2})$ is unique provided it is ordered in a normal way: in
general, the operator-valued vector $\Phi _{m}^{(l)}(\lambda _{1},\ldots
,\lambda _{m})$ is composite of normal-ordered monomials. A monomial is
normally ordered if in it all elements $\mathcal{B}^{(l)}$ , $\mathcal{B}%
^{\ast (l)}$ and $B_{N_{l}-1}$ are on the left, and all elements $\mathcal{C}%
^{(l)}$, $\mathcal{C}^{\ast (l)}$ and $C_{N_{l}-1}$ are on the right of all
elements $A_{1}^{(l)}$, $\mathcal{D}^{(l)}$ and $A_{3}^{(l)}$. Moreover, all
elements of one given type have standard ordering, i.e., $T_{\alpha
_{1}\beta _{1}}^{(l)}(\lambda _{1})T_{\alpha _{2}\beta _{2}}^{(l)}(\lambda
_{2})\cdots T_{\alpha _{mn}\beta _{m}}^{(l)}(\lambda _{m})$.

Now we recall the intertwining relation (\ref{fcr.5}) to solve the equations
of the block $B[1,5]$ in order to get the following commutation relation%
\begin{eqnarray}
\mathcal{B}^{(l)}(\lambda )\otimes \mathcal{B}^{(l)}(\mu ) &=&\left( 
\mathcal{B}^{(l)}(\mu )\otimes \mathcal{B}^{(l)}(\lambda )-\frac{\hat{Y}%
_{N_{l}2}^{(l)}(\mu -\lambda )}{y_{N_{l}N_{l}}^{(l)}(\mu -\lambda )}%
B_{N_{l}-1}(\mu )A_{1}^{(l)}(\lambda )\right) \frac{S^{(l+1)}(\lambda -\mu )%
}{x_{1}^{(l+1)}(\lambda -\mu )}  \notag \\
&&+B_{N_{l}-1}(\lambda )A_{1}^{(l)}(\mu )\frac{\hat{Y}_{N_{l}2}(\lambda -\mu
)}{y_{N_{l}N_{l}}(\lambda -\mu )},  \label{two.2}
\end{eqnarray}%
from which we can see that (\ref{two.1}) will be normally ordered if only if
it satisfies the property%
\begin{equation}
\Phi _{2}^{(l)}(\lambda _{1},\lambda _{2})=\Phi _{2}^{(l)}(\lambda
_{2},\lambda _{1})\frac{S^{(l+1)}(\lambda _{1}-\lambda _{2})}{%
x_{1}^{(l+1)}(\lambda _{1}-\lambda _{2})}  \label{two.3}
\end{equation}%
This condition fixes $\Gamma (\lambda _{1},\lambda _{2})$ and our vector for
the two-particle state has the form%
\begin{equation}
\Phi _{2}^{(l)}(\lambda _{1},\lambda _{2})=\mathcal{B}^{(l)}(\lambda
_{1})\otimes \mathcal{B}^{(l)}(\lambda _{2})+B_{N_{l}-1}(\lambda
_{1})A_{1}^{(l)}(\lambda _{2})\frac{Y_{N_{l}2}^{(l)}(\lambda _{1}-\lambda
_{2})}{y_{N_{l}N_{l}}^{(l)}(\lambda _{1}-\lambda _{2})}  \label{two.4}
\end{equation}%
Here we notice that the condition (\ref{two.3}) must be generalized to
include multi-particle state and it will play the main role in the
elimination rules of the unwanted terms.

The action of the transfer matrix on this vector is more laborious. In
addition to (\ref{fcr.8}), (\ref{fcr.10}) and (\ref{one.2}) we appeal to (%
\ref{fcr.5}) to derive nine \ further commutation relations.

Due to the presence of the scalar $B_{N_{l}-1}$ in (\ref{two.4}) we have
solve the block $B[1,7]$ of (\ref{fcr.5})\ in order to get its commutation
relations with $A_{1}$ and $A_{3}$ in a given \emph{layer} $l$ $\ $ 
\begin{eqnarray}
A_{1}^{(l)}(\lambda )B_{N_{l}-1}(\mu ) &=&\frac{x_{1}^{(l)}(\mu -\lambda )}{%
y_{N_{l}N_{l}}(\mu -\lambda )}B_{N_{l}-1}(\mu )A_{1}^{(l)}(\lambda )-\frac{%
y_{1N_{l}}^{(l)}(\mu -\lambda )}{y_{N_{l}N_{l}}^{(l)}(\mu -\lambda )}%
B_{N_{l}-1}(\lambda )A_{1}^{(l)}(\mu )  \notag \\
&&-\mathcal{B}^{(l)}(\lambda )\otimes \mathcal{B}^{(l)}(\mu )\frac{\hat{Y}%
_{2N_{l}}^{(l)}(\mu -\lambda )}{y_{N_{l}N_{l}}^{(l)}(\mu -\lambda )},
\label{two.5}
\end{eqnarray}%
and%
\begin{eqnarray}
A_{3}^{(l)}(\lambda )B_{N_{l}-1}(\mu ) &=&\frac{x_{1}^{(l)}(\lambda -\mu )}{%
y_{N_{l}N_{l}}(\lambda -\mu )}B_{N_{l}-1}(\mu )A_{3}^{(l)}(\lambda )-\frac{%
y_{N_{l}1}^{(l)}(\lambda -\mu )}{y_{N_{l}N_{l}}^{(l)}(\lambda -\mu )}%
B_{N_{l}-1}(\lambda )A_{3}^{(l)}(\mu )  \notag \\
&&-\frac{\hat{Y}_{N_{l}2}^{(l)}(\lambda -\mu )}{y_{N_{l}N_{l}}^{(l)}(\lambda
-\mu )}\mathcal{B}^{\ast (l)}(\lambda )\otimes \mathcal{B}^{\ast (l)}(\mu ),
\label{two.6}
\end{eqnarray}%
The block $B[2,6]$, $B[4,6]$ and $B[4,8]$ can be used to derive the relation 
\begin{eqnarray}
\mathcal{D}^{(l)}(\lambda )B_{N_{l}-1}(\mu ) &=&z^{(l)}(\lambda -\mu
)z^{(l)}(\mu -\lambda )B_{N_{l}-1}(\mu )\mathcal{D}^{(l)}(\lambda )+\frac{%
x_{4}^{(l)}(\lambda -\mu )^{2}}{x_{2}^{(l)}(\lambda -\mu )^{2}}%
B_{N_{l}-1}(\lambda )\mathcal{D}^{(l)}(\mu )  \notag \\
&&-\frac{x_{4}^{(l)}(\lambda -\mu )}{x_{2}^{(l)}(\lambda -\mu )}\left[ 
\mathcal{B}^{(l)}(\lambda )\otimes \mathcal{B}^{\ast (l)}(\mu )-\mathcal{B}%
^{\ast (l)}(\lambda )\otimes \mathcal{B}^{(l)}(\mu )\right] .  \label{two.7}
\end{eqnarray}%
Again, we must work out (\ref{two.7} in order to get the commutation
relations of $B_{N_{l}-1}$ with the scalar $\mathrm{Tr}_{a}[\mathcal{D}%
^{(l)}(\lambda )]$%
\begin{eqnarray}
\mathrm{Tr}_{a}[\mathcal{D}^{(l)}(\lambda )]B_{N_{l}-1}(\mu )
&=&z^{(l)}(\lambda -\mu )z^{(l)}(\mu -\lambda )B_{N-1}(\mu )\mathrm{Tr}_{a}[%
\mathcal{D}^{(l)}(\lambda )]+\frac{x_{4}^{(l)}(\lambda -\mu )^{2}}{%
x_{2}^{(l)}(\lambda -\mu )^{2}}B_{N-1}(\lambda )\mathrm{Tr}_{a}[\mathcal{D}%
^{(l)}(\mu )]  \notag \\
&&-\frac{x_{4}^{(l)}(\lambda -\mu )}{x_{2}^{(l)}(\lambda -\mu )}(\mathcal{B}%
^{(l)}(\lambda )\mathcal{B}^{\ast (l)}(\mu )-\mathrm{Tr}_{a}[\mathcal{B}%
^{\ast (l)}(\lambda )\mathcal{B}^{(l)}(\mu )])  \label{two.8}
\end{eqnarray}%
where we have used the identity 
\begin{equation}
1-\frac{x_{4}^{(l)}(\lambda -\mu )}{x_{2}^{(l)}(\lambda -\mu )}\frac{%
x_{3}^{(l)}(\lambda -\mu )}{x_{2}^{(l)}(\lambda -\mu )}=z^{(l)}(\lambda -\mu
)z^{(l)}(\mu -\lambda )  \label{two.8a}
\end{equation}%
Note also that these scalar relations will survive the reduction mechanism
for the six-vertex models previously presented.

Since we need a second commutation step in order to that $\tau _{L}^{(l)}$
hits the reference state, we compute the commutation relations among
creation and annihilation operators 
\begin{equation}
\mathcal{C}^{(l)}(\lambda )\otimes \mathcal{B}^{(l)}(\mu )=\mathcal{B}%
^{(l)}(\mu )\otimes \mathcal{C}^{(l)}(\lambda )-\frac{x_{4}^{(l)}(\lambda
-\mu )}{x_{2}^{(l)}(\lambda -\mu )}\left[ A_{1}^{(l)}(\lambda )D^{(l)}(\mu
)-A_{1}^{(l)}(\mu )\mathcal{D}^{(l)}(\lambda ))\right]  \label{two.9}
\end{equation}%
\begin{eqnarray}
\mathcal{C}^{(l)}(\lambda )B_{N_{l}-1}(\mu ) &=&\frac{x_{2}^{(l)}(\mu
-\lambda )}{y_{N_{l}N_{l}}^{(l)}(\mu -\lambda )}B_{N_{l}-1}(\mu )\mathcal{C}%
^{(l)}(\lambda )+\frac{x_{3}^{(l)}(\mu -\lambda )}{y_{N_{l}N_{l}}^{(l)}(\mu
-\lambda )}\mathcal{B}^{\ast (l)}(\lambda )A_{1}^{(l)}(\mu )  \notag \\
&&-\frac{y_{1N_{l}}^{(l)}(\mu -\lambda )}{y_{N_{l}N_{l}}^{(l)}(\mu -\lambda )%
}\mathcal{B}^{\ast (l)}(\mu )A_{1}^{(l)}(\lambda )-\mathcal{D}^{(l)}\otimes 
\mathcal{B}^{(l)}(\lambda )C_{N_{l}-1}(\mu )\frac{\hat{Y}_{2N_{l}}^{(l)}(\mu
-\lambda )}{y_{N_{l}N_{l}}^{(l)}(\mu -\lambda )},  \label{two.10}
\end{eqnarray}%
both relations were obtained from the blocks $B[2,2]$ and $B[4,7]$,
respectively. \ We also need the commutation of $B_{N_{l}-1}(\lambda )$ with
the other creation operators%
\begin{equation}
B_{N_{l}-1}(\lambda )\mathcal{B}^{(l)}(\mu )=z^{(l)}(\mu -\lambda )\mathcal{B%
}^{(l)}(\mu )B_{N_{l}-1}(\lambda )-\frac{x_{4}^{(l)}(\mu -\lambda )}{%
x_{2}^{(l)}(\mu -\lambda )}\mathcal{B}^{(l)}(\lambda )B_{N_{l}-1}(\mu ),
\end{equation}%
and%
\begin{equation}
\mathcal{B}^{\ast (l)}(\lambda )B_{N_{l}-1}(\mu )=z^{(l)}(\lambda -\mu
)B_{N_{l}-1}(\mu )\mathcal{B}^{\ast (l)}(\lambda )-\frac{x_{4}^{(l)}(\lambda
-\mu )}{x_{2}^{(l)}(\lambda -\mu )}B_{N_{l}-1}(\lambda )\mathcal{B}^{\ast
(l)}(\mu ).  \label{two.11}
\end{equation}%
which are obtained from the blocks $B[1,6]$, $B[1,8]$ and $B[1,10]$.

Here we observe that the final action of $\tau _{L}^{(l)}(\lambda )$ on
normally ordered vectors must be normal ordered. This implies an increasing
use of commutation relations needed for the eigenvalue problem. For
instance, the action of the scalar $A_{1}^{(l)}(\lambda )$ on the vector $%
\Phi _{2}^{(l)}(\lambda _{1},\lambda _{2})$ has its normal ordered form
given by%
\begin{eqnarray}
A_{1}^{(l)}(\lambda )\Phi _{2}^{(l)}(\lambda _{1},\lambda _{2})
&=&\prod\limits_{k=1}^{2}z^{(l)}(\lambda _{k}-\lambda )\Phi
_{2}^{(l)}(\lambda _{1},\lambda _{2})A_{1}^{(l)}(\lambda )  \notag \\
&&-\frac{x_{3}^{(l)}(\lambda _{1}-\lambda )}{x_{2}^{(l)}(\lambda
_{1}-\lambda )}z^{(l)}(\lambda _{21})\mathcal{B}^{(l)}(\lambda )\otimes 
\mathcal{B}^{(l)}(\lambda _{2})A_{1}^{(l)}(\lambda _{1})  \notag \\
&&-\frac{x_{3}^{(l)}(\lambda _{2}-\lambda )}{x_{2}^{(l)}(\lambda
_{2}-\lambda )}z^{(l)}(\lambda _{12})\mathcal{B}^{(l)}(\lambda )\otimes 
\mathcal{B}^{(l)}(\lambda _{1})A_{1}^{(l)}(\lambda _{2})\frac{%
S_{12}^{(l+1)}(\lambda _{1}-\lambda _{2})}{x_{1}^{(l+1)}(\lambda
_{1}-\lambda _{2})}  \notag \\
&&+B_{N_{l}-1}(\lambda )\mathbf{G}_{21}^{(l)}(\lambda ,\lambda _{1},\lambda
_{2})A_{1}^{(l)}(\lambda _{1})A_{1}^{(l)}(\lambda _{2})+\cdots
\label{two.12}
\end{eqnarray}%
where we have defined a row vector with $N_{l+1}^{2}$ entries%
\begin{equation}
\mathbf{G}_{21}^{(l)}(\lambda ,\lambda _{1},\lambda _{2})=\frac{%
x_{3}^{(l)}(\lambda _{2}-\lambda )}{x_{2}^{(l)}(\lambda _{2}-\lambda )}\frac{%
\hat{Y}_{N_{l}\ 2}^{(l)}(\lambda -\lambda _{1})}{y_{N_{l}N_{l}}^{(l)}(%
\lambda -\lambda _{1})}\frac{S^{(l+1)}(\lambda _{1}-\lambda )}{%
x_{2}^{(l+1)}(\lambda _{1}-\lambda )}+\frac{y_{1N_{l}}^{(l)}(\lambda
_{1}-\lambda )}{y_{N_{l}N_{l}}^{(l)}(\lambda _{1}-\lambda )}\frac{\hat{Y}%
_{N_{l}\ 2}^{(l)}(\lambda _{1}-\lambda _{2})}{y_{N_{l}N_{l}}^{(l)}(\lambda
_{1}-\lambda _{2})}  \label{two.13}
\end{equation}%
which satisfies the cyclic permutation property%
\begin{equation}
\mathbf{G}_{21}^{(l)}(\lambda ,\lambda _{1},\lambda _{2})=\mathbf{G}%
_{21}^{(l)}(\lambda ,\lambda _{2},\lambda _{1})\frac{S^{(l+1)}(\lambda
_{1}-\lambda _{2})}{x_{1}^{(l+1)}(\lambda _{1}-\lambda _{2})}  \label{two.14}
\end{equation}%
Observe the mix of \emph{layer}s with respect the Boltzmann weights of $%
\mathbf{G}_{21}^{(l)}$ which really acts on the $l+1$ \emph{layer}.

The action of the scalar $A_{3}^{(l)}(\lambda )$ on the vector $\Phi
_{2}^{(l)}(\lambda _{1},\lambda _{2})$ has a very similar normal ordered form%
\begin{eqnarray}
A_{3}^{(l)}(\lambda )\Phi _{2}^{(l)}(\lambda _{1},\lambda _{2})
&=&\prod\limits_{k=1}^{2}\frac{x_{2}^{(l)}(\lambda -\lambda _{k})}{%
y_{N_{l}N_{l}}^{(l)}(\lambda -\lambda _{k})}\ \Phi _{2}^{(l)}(\lambda
_{1},\lambda _{2})A_{3}^{(l)}(\lambda )  \notag \\
&&-z^{(l)}(\lambda _{21})\frac{\hat{Y}_{N_{l}2}^{(l)}(\lambda -\lambda _{2})%
}{y_{N_{l}Nl}(\lambda -\lambda _{2})}\ \mathcal{B}^{\ast (l)}(\lambda
)\otimes \mathcal{B}^{(l)}(\lambda _{1})\otimes \mathcal{D}^{(l)}(\lambda
_{2})  \notag \\
&&-z^{(l)}(\lambda _{12})\frac{\hat{Y}_{N_{l}2}^{(l)}(\lambda -\lambda _{1})%
}{y_{N_{l}N_{l}}^{(l)}(\lambda -\lambda _{1})}\ \mathcal{B}^{\ast
(l)}(\lambda )\otimes \mathcal{B}^{(l)}(\lambda _{2})\otimes \mathcal{D}%
^{(l)}(\lambda _{1})\frac{S_{12}^{(l+1)}(\lambda _{1}-\lambda _{2})}{%
x_{1}^{(l+1)}(\lambda _{1}-\lambda _{2})}  \notag \\
&&+B_{N_{l}-1}(\lambda )\mathbf{H}_{21}^{(l)}(\lambda ,\lambda _{1},\lambda
_{2})\mathcal{D}^{(l)}(\lambda _{1})\otimes \mathcal{D}^{(l)}(\lambda
_{2})+\cdots  \label{two.15}
\end{eqnarray}%
where the vector $\mathbf{H}_{21}^{(l)}$ is given by 
\begin{equation}
\mathbf{H}_{21}^{(l)}(\lambda ,\lambda _{1},\lambda _{2})=\frac{%
y_{N_{l}1}^{(l)}(\lambda -\lambda _{1})}{y_{N_{l}N_{l}}^{(l)}(\lambda
-\lambda _{1})}\frac{\hat{Y}_{N_{l}\ 2}^{(l)}(\lambda _{1}-\lambda _{2})}{%
y_{N_{l}\ N_{l}}^{(l)}(\lambda _{1}-\lambda _{2})}-\frac{x_{3}^{(l)}(\lambda
-\lambda _{1})}{y_{N_{l}N_{l}}^{(l)}(\lambda -\lambda _{1})}\frac{\hat{Y}%
_{N_{l}2}^{(l)}(\lambda -\lambda _{2})}{y_{N_{l}N_{l}}^{(l)}(\lambda
-\lambda _{2})}  \label{two.16}
\end{equation}%
and satisfies the cyclic permutation property%
\begin{equation}
\mathbf{H}_{21}^{(l)}(\lambda ,\lambda _{1},\lambda _{2})=\mathbf{H}%
_{21}^{(l)}(\lambda ,\lambda _{2},\lambda _{1})\frac{S^{(l+1)}(\lambda
_{1}-\lambda _{2})}{x_{1}^{(l+1)}(\lambda _{1}-\lambda _{2})}.
\label{two.17}
\end{equation}%
Finally, the action of the scalar $\mathrm{Tr}_{a}[\mathcal{D}^{(l)}(\lambda
)]$ on the vector $\Phi _{2}^{(l)}(\lambda _{1},\lambda _{2})$ is a little
bit different%
\begin{eqnarray}
&&\mathrm{Tr}_{a}[\mathcal{D}^{(l)}(\lambda )]\Phi _{2}^{(l)}(\lambda
_{1},\lambda _{2})=\Phi _{2}^{(l)}(\lambda _{1},\lambda _{2})\mathrm{Tr}_{a}[%
\frac{\mathcal{L}_{a2}^{(l+1)}(\lambda -\lambda _{2})}{x_{2}^{(l+1)}(\lambda
-\lambda _{2})}\frac{\mathcal{L}_{a1}^{(l+1)}(\lambda -\lambda _{1})}{%
x_{2}^{(l+1)}(\lambda -\lambda _{1)}}\mathcal{D}^{(l)}(\lambda )]  \notag \\
&&-z^{(l)}(\lambda _{1}-\lambda _{2})\frac{x_{4}^{(l)}(\lambda -\lambda _{1})%
}{x_{2}^{(l)}(\lambda -\lambda _{1})}B^{(l)}(\lambda )\otimes \mathcal{B}%
^{(l)}(\lambda _{2})[\mathcal{D}^{(l)}(\lambda _{1})\otimes \mathbf{1}^{(l)}]%
\frac{\mathcal{R}^{(l+1)}(\lambda _{1}-\lambda _{2})}{x_{1}^{(l+1)}(\lambda
_{1}-\lambda _{2})}  \notag \\
&&-z^{(l)}(\lambda _{2}-\lambda _{1})\frac{x_{4}^{(l)}(\lambda -\lambda _{2})%
}{x_{2}^{(l)}(\lambda -\lambda _{2})}B^{(l)}(\lambda )\otimes \mathcal{B}%
^{(l)}(\lambda _{1})[\mathcal{D}^{(l)}(\lambda _{2})\otimes \mathbf{1}^{(l)}]
\notag \\
&&+z^{(l)}(\lambda _{2}-\lambda _{1})\frac{\hat{Y}_{N_{l}\ 2}^{(l)}(\lambda
-\lambda _{1})}{y_{N_{l}N_{l}}^{(l)}(\lambda -\lambda _{1})}[B^{\ast
(l)}(\lambda )\otimes \mathcal{B}^{(l)}(\lambda _{2})\otimes \mathbf{1}%
^{(l)}]A_{1}^{(l)}(\lambda _{1})  \notag \\
&&+z^{(l)}(\lambda _{1}-\lambda _{2})\frac{\hat{Y}_{N_{l}\ 2}^{(l)}(\lambda
-\lambda _{2})}{y_{N_{l}N_{l}}^{(l)}(\lambda -\lambda _{2})}[B^{\ast
(l)}(\lambda )\otimes \mathcal{B}^{(l)}(\lambda _{1})\otimes \mathbf{1}%
^{(l)}]\frac{\mathcal{R}_{12}^{(l+1)}(\lambda _{1}-\lambda _{2})}{%
x_{1}^{(l+1)}(\lambda _{1}-\lambda _{2})}A_{1}^{(l)}(\lambda _{2})  \notag \\
&&+B_{N_{l}-1}(\lambda )\mathbf{Y}_{21}^{(l)}(\lambda ,\lambda _{1},\lambda
_{2})[\mathcal{D}^{(l)}(\lambda _{2})\otimes \mathbf{1}^{(l)}]A_{1}^{(l)}(%
\lambda _{1})  \notag \\
&&+B_{N_{l}-1}(\lambda )\mathbf{Y}_{21}^{(l)}(\lambda ,\lambda _{2},\lambda
_{1})[\mathcal{D}^{(l)}(\lambda _{1})\otimes \mathbf{1}^{(l)}]A_{1}^{(l)}(%
\lambda _{2})\frac{\mathcal{R}_{12}^{(l+1)}(\lambda _{1}-\lambda _{2})}{%
x_{1}^{(l+1)}(\lambda _{1}-\lambda _{2})}+\cdots  \label{two.18}
\end{eqnarray}%
where the vector $\mathbf{Y}_{21}^{(l)}$ is given by%
\begin{equation}
\mathbf{Y}_{21}^{(l)}(\lambda ,\lambda _{1},\lambda _{2})=[z^{(l)}(\lambda
-\lambda _{1})\frac{x_{4}^{(l)}(\lambda -\lambda _{2})}{x_{2}^{(l)}(\lambda
-\lambda _{2})}-\frac{x_{4}^{(l)}(\lambda -\lambda _{1})}{%
x_{2}^{(l)}(\lambda -\lambda _{1})}\frac{x_{4}^{(l)}(\lambda _{1}-\lambda
_{2})}{x_{2}^{(l)}(\lambda _{1}-\lambda _{2})}]\frac{\hat{Y}%
_{N_{l}2}^{(l)}(\lambda -\lambda _{1})}{y_{N_{l}N_{l}}^{(l)}(\lambda
-\lambda _{1})}  \label{two.19}
\end{equation}%
and also satisfies the cyclic permutation property%
\begin{equation}
\mathbf{Y}_{21}^{(l)}(\lambda ,\lambda _{1},\lambda _{2})=\mathbf{Y}%
_{21}^{(l)}(\lambda ,\lambda _{2},\lambda _{1})\frac{S^{(l+1)}(\lambda
_{1}-\lambda _{2})}{x_{1}^{(l+1)}(\lambda _{1}-\lambda _{2})}  \label{two.20}
\end{equation}%
The ellipses in these expressions denote normally ordered terms containing
factors of the type $\mathcal{C}^{(l)}$, $\mathcal{C}^{\ast (l)}$ and $%
C_{N_{l}-1}$.

\ It should be worth note that we have used some matrix identities to derive
(\ref{two.12})-(\ref{two.18}) 
\begin{eqnarray}
\qquad \frac{S^{(l+1)}(\lambda _{ab})}{x_{1}^{(l+1)}(\lambda _{ab})}\frac{%
S^{(l+1)}(\lambda _{ba})}{x_{1}^{(l+1)}(\lambda _{ba})} &=&I,  \notag \\
\frac{x_{3}^{(l)}(\lambda _{cb})}{x_{2}^{(l)}(\lambda _{cb})}\frac{%
S^{(l+1)}(\lambda _{ab})}{x_{2}^{(l+1)}(\lambda _{ab})}-\frac{\hat{Y}%
_{2N}^{(l)}(\lambda _{ab})}{y_{NN}^{(l)}(\lambda _{ab})}\frac{\hat{Y}%
_{N2}^{(l)}(\lambda _{ac})}{y_{NN}^{(l)}(\lambda _{ac})} &=&\frac{%
x_{3}^{(l)}(\lambda _{ab})}{x_{2}^{(l)}(\lambda _{ab})}\frac{%
x_{3}^{(l)}(\lambda _{ca})}{x_{2}^{(l)}(\lambda _{ca})}I+\frac{%
x_{3}^{(l)}(\lambda _{cb})}{x_{2}^{(l)}(\lambda _{cb})}\frac{%
S^{(l+1)}(\lambda _{ac})}{x_{2}^{(l+1)}(\lambda _{ac})}  \notag \\
\frac{x_{2}^{(l)}(\lambda _{ab})}{y_{NN}^{(l)}(\lambda _{ab})}\frac{\hat{Y}%
_{N_{l}2}^{(l)}(\lambda _{ac})}{y_{N_{l}N_{l}}^{(l)}(\lambda _{ac})}+\frac{%
x_{3}^{(l)}(\lambda _{cb})}{x_{2}^{(l)}(\lambda _{cb})}\frac{\hat{Y}%
_{N_{l}2}^{(l)}(\lambda _{ab})}{y_{N_{l}N_{l}}^{(l)}(\lambda _{ab})}
&=&z^{(l)}(\lambda _{cb})\frac{\hat{Y}_{N_{l}2}^{(l)}(\lambda _{ac})}{%
y_{N_{l}N_{l}}^{(l)}(\lambda _{ac})},\qquad (a\neq b\neq c)  \label{two.20a}
\end{eqnarray}%
where $I$ is a $N_{l+1}^{2}$ by $N_{l+1}^{2}$ matrix identity and $\lambda
_{ab}=\lambda _{a}-\lambda _{b}$.

Let us begin with the eigenvalue problem for the two-particle state in a
homogeneous $L$ site lattice which is defined by the linear combination%
\begin{equation}
\Psi _{2}(\lambda _{1},\lambda _{2})=\Phi _{2}(\lambda _{1},\lambda _{2})%
\mathcal{F}_{2}\left\vert 0_{L}\right\rangle  \label{two.21}
\end{equation}%
where $\Phi _{2}$ is given by (\ref{two.4})\ and the vector $\mathcal{F}_{2}$
has components $f^{\alpha \beta }\in C$, ($\alpha ,\beta =1,...,N_{l+1}$).

The action of $\tau _{L}(\lambda )$ on $\Psi _{2}(\lambda _{1},\lambda _{2})$
can now be computed \ using (\ref{two.12}), (\ref{two.15}) and (\ref{two.18}%
): 
\begin{eqnarray}
\tau _{L}(\lambda )\Psi _{2}(\lambda _{1},\lambda _{2}) &=&X_{1}(\lambda
)\prod\limits_{k=1}^{2}z(\lambda _{k}-\lambda )\Psi _{2}(\lambda
_{1},\lambda _{2})+X_{3}(\lambda )\prod\limits_{k=1}^{2}\frac{x_{2}(\lambda
-\lambda _{k})}{y_{NN}(\lambda -\lambda _{k})}\ \Psi _{2}(\lambda
_{1},\lambda _{2})  \notag \\
&&+X_{2}(\lambda )\Phi _{2}(\lambda _{1},\lambda _{2})\frac{1}{%
X_{2}^{(1)}(\lambda )}\left\{ \tau _{2}^{(1)}(\lambda )\mathcal{F}%
_{2}\right\} \left\vert 0_{L}\right\rangle  \notag \\
&&-\frac{x_{3}(\lambda _{10})}{x_{2}(\lambda _{10})}\mathcal{B}(\lambda
)\otimes \Phi _{1}(\lambda _{2})[X_{1}(\lambda _{1})z(\lambda
_{21})-X_{2}(\lambda _{1})z(\lambda _{12})\frac{\mathcal{R}^{(1)}(\lambda
_{12})}{x_{1}^{(1)}(\lambda _{12})}]\mathcal{F}_{2}\left\vert
0_{L}\right\rangle  \notag \\
&&-\frac{x_{3}(\lambda _{20})}{x_{2}(\lambda _{20})}\mathcal{B}(\lambda
)\otimes \Phi _{1}(\lambda _{1})[X_{1}(\lambda _{2})z(\lambda _{12})\frac{%
S_{12}^{(1)}(\lambda _{12})}{x_{1}^{(1)}(\lambda _{12})}-X_{2}(\lambda
_{2})z(\lambda _{21})]\mathcal{F}_{2}\left\vert 0_{L}\right\rangle  \notag \\
&&-\frac{\hat{Y}_{N2}(\lambda _{02})}{y_{NN}(\lambda _{02})}\ \mathcal{B}%
^{\ast }(\lambda )\otimes \Phi _{1}(\lambda _{1})\otimes 1[X_{2}(\lambda
_{2})z(\lambda _{21})-X_{1}(\lambda _{2})z(\lambda _{12})\frac{\mathcal{R}%
_{12}^{(1)}(\lambda _{12})}{x_{1}^{(1)}(\lambda _{12})}]\mathcal{F}%
_{2}\left\vert 0_{L}\right\rangle  \notag \\
&&-\frac{\hat{Y}_{N2}(\lambda _{01})}{y_{NN}(\lambda _{01})}\ \mathcal{B}%
^{\ast }(\lambda )\otimes \Phi _{1}(\lambda _{2})\otimes 1[X_{2}(\lambda
_{1})z(\lambda _{12})\frac{S_{12}^{(1)}(\lambda _{12})}{x_{1}^{(1)}(\lambda
_{12})}-X_{1}(\lambda _{1})z(\lambda _{21})]\mathcal{F}_{2}\left\vert
0_{L}\right\rangle  \notag \\
&&+B_{N-1}(\lambda )\{\mathbf{G}_{21}(\lambda ,\lambda _{1},\lambda
_{2})X_{1}(\lambda _{1})X_{1}(\lambda _{2})+\mathbf{H}_{21}(\lambda ,\lambda
_{1},\lambda _{2})X_{2}(\lambda _{1})X_{2}(\lambda _{2})  \notag \\
&&+\mathbf{Y}_{21}(\lambda ,\lambda _{1},\lambda _{2})X_{2}(\lambda
_{2})X_{1}(\lambda _{1})+\mathbf{Y}_{21}(\lambda ,\lambda _{2},\lambda
_{1})X_{2}(\lambda _{1})X_{1}(\lambda _{2})\frac{\mathcal{R}%
_{12}^{(1)}(\lambda _{12})}{x_{1}^{(1)}(\lambda _{12})}\}\mathcal{F}%
_{2}\left\vert 0_{L}\right\rangle  \notag \\
&&  \label{two.21b}
\end{eqnarray}%
where we have used the rapidity difference notation with $\lambda
_{0}=\lambda $ and the definitions%
\begin{equation}
\Phi _{1}(\lambda _{i})=\mathcal{B}(\lambda _{i}),\qquad X_{i}^{(1)}(\lambda
)=\prod\limits_{k=1}^{2}x_{i}^{(1)}(\lambda -\lambda _{k})\qquad (i=1,2)
\label{two.21c}
\end{equation}%
Moreover, the unwanted terms were combined and we are presenting the
two-site inhomogeneous transfer matrix for the first \emph{layer}%
\begin{equation}
\tau _{2}^{(1)}(\lambda )=\mathrm{Tr}_{a}\left[ \mathcal{L}%
_{a2}^{(1)}(\lambda -\lambda _{2})\mathcal{L}_{a1}^{(1)}(\lambda -\lambda
_{1})\right] .  \label{two.22}
\end{equation}

Before we see how the unwanted terms can be canceled, let us first to
consider the eigenvalue problem%
\begin{equation}
\tau _{2}^{(1)}(\lambda )\mathcal{F}_{2}=\Lambda _{2}^{(1)}(\lambda
,\{\lambda _{i}\}|\{\lambda _{i}^{(1)}\})\mathcal{F}_{2}\qquad (i=1,2),
\label{two.23}
\end{equation}%
where our choice for the vector $\mathcal{F}_{2}$ is implicit 
\begin{equation}
\mathcal{F}_{2}=\Phi _{2}^{(1)}(\lambda _{1}^{(1)},\lambda
_{2}^{(1)})\left\vert 0_{2}\right\rangle ^{(1)}.  \label{two.24}
\end{equation}%
and above $\Phi _{2}^{(1)}$ is given by (\ref{two.4}) and $\left\vert
0_{2}\right\rangle ^{(1)}$ by (\ref{eig.8}). From these results we have an
incomplete expression for the eigenvalue 
\begin{equation}
\Lambda _{L}(\lambda |\{\lambda _{1},\lambda _{2}\})=X_{1}(\lambda
)\prod\limits_{k=1}^{2}z(\lambda _{k}-\lambda )+X_{3}(\lambda
)\prod\limits_{k=1}^{2}\frac{x_{2}(\lambda -\lambda _{k})}{y_{NN}(\lambda
-\lambda _{k})}+X_{2}(\lambda )\frac{\Lambda _{2}^{(1)}(\lambda ,\{\lambda
_{i}\}|\{\lambda _{i}^{(1)}\})}{X_{2}^{(1)}(\lambda )}  \label{two.25}
\end{equation}%
because $\Lambda _{2}^{(1)}(\lambda ,\{\lambda _{i}\}|\{\lambda
_{i}^{(1)}\}) $ is still unknown.

\bigskip At this point it is worth giving further information about the
shorthand notation used in this text:

\begin{itemize}
\item $\tau _{L}(\lambda )$ is a row-to-row homogeneous transfer matrix in a 
$L$ site lattice and $\Lambda _{m}(\lambda |\{\lambda _{i}\})$ is the
eigenvalue of $\tau _{L}(\lambda )$ for the $m$-particle state with
rapidities $\lambda _{i}$, $i=1,2,...m$. Acting with $\tau _{L}(\lambda )$
on its reference state $\left\vert 0_{L}\right\rangle $ we have the factors%
\begin{equation}
X_{a}(\lambda )=[x_{a}(\lambda )]^{L},\quad a=1,2\qquad \mathrm{and}\qquad
X_{3}(\lambda )=[y_{NN}(\lambda )]^{L}  \label{two.26}
\end{equation}

\item $\tau _{m}^{(l)}(\lambda )\circeq \tau _{m}^{(l)}(\lambda ,\{\lambda
_{i}^{(l-1)}\})$ is a row-to-row inhomogeneous transfer matrix in a $m$ site
lattice with inhomogeneous parameters $\lambda _{i}^{(l-1)},i=1,2,...,m.$
Its eigenvalue for the $m$-particle state is denoted by $\Lambda
_{m}^{(l)}(\lambda ,\{\lambda _{i}^{(l-1)}\}|\{\lambda _{i}^{(l)}\})$ where $%
\lambda _{i}^{(l)}$ are rapidities of the particles and $\lambda
_{i}^{(l-1)} $ are inhomogeneous parameters of the corresponding lattice.
Acting with $\tau _{m}^{(l)}(\lambda )$ with $l\geq 1$ on its reference
state\ $\left\vert 0_{m}\right\rangle ^{(l)}$ we have the factors 
\begin{eqnarray}
X_{a}^{(l)}(\lambda ) &\circeq &X_{a}^{(l)}(\lambda ,\{\lambda
_{i}^{(l-1)}\})=\prod_{k=1}^{m}x_{a}^{(l)}(\lambda -\lambda
_{k}^{(l-1)}),\quad a=1,2  \notag \\
X_{3}^{(l)}(\lambda ) &\circeq &X_{3}^{(l)}(\lambda ,\{\lambda
_{i}^{(l-1)}\})=\prod_{k=1}^{m}y_{N_{l}N_{l}}^{(l)}(\lambda -\lambda
_{k}^{(l-1)}),\quad i=1,...,m.  \label{two.27}
\end{eqnarray}%
where $m$ is the particle (site) number in the \emph{layer} $l$ and $\lambda
_{k}^{(0)}=\lambda _{k}$.
\end{itemize}

In respect to the unwanted terms of (\ref{two.21b}), the main motivation to
write this section, we first recall the relation (\ref{two.3}) in its
generalized form \ by using the cyclic permutations of the factors in the
normal ordered vector for the $m$-particle state in the $l$-\textit{th} 
\emph{layer} 
\begin{equation}
\Phi _{m}^{(l)}(\lambda _{1},\lambda _{2},...,\lambda _{m})=\Phi
_{m}^{(l)}(\lambda _{2},...,\lambda _{m},\lambda _{1})\frac{%
S_{12}^{(l+1)}(\lambda _{1}-\lambda _{2})}{x_{1}^{(l+1)}(\lambda
_{1}-\lambda _{2})}\frac{S_{23}^{(l+1)}(\lambda _{1}-\lambda _{3})}{%
x_{1}^{(l+1)}(\lambda _{1}-\lambda _{3})}\cdots \frac{S_{m-1,m}^{(l+1)}(%
\lambda _{1}-\lambda _{m})}{x_{1}^{(l+1)}(\lambda _{1}-\lambda _{m})}
\label{two.29}
\end{equation}

\bigskip Now we define an operator $M$ by%
\begin{equation}
M_{m}^{(l+1)}(\lambda ,\{\lambda _{i}\})=\mathrm{Tr}_{a}\left[ \frac{%
S_{a1}^{(l+1)}(\lambda _{1}-\lambda _{2})}{x_{1}^{(l+1)}(\lambda
_{1}-\lambda _{2})}\frac{S_{a2}^{(l+1)}(\lambda _{1}-\lambda _{3})}{%
x_{1}^{(l+1)}(\lambda _{1}-\lambda _{3})}\cdots \frac{S_{a,m}^{(l+1)}(%
\lambda _{1}-\lambda _{m})}{x_{1}^{(l+1)}(\lambda _{1}-\lambda _{m})}\right]
\label{two.30}
\end{equation}%
and we remark that $M$ is the normalized permutation of \ a $m$ site
inhomogeneous transfer matrix 
\begin{equation}
\tau _{m}^{(l+1)}(\lambda ,\{\lambda _{i}\})=\mathrm{Tr}_{a}\left[ \mathcal{L%
}_{am}^{(l+1)}(\lambda -\lambda _{m})\mathcal{L}_{am-1}^{(l+1)}(\lambda
-\lambda _{m-1})\cdots \mathcal{L}_{a1}^{(l+1)}(\lambda -\lambda _{1})\right]
\label{two.31}
\end{equation}%
Thus, we can write the generalization of (\ref{two.3}) in a more convenient
form%
\begin{equation}
\Phi _{m}^{(l)}(\lambda _{1},\lambda _{2},...,\lambda _{m})=\Phi
_{m}^{(l)}(\lambda _{2},...,\lambda _{m},\lambda _{1})M_{m}^{(l+1)}(\lambda
_{1},\{\lambda _{i}\})  \label{two.32}
\end{equation}%
where $i=1,...,m$.

\bigskip Using these expressions with $l=1$ and $m=2$, one can see that the
unwanted term $\mathcal{B}(\lambda )\otimes \Phi _{1}(\lambda _{2})$ in (\ref%
{two.21b}) can be eliminated by the condition 
\begin{equation}
\left( z(\lambda _{2}-\lambda _{1})X_{1}(\lambda _{1})-z(\lambda
_{1}-\lambda _{2})X_{2}(\lambda _{1})\frac{\tau _{2}^{(1)}(\lambda _{1})}{%
x_{1}^{(1)}(\lambda _{1}-\lambda _{2})}\right) \mathcal{F}_{2}=0
\label{two.33}
\end{equation}%
and from the fact that $\mathcal{B}(\lambda )\otimes \Phi _{1}(\lambda _{1})$
is a permutation of $\mathcal{B}(\lambda )\otimes \Phi _{1}(\lambda _{2})$ \
by the interchange $\lambda _{1}\leftrightarrow \lambda _{2}$, one can use (%
\ref{two.32}) with $l=1$ and $m=2$, in order to get a second elimination
condition%
\begin{equation}
M_{2}^{(1)}(\lambda _{1},\{\lambda _{i}\})\left( z(\lambda _{1}-\lambda
_{2})X_{1}(\lambda _{2})-z(\lambda _{2}-\lambda _{1})X_{2}(\lambda _{2})%
\frac{\tau _{2}^{(1)}(\lambda _{2})}{x_{1}^{(1)}(\lambda _{2}-\lambda _{1})}%
\right) \mathcal{F}_{2}=0  \label{two.34}
\end{equation}%
Consequently, we have the following eigenvalue problems 
\begin{equation}
\tau _{2}^{(1)}(\lambda _{a})\mathcal{F}_{2}=x_{1}^{(1)}(\lambda
_{a}-\lambda _{b})\frac{z(\lambda _{b}-\lambda _{a})}{z(\lambda _{a}-\lambda
_{b})}\frac{X_{1}(\lambda _{a})}{X_{2}(\lambda _{a})}\mathcal{F}_{2}\qquad
(a\neq b=1,2)  \label{two.35}
\end{equation}%
which should be seen as a generalization of the Bethe equations since this
restriction alone eliminate all unwanted terms. To see this, one can use the
same statements in $\mathcal{B}^{\ast }(\lambda )\otimes \Phi _{1}(\lambda
_{1})\otimes 1$ and $\mathcal{B}^{\ast }(\lambda )\otimes \Phi _{1}(\lambda
_{2})\otimes 1$ terms of (\ref{two.21b}) in order to get the same pair of
equations (\ref{two.35}). \ Finally, substituting (\ref{two.35}) into the $%
B_{N-1}(\lambda )$ unwanted term of (\ref{two.21b}) one can see that it is
also canceled out.

It is now clear that the eigenvalue (\ref{two.25}) as well as the Bethe
equations are written in terms of the eigenvalue of $\tau _{2}^{(1)}(\lambda
)$ :%
\begin{equation}
\frac{X_{1}(\lambda _{a})}{X_{2}(\lambda _{a})}=\frac{z(\lambda _{a}-\lambda
_{b})}{z(\lambda _{b}-\lambda _{a})}\frac{\Lambda _{2}^{(1)}(\lambda
_{a},\{\lambda _{i}\}|\{\lambda _{i}^{(1)}\})}{x_{1}^{(1)}(\lambda
_{a}-\lambda _{b})}\qquad \quad (a\neq b=1,2)  \label{two.36}
\end{equation}%
where $\Lambda _{2}^{(1)}(\lambda _{a},\{\lambda _{i}\}|\{\lambda
_{i}^{(1)}\})$ is the residue of $\Lambda _{2}^{(1)}(\lambda ,\{\lambda
_{i}\}|\{\lambda _{i}^{(1)}\})$ at $\lambda =\lambda _{a}$. Consequently, we
have to consider this second eigenvalue problem (\ref{two.23}) in order to
fix the results of the first eigenvalue problem. Indeed it is a simple task
because all computation was already made and the result follows directly
read from (\ref{two.21b}) with the following trivial modifications:

\begin{itemize}
\item Changing the vertex models: \ (B$_{n}^{(1)},$C$_{n}^{(1)},$D$%
_{n}^{(1)},$A$_{2n}^{(2)},$A$_{2n-1}^{(2)}$)$\rightarrow $(B$_{n-1}^{(1)},$C$%
_{n-1}^{(1)},$D$_{n-1}^{(1)},$A$_{2(n-1)}^{(2)},$A$_{2(n-1)-1}^{(2)}$),
putting the label $l=1$ in the new Boltzmann weights.

\item Changing the lattice: $L$ site homogeneous lattice ($\tau _{L}(\lambda
)$) $\rightarrow $ $2$ -site inhomogeneous lattice ($\tau _{2}^{(1)}(\lambda
)$).

\item Changing the two-particle state: $\Phi _{2}(\lambda _{1},\lambda _{2})%
\mathcal{F}_{2}\left\vert 0_{L}\right\rangle \rightarrow \Phi
_{2}^{(1)}(\lambda _{1}^{(1)},\lambda _{2}^{(1)})\mathcal{F}%
_{2}^{(1)}\left\vert 0_{2}\right\rangle ^{(1)}$.
\end{itemize}

After these modifications we still have to consider a third eigenvalue
problem by repeating this changing procedure by increasing by one unity the
value of the label $l$ but keep the size of the last lattice. Next, we
repeat the last procedure and so on. Of course, such a procedure must have
an end.

In the nested Bethe ansatz procedure the end is obtained in the \emph{layer}
for which the eigenvalue and the Bethe equations are fixed. \ It means that
we are work with the models in the last \emph{layer}.

The value $l=n-1$ defines the last \emph{layer} as being the B$_{1}^{(1)}$
and A$_{2}^{(2)}$ vertex models ($N_{l}$ odd) and as the C$_{2}^{(1)},$D$%
_{2}^{(1)}$.an A$_{3}^{(2)}$ vertex models ($N_{l}$ even). Thus they must be
considered separately:

\subsection{B$_{n}^{(1)}$ and A$_{2n}^{(2)}$ two-particle state}

We already mentioned that these models possess a limit via the reduction to
scalar status. It means that in the last \emph{layer} we are working with (%
\ref{fcr.5a}) and (\ref{fcr.5b}) with the Boltzmann weights of the B$%
_{1}^{(1)}$ and A$_{2}^{(2)}$ vertex models. The explicit expression for the
eigenvalue problem of the $L$ site homogeneous transfer matrix for these
nineteen-vetex models was already presented in the reference \cite{ALS}.
However, here this expression will be presented by a reduction procedure
from our general formulation.

Consequently the eigenvalue expression (\ref{two.21b}) is reduced to 
\begin{eqnarray}
\tau _{2}(\lambda )\Psi _{2}(\lambda _{1},\lambda _{2}) &=&(X_{1}(\lambda
)\prod\limits_{k=1}^{2}z(\lambda _{k}-\lambda )+X_{2}(\lambda
)\prod\limits_{k=1}^{2}\frac{z(\lambda -\lambda _{k})}{\omega (\lambda
-\lambda _{k})}+X_{3}(\lambda )\prod\limits_{k=1}^{2}\frac{x_{2}(\lambda
-\lambda _{k})}{y_{33}(\lambda -\lambda _{k})})\Psi _{2}(\lambda
_{1},\lambda _{2})  \notag \\
&&-\frac{x_{3}(\lambda _{10})}{x_{2}(\lambda _{10})}\left( X_{1}(\lambda
_{1})z(\lambda _{21})-X_{2}(\lambda _{1})\frac{z(\lambda _{12})}{\omega
(\lambda _{12})}\right) B_{1}(\lambda )B_{1}(\lambda _{2})\left\vert
0_{2}\right\rangle  \notag \\
&&-\frac{x_{3}(\lambda _{20})}{x_{2}(\lambda _{20})}\left( X_{1}(\lambda
_{2})\frac{z(\lambda _{12})}{\omega (\lambda _{12})}-X_{2}(\lambda
_{2})z(\lambda _{21})\right) B_{1}(\lambda )B_{1}(\lambda _{2})\left\vert
0_{2}\right\rangle  \notag \\
&&-\frac{y_{32}(\lambda _{02})}{y_{33}(\lambda _{02})}\ \left( X_{2}(\lambda
_{2})z(\lambda _{21})-X_{1}(\lambda _{2})\frac{z(\lambda _{12})}{\omega
(\lambda _{12})}\right) B_{3}(\lambda )B_{1}(\lambda _{1})\left\vert
0_{2}\right\rangle  \notag \\
&&-\frac{y_{32}(\lambda _{01})}{y_{33}(\lambda _{01})}\ \left( X_{2}(\lambda
_{1})\frac{z(\lambda _{12})}{\omega (\lambda _{12})}-X_{1}(\lambda
_{1})z(\lambda _{21})\right) B_{3}(\lambda )B_{1}(\lambda _{2})\left\vert
0_{2}\right\rangle  \notag \\
&&+B_{2}(\lambda )\left\{ G_{21}(\lambda ,\lambda _{1},\lambda
_{2})X_{1}(\lambda _{1})X_{1}(\lambda _{2})+H_{21}(\lambda ,\lambda
_{1},\lambda _{2})X_{2}(\lambda _{1})X_{2}(\lambda _{2})\right.  \notag \\
&&+\left. Y_{21}(\lambda ,\lambda _{1},\lambda _{2})X_{2}(\lambda
_{2})X_{1}(\lambda _{1})+Y_{21}(\lambda ,\lambda _{2},\lambda
_{1})X_{2}(\lambda _{1})X_{1}(\lambda _{2})\frac{1}{\omega (\lambda _{12})}%
\right\} \left\vert 0_{2}\right\rangle  \label{two.37}
\end{eqnarray}%
where we have omitted the label $l=n-1$.

In (\ref{two.37}) we have used the function $\omega (\lambda )$ \ which is
the reduction of the objects $S$, $\mathcal{R}$ and \textrm{Tr}$_{a}[%
\mathcal{L]}$ present in (\ref{two.21b}) 
\begin{eqnarray}
\frac{1}{\omega (\lambda )} &=&\frac{s(\lambda )}{x_{1}(\lambda )}=\frac{%
\mathcal{R}(\lambda )}{x_{1}(\lambda )}=\frac{\mathrm{Tr}_{a}[\mathcal{L}%
(\lambda )]}{x_{1}(\lambda )}=\frac{1}{x_{1}(\lambda )}\left( y_{22}(\lambda
)-\frac{y_{23}(\lambda )y_{32}(\lambda )}{y_{33}(\lambda )}\right) .  \notag
\\
&&  \notag \\
\omega (\lambda )\omega (-\lambda ) &=&1.  \label{two.38}
\end{eqnarray}%
Therefore, for the last \emph{layer} the eigenvalue is

\begin{equation}
\Lambda _{2}(\lambda ,\{\lambda _{i}^{(n-2)}\}|\{\lambda
_{i}\})=X_{1}(\lambda )\prod\limits_{k=1}^{2}z(\lambda _{k}-\lambda
)+X_{3}(\lambda )\prod\limits_{k=1}^{2}\frac{x_{2}(\lambda -\lambda _{k})}{%
y_{33}(\lambda -\lambda _{k})}+X_{2}(\lambda )\prod\limits_{k=1}^{2}\frac{%
z(\lambda -\lambda _{k})}{\omega (\lambda -\lambda _{k})}  \label{two.39}
\end{equation}%
provided that%
\begin{equation}
\frac{X_{1}(\lambda _{a})}{X_{2}(\lambda _{a})}=\frac{z(\lambda _{a}-\lambda
_{b})}{z(\lambda _{b}-\lambda _{v})}\omega (\lambda _{b}-\lambda _{a}),\quad
a\neq b=1,2  \label{two.41}
\end{equation}%
The factors $X_{a}$ are given by%
\begin{equation}
X_{3}(\lambda )=\prod\limits_{k=1}^{2}y_{33}(\lambda -\lambda
_{k}^{n-2})\qquad ,X_{a}(\lambda )=\prod\limits_{k=1}^{2}x_{a}(\lambda
-\lambda _{k}^{n-2}),\qquad (a=1,2)  \label{two.40}
\end{equation}%
where the inhomogeneity parameter \{$\lambda _{i}^{n-2}$\} make the link
with the next to the last \emph{layer}.

Let us go back to the \emph{ground} in order to write the full nest through
of a sequence of terms where the models are explicitly identified:%
\begin{eqnarray*}
\Lambda _{L}(\lambda |\{\lambda _{1},\lambda _{2}\}) &=&X_{1}(\lambda
)\prod\limits_{k=1}^{2}z(\lambda _{k}-\lambda )+X_{3}(\lambda
)\prod\limits_{k=1}^{2}\frac{x_{2}(\lambda -\lambda _{k})}{y_{NN}(\lambda
-\lambda _{k})}+X_{2}(\lambda )\frac{\Lambda _{2}^{(1)}(\lambda ,\{\lambda
_{i}\}|\{\lambda _{i}^{(1)}\})}{X_{2}^{(1)}(\lambda )} \\
(0) &\in &(\mathrm{B}_{n}^{(1)},\mathrm{A}_{2n}^{(2)})\qquad \mathrm{and}%
\qquad (1)\in (\mathrm{B}_{n-1}^{(1)},\mathrm{A}_{2(n-1)}^{(2)})
\end{eqnarray*}%
\begin{eqnarray*}
\frac{\Lambda _{2}^{(l)}(\lambda ,\{\lambda _{i}^{(l-1)}\}|\{\lambda
_{i}^{(l)}\})}{X_{2}^{(l)}(\lambda )} &=&\frac{X_{1}^{(l)}(\lambda )}{%
X_{2}^{(l)}(\lambda )}\prod\limits_{k=1}^{2}z^{(l)}(\lambda
_{k}^{(l)}-\lambda )+\frac{X_{3}^{(l)}(\lambda )}{X_{2}^{(l)}(\lambda )}%
\prod\limits_{k=1}^{2}\frac{x_{2}^{(l)}(\lambda -\lambda _{k}^{(l)})}{%
y_{N_{l}N_{l}}(\lambda -\lambda _{k}^{(l)})} \\
&&+\frac{\Lambda _{2}^{(l+1)}(\lambda ,\{\lambda _{i}^{(l)}\}|\{\lambda
_{i}^{(l+1)}\})}{X_{2}^{(l+1)}(\lambda )} \\
(l) &\in &(\mathrm{B}_{n-l}^{(1)},\mathrm{A}_{2(n-l)}^{(2)})\qquad \mathrm{%
and}\qquad (l+1)\in (\mathrm{B}_{n-l-1}^{(1)},\mathrm{A}_{2(n-l-1)}^{(2)}) \\
l &=&1,2,...,n-2
\end{eqnarray*}%
\begin{eqnarray}
\frac{\Lambda _{2}^{(n-1)}(\lambda ,\{\lambda _{i}^{(n-2)}\}|\{\lambda
_{i}^{(n-1)}\})}{X_{2}^{(n-1)}(\lambda )} &=&\frac{X_{1}^{(n-1)}(\lambda )}{%
X_{2}^{(n-1)}(\lambda )}\prod\limits_{k=1}^{2}z^{(n-1)}(\lambda
_{k}^{(n-1)}-\lambda )  \notag \\
&&+\frac{X_{3}^{(n-1)}(\lambda )}{X_{2}^{(n-1)}(\lambda )}%
\prod\limits_{k=1}^{2}\frac{x_{2}^{(n-1)}(\lambda -\lambda _{k}^{(n-1)})}{%
y_{33}^{(n-1)}(\lambda -\lambda _{k}^{(n-1)})}  \notag \\
&&+\prod\limits_{k=1}^{2}\frac{z^{(n-1)}(\lambda -\lambda _{k}^{(n-1)})}{%
\omega (\lambda -\lambda _{k}^{(n-1)})}  \notag \\
(n-1) &\in &(\mathrm{B}_{1}^{(1)},\mathrm{A}_{2}^{(2)})  \label{two.41a}
\end{eqnarray}%
Remember that the inhomogeneity parameters $\{\lambda _{i}^{(l-1)}\}$ are
implicit in (\ref{two.41a}) through of the definition of $%
X_{a}^{(l)}(\lambda )$,$\ a=1,2,3.$ The function $\omega (\lambda -\lambda
_{k}^{(n-1)})$ is given by (\ref{two.38}).

The corresponding Bethe equations are%
\begin{equation*}
\frac{X_{1}(\lambda _{a})}{X_{2}(\lambda _{a})}=\frac{z(\lambda _{a}-\lambda
_{b})}{z(\lambda _{b}-\lambda _{a})}\frac{\Lambda _{2}^{(1)}(\lambda
_{a},\{\lambda _{i}\}|\{\lambda _{i}^{(1)}\})}{x_{1}^{(1)}(\lambda
_{a}-\lambda _{b})},\quad a\neq b=1,2
\end{equation*}%
\begin{eqnarray*}
\frac{X_{1}^{(l)}(\lambda _{a}^{(l)})}{X_{2}^{(l)}(\lambda _{a}^{(l)})} &=&%
\frac{z^{(l)}(\lambda _{a}^{(l)}-\lambda _{b}^{(l)})}{z^{(l)}(\lambda
_{b}^{(l)}-\lambda _{a}^{(l)})}\frac{\Lambda _{2}^{(l+1)}(\lambda
_{a}^{(l)},\{\lambda _{i}^{(l)}\}|\{\lambda _{i}^{(l+1)}\})}{%
x_{1}^{(l+1)}(\lambda _{a}^{(l)}-\lambda _{b}^{(l)})},\quad a\neq b=1,2 \\
l &=&1,2,...,n-2
\end{eqnarray*}%
\begin{equation}
\frac{X_{1}^{(n-1)}(\lambda _{a}^{(n-1)})}{X_{2}^{(n-1)}(\lambda
_{a}^{(n-1)})}=\frac{z^{(n-1)}(\lambda _{a}^{(n-1)}-\lambda _{b}^{(n-1)})}{%
z^{(n-1)}(\lambda _{b}^{(n-1)}-\lambda _{a}^{(n-1)})}\omega (\lambda
_{b}^{(n-1)}-\lambda _{a}^{(n-1)}),\quad a\neq b=1,2  \label{two.44}
\end{equation}%
where 
\begin{eqnarray}
\Lambda _{2}^{(l+1)}(\lambda _{a}^{(l)},\{\lambda _{i}^{(l)}\}|\{\lambda
_{i}^{(l+1)}\}) &=&\underset{\lambda =\lambda _{a}^{(l)}}{\mathrm{res}}%
\Lambda _{2}^{(l+1)}(\lambda ,\{\lambda _{i}^{(l)}\}|\{\lambda
_{i}^{(l+1)}\})  \notag \\
&=&x_{1}^{(l+1)}(0)x_{1}^{(l+1)}(\lambda _{a}^{(l)}-\lambda
_{b}^{(l)})\prod\limits_{k=1}^{2}z^{(l+1)}(\lambda _{k}^{(l+1)}-\lambda
_{a}^{(l)})  \notag \\
l &=&0,1,...,n-2,\quad a\neq b=1,2.  \label{two.44a}
\end{eqnarray}

It is curious that the solution of the eigenvalue problem of the $L$ site
homogeneous transfer matrix for the two-particle state is given in terms of
the eigenvalue problems of the two site inhomogeneous transfer matrices for
two-particle state. We remark \ the participation of $n$ different models in
the construction of the nested Bethe ansatz for the B$_{n}^{(1)}$ and A$%
_{2n}^{(1)}$ vertex models.

\subsection{C$_{n}$ , D$_{n}$ and A$_{2n-1}^{(2)}$ two-particle state}

For $N_{l}=4,6,...$ the last \emph{layer} involves the C$_{2}^{(1)},$D$%
_{2}^{(1)}$ and A$_{3}^{(2)}$ vertex models for which their nests are not
complete. \ Indeed, just more one \emph{layer} is necessary in order to
complete them. To do this we recall (\ref{fcr.5c}) and (\ref{fcr.5d}) in
order to include the C$_{1}^{(1)},$D$_{1}^{(1)}$ and A$_{1}^{(2)}$ vertex
models in our discussion.

For the C$_{1}$ and A$_{1}^{(2)}$ vertex models, the $L$ site homogeneous
transfer matrix is the trace of (\ref{fcr.5d}) 
\begin{equation}
\tau _{L}(\lambda )=A_{1}(\lambda )+A_{3}(\lambda )  \label{two.46}
\end{equation}%
and the reference state%
\begin{equation}
\left\vert 0_{L}\right\rangle =\left( 
\begin{array}{c}
1 \\ 
0%
\end{array}%
\right) _{1}\otimes \left( 
\begin{array}{c}
1 \\ 
0%
\end{array}%
\right) _{2}\otimes \cdots \otimes \left( 
\begin{array}{c}
1 \\ 
0%
\end{array}%
\right) _{L}  \label{two.47}
\end{equation}%
is a highest vector of (\ref{fcr.5d})%
\begin{eqnarray}
A_{1}(\lambda )\left\vert 0_{L}\right\rangle &=&X_{1}(\lambda )\left\vert
0_{L}\right\rangle ,\quad A_{3}(\lambda )\left\vert 0_{L}\right\rangle
=X_{3}(\lambda )\left\vert 0_{L}\right\rangle  \notag \\
C_{1}(\lambda )\left\vert 0_{L}\right\rangle &=&0,\quad B_{1}(\lambda
)\left\vert 0_{L}\right\rangle \neq \left\{ 0,\left\vert 0_{L}\right\rangle
\right\}  \notag \\
X_{1}(\lambda ) &=&[x_{1}(\lambda )]^{L},\quad X_{3}(\lambda
)=[y_{22}(\lambda )]^{L}  \label{two.48}
\end{eqnarray}%
The two-particle state is a new state without any relation with (\ref{two.4}%
) which can be defined by%
\begin{equation}
\Psi _{2}(\lambda _{1},\lambda _{2})=\Psi _{2}(\lambda _{2},\lambda
_{1})=B_{1}(\lambda _{1})B_{1}(\lambda _{2})  \label{two.49}
\end{equation}%
The action of $\tau _{L}(\lambda )$ on this state can be computed using the
commutation relations (\ref{two.5})-(\ref{two.7}) in their reduced form 
\begin{eqnarray}
\tau _{L}(\lambda )\Psi _{2}(\lambda _{1},\lambda _{2}) &=&\left(
X_{1}(\lambda )\prod\limits_{k=1}^{2}\frac{x_{1}(\lambda _{k}-\lambda )}{%
y_{22}(\lambda _{k}-\lambda )}+X_{3}(\lambda )\prod\limits_{k=1}^{2}\frac{%
x_{1}(\lambda -\lambda _{k})}{y_{22}(\lambda -\lambda _{k})}\right) \Psi
_{2}(\lambda _{1},\lambda _{2})  \notag \\
&&-\frac{y_{12}(\lambda _{1}-\lambda )}{y_{22}(\lambda _{1}-\lambda )}\left(
X_{1}(\lambda _{1})\frac{x_{1}(\lambda _{2}-\lambda _{1})}{y_{22}(\lambda
_{2}-\lambda _{1})}-X_{3}(\lambda _{1})\frac{x_{1}(\lambda _{1}-\lambda _{2})%
}{y_{22}(\lambda _{1}-\lambda _{2})}\right) B_{1}(\lambda )B(\lambda
_{2})\left\vert 0_{L}\right\rangle  \notag \\
&&-\frac{y_{12}(\lambda _{2}-\lambda )}{y_{22}(\lambda _{2}-\lambda )}\left(
X_{1}(\lambda _{2})\frac{x_{1}(\lambda _{1}-\lambda _{2})}{y_{22}(\lambda
_{1}-\lambda _{2})}-X_{3}(\lambda _{2})\frac{x_{1}(\lambda _{2}-\lambda _{1})%
}{y_{22}(\lambda _{2}-\lambda _{1})}\right) B_{1}(\lambda )B(\lambda
_{1})\left\vert 0_{L}\right\rangle  \notag \\
&&  \label{two.50}
\end{eqnarray}%
where we have used the property of the definition (\ref{two.49}) and the
identity%
\begin{equation}
\frac{y_{12}(\lambda )}{y_{22}(\lambda )}+\frac{y_{21}(-\lambda )}{%
y_{22}(-\lambda )}=0.  \label{two.51}
\end{equation}%
The eigenvalue is%
\begin{equation}
\Lambda _{L}(\lambda |\lambda _{1},\lambda _{2})=X_{1}(\lambda
)\prod\limits_{k=1}^{2}\frac{x_{1}(\lambda _{k}-\lambda )}{y_{22}(\lambda
_{k}-\lambda )}+X_{3}(\lambda )\prod\limits_{k=1}^{2}\frac{x_{1}(\lambda
-\lambda _{k})}{y_{22}(\lambda -\lambda _{k})}  \label{two.52}
\end{equation}%
provided that%
\begin{equation}
\frac{X_{1}(\lambda _{a})}{X_{3}(\lambda _{a})}=\frac{x_{1}(\lambda
_{a}-\lambda _{b})}{x_{1}(\lambda _{b}-\lambda _{a})}\frac{y_{22}(\lambda
_{b}-\lambda _{a})}{y_{22}(\lambda _{a}-\lambda _{b})},\quad a\neq b=1,2.
\label{two.53}
\end{equation}

We now turn to the diagonalization problem of the D$_{2}^{(1)}$ vertex
model. It turns out, however, that the Lax operator of this model can
decomposed in terms of Lax operator for the six-vertex model associated with
the A$_{1}^{(1)}$ Lie algebra. \ It means that from the isomorphism\ D$%
_{2}^{(1)}=$A$_{1}^{(1)}\oplus $A$_{1}^{(1)}$, we can write%
\begin{equation}
\mathcal{L}^{D_{2}}(\lambda )=\mathcal{L}_{+}^{A_{1}}(\lambda )\otimes 
\mathcal{L}_{-}^{A_{1}}(\lambda )  \label{two.54}
\end{equation}%
Here, a more careful analysis is required. First we identify the Lax
operator $\mathcal{L}^{D_{2}}$ with the corresponding $\mathcal{R}$-matrix (%
\ref{mod.1}) and then we make a sign transformation in the Boltzmann weights 
$y_{\alpha \beta }(\lambda )$ that preserves the spectrum of the transfer
matrix associated i.e., $y_{\alpha \beta }(\lambda )\rightarrow -y_{\alpha
\beta }(\lambda )$ for $\alpha \neq \beta $ and $\alpha \neq \beta
^{^{\prime }}$. \ Now is not difficult to verify the isomorphism (\ref%
{two.54}) where $\mathcal{L}^{A_{1}}(\lambda )$ is identified with the $%
\mathcal{R}$-matrix of the A$_{1}^{(1)}$ vertex model listed in the our
appendix.

Consequently, the eigenvalues of the model D$_{2}^{(1)}$ are given in terms
of the product of the eigenvalues of two A$_{1}^{(1)}$ six-vertex models: 
\begin{equation}
\Lambda _{L}^{D_{2}}(\lambda |\lambda _{1}^{\pm },\lambda _{2}^{\pm
})=\Lambda _{L}^{+}(\lambda |\lambda _{1}^{+},\lambda _{2}^{+})\Lambda
_{L}^{-}(\lambda |\lambda _{1}^{-},\lambda _{2}^{-})  \label{two.56}
\end{equation}%
where 
\begin{equation}
\Lambda _{L}^{\pm }(\lambda |\lambda _{1},\lambda _{2})=X_{1}(\lambda
)\prod\limits_{k=1}^{2}z(\lambda _{k}^{\pm }-\lambda )+X_{2}(\lambda
)\prod\limits_{k=1}^{2}z(\lambda -\lambda _{k}^{\pm })  \label{two.57}
\end{equation}%
provided that%
\begin{equation}
\frac{X_{1}(\lambda _{a}^{\pm })}{X_{2}(\lambda _{a}^{\pm })}=\frac{%
z(\lambda _{a}^{\pm }-\lambda _{b}^{\pm })}{z(\lambda _{b}^{\pm }-\lambda
_{a}^{\pm })},\quad a\neq b=1,2.  \label{two.58}
\end{equation}%
Here \{$\lambda _{1}^{+},\lambda _{2}^{+}$\} and \{$\lambda _{1}^{-},\lambda
_{2}^{-}$\} are rapidities of the two-particle state related to each one of
the two six-vertex models. The algebraic Bethe ansatz for A$_{n}^{(1)}$
vertex models is discussed in the section $8$ of this paper.

From these results we can see that the two particle state $\Psi _{2}(\lambda
_{1},\lambda _{2})$ is an eigenstate of the homogeneous transfer matrix $%
\Lambda _{L}(\lambda )$ with eigenvalue%
\begin{eqnarray}
\Lambda _{L}(\lambda |\{\lambda _{i}\}) &=&X_{1}(\lambda
)\prod\limits_{k=1}^{2}z(\lambda _{k}-\lambda )+X_{3}(\lambda
)\prod\limits_{k=1}^{2}\frac{x_{2}(\lambda -\lambda _{k})}{y_{NN}(\lambda
-\lambda _{k})}  \notag \\
&&+X_{2}(\lambda )\left( \sum_{l=1}^{n-2}G_{2}^{(l)}(\lambda ,\{\lambda
_{i}^{(l-1)}\}|\{\lambda _{i}^{(l)}\})+\mathcal{T}\right)
\end{eqnarray}%
where%
\begin{equation}
G_{2}^{(l)}(\lambda ,\{\lambda _{i}^{(l-1)}\}|\{\lambda _{i}^{(l)}\})=\frac{%
X_{1}^{(l)}(\lambda )}{X_{2}^{(l)}(\lambda )}\prod\limits_{k=1}^{2}z^{(l)}(%
\lambda _{k}^{(l)}-\lambda )+\frac{X_{3}^{(l)}(\lambda )}{%
X_{2}^{(l)}(\lambda )}\prod\limits_{k=1}^{2}\frac{x_{2}^{(l)}(\lambda
-\lambda _{k}^{(l)})}{y_{N_{l}N_{l}}^{(l)}(\lambda -\lambda _{k}^{(l)})}
\end{equation}%
and the last terms $\mathcal{T}$ depend on the model: 
\begin{equation}
\mathcal{T}=G_{2}^{(n-1)}(\lambda ,\{\lambda _{i}^{(n-2)}\}|\{\lambda
_{i}^{(n-1)}\})+\frac{X_{1}^{(n)}(\lambda )}{X_{2}^{(n)}(\lambda )}%
\prod\limits_{k=1}^{2}\frac{x_{1}^{(n)}(\lambda _{k}^{(n)}-\lambda )}{%
y_{22}^{(n)}(\lambda _{k}^{(n)}-\lambda )}+\frac{X_{3}^{(n)}(\lambda )}{%
X_{2}^{(n)}(\lambda )}\prod\limits_{k=1}^{2}\frac{x_{1}^{(n)}(\lambda
-\lambda _{k}^{(n)})}{y_{22}^{(n)}(\lambda -\lambda _{k}^{(n)})}
\end{equation}%
for the C$_{n}^{(1)}$ and A$_{2n-1}^{(2)}$ vertex models and

\begin{eqnarray}
\mathcal{T} &=&[X_{1}^{(n-1)}(\lambda
)\prod\limits_{k=1}^{2}z^{(n-1)}(\lambda _{k}^{+}-\lambda
)+X_{2}^{(n-1)}(\lambda )\prod\limits_{k=1}^{2}z^{(n-1)}(\lambda -\lambda
_{k}^{+})]  \notag \\
&&\times \lbrack X_{1}^{(n-1)}(\lambda
)\prod\limits_{k=1}^{2}z^{(n-1)}(\lambda _{k}^{-}-\lambda
)+X_{2}^{(n-1)}(\lambda )\prod\limits_{k=1}^{2}z^{(n-1)}(\lambda -\lambda
_{k}^{-})]  \notag \\
(n-1) &\in &\mathrm{A}_{1}^{(1)}  \label{two.62}
\end{eqnarray}%
for D$_{n}^{(1)}$ vertex model. The last row in (\ref{two.62}) is to
remember that the Boltzmann weights are those presented in the appendix.

The sequence of the Bethe equations has the form 
\begin{equation*}
\frac{X_{1}(\lambda _{a})}{X_{2}(\lambda _{a})}=\frac{z(\lambda _{a}-\lambda
_{b})}{z(\lambda _{b}-\lambda _{a})}x_{1}^{(1)}(0)\prod%
\limits_{k=1}^{2}z^{(1)}(\lambda _{k}^{(1)}-\lambda _{a})\quad a\neq b=1,2
\end{equation*}%
\begin{eqnarray}
\frac{X_{1}^{(l)}(\lambda _{a}^{(l)})}{X_{2}^{(l)}(\lambda _{a}^{(l)})} &=&%
\frac{z^{(l)}(\lambda _{a}^{(l)}-\lambda _{b}^{(l)})}{z^{(l)}(\lambda
_{b}^{(l)}-\lambda _{a}^{(l)})}x_{1}^{(l+1)}(0)\prod%
\limits_{k=1}^{2}z^{(l+1)}(\lambda _{k}^{(l+1)}-\lambda _{a}^{(l)}),\quad
a\neq b=1,2  \notag \\
l &=&1,2,...,n-2.  \label{two.63}
\end{eqnarray}%
which will end in two different ways:%
\begin{eqnarray}
\prod\limits_{k=1}^{2}\frac{x_{1}^{(n-1)}(\lambda _{a}^{\pm }-\lambda
_{k}^{(n-2)})}{x_{2}^{(n-1)}(\lambda _{a}^{\pm }-\lambda _{k}^{(n-2)})} &=&%
\frac{z^{(n-1)}(\lambda _{a}^{\pm }-\lambda _{b}^{\pm })}{z^{(n-1)}(\lambda
_{b}^{\pm }-\lambda _{a}^{\pm })},\quad a\neq b=1,2  \notag \\
(n-1) &\in &\mathrm{A}_{1}^{(1)}  \label{two.64}
\end{eqnarray}%
for D$_{n}^{(1)}$ and more two equations 
\begin{eqnarray*}
\frac{X_{1}^{(n-1)}(\lambda _{a}^{(n-1)})}{X_{2}^{(n-1)}(\lambda
_{a}^{(n-1)})} &=&\frac{z^{(n-1)}(\lambda _{a}^{(n-1)}-\lambda _{b}^{(n-1)})%
}{z^{(n-1)}(\lambda _{b}^{(n-1)}-\lambda _{a}^{(n-1)})}x_{1}^{(n)}(0)\prod%
\limits_{k=1}^{2}z^{(n)}(\lambda _{k}^{(n)}-\lambda _{a}^{(n-1)}),\quad
a\neq b=1,2 \\
(n-1) &\in &(\mathrm{C}_{2}^{(1)},\mathrm{A}_{3}^{(2)})
\end{eqnarray*}%
\begin{eqnarray}
\frac{X_{1}^{(n)}(\lambda _{a}^{(n)})}{X_{3}^{(n)}(\lambda _{a}^{(n)})} &=&%
\frac{x_{1}^{(n)}(\lambda _{a}^{(n)}-\lambda _{b}^{(n)})}{%
x_{1}^{(n)}(\lambda _{b}^{(n)}-\lambda _{a}^{(n)})}\frac{y_{22}^{(n)}(%
\lambda _{b}^{(n)}-\lambda _{a}^{(n)})}{y_{22}^{(n)}(\lambda
_{a}^{(n)}-\lambda _{b}^{(n)})},\quad a\neq b=1,2.  \notag \\
(n) &\in &(\mathrm{C}_{1}^{(1)},\mathrm{A}_{1}^{(2)})  \label{two.65}
\end{eqnarray}%
for C$_{n}^{(1)}$ and A$_{2n-1}^{(2)}$ due to their continuation for C$%
_{1}^{(1)}$ and A$_{1}^{(2)}$ respectively. Here we are substituting the
residue expressions for the eigenvalues given by (\ref{two.44a}).

The nested Bethe ansatz for one and two-particle state presented here
contain all information necessary to known what happens when a
multi-particle state is considered. The sequence of terms in the eigenvalues
and in the Bethe equations are common for all models and differences will
appear only in the last terms.

\section{The multi-particle Bethe state}

Generalization of previous results in order to consider Bethe states with
more the two particles follows from \cite{Martins4} where the vector $\Phi
_{m}^{(l)}$ was defined through the following recurrence formula: 
\begin{eqnarray}
&&\left. \Phi _{m}^{(l)}(\lambda _{1},...,\lambda _{m})=\mathcal{B}%
^{(l)}(\lambda _{1})\otimes \Phi _{m-1}^{(l)}(\lambda _{2},...,\lambda
_{m})\right.  \notag \\
&&\left. -B_{N_{l}-1}(\lambda _{1})\sum_{j=2}^{m}\frac{\hat{Y}_{N_{l}\
2}^{(l)}(\lambda _{1}-\lambda _{j})}{y_{N_{l}N_{l}}^{(l)}(\lambda
_{1}-\lambda _{j})}\otimes \Phi _{m-2}^{(l)}(\overset{\wedge }{\lambda }%
_{j})\prod_{k=2}^{j-1}\frac{S_{k,k+1}^{(l+1)}(\lambda _{k}-\lambda _{j})}{%
x_{1}^{(l+1)}(\lambda _{k}-\lambda _{j})}\prod\limits_{\substack{ k=2  \\ %
k\neq j}}^{m}z^{(l)}(\lambda _{k}-\lambda _{j})A_{1}^{(l)}(\lambda
_{j})\right.  \label{mut.1}
\end{eqnarray}%
with the initial condition $\Phi _{0}^{(l)}=1,\quad \Phi _{1}^{(l)}(\lambda
)=\mathcal{B}^{(l)}(\lambda )$. Here we have used the notation $(\overset{%
\wedge }{\lambda }_{j})$ to indicate the absence of the spectral parameter $%
\lambda _{j}$.

In order to proceed with the Bethe ansatz construction we must compute the
action of the diagonal operators $A_{i}^{(l)}(\lambda ),i=1,3$ \ and $%
\mathrm{Tr}_{a}[\mathcal{D}^{(l)}(\lambda )]$ of the monodromy matrix on the
vectors $\Phi _{m}^{(l)}$. This procedure is really very laborious due to
the normal-ordered condition but one can do it recursively. Acting with $%
A_{1}^{(l)}(\lambda )$ \ on (\ref{mut.1}) we have the following
normal-ordered expression

\begin{eqnarray}
&&A_{1}^{(l)}(\lambda )\Phi _{m}^{(l)}(\lambda _{1},...,\lambda
_{n})=\prod_{k=1}^{m}z^{(l)}(\lambda _{k}-\lambda )\Phi _{m}^{(l)}(\lambda
_{1},...,\lambda _{m})A_{1}^{(l)}(\lambda )  \notag \\
&&-\sum_{j=1}^{m}\frac{x_{3}^{(l)}(\lambda _{j}-\lambda )}{%
x_{2}^{(l)}(\lambda _{j}-\lambda )}\mathcal{B}^{(l)}(\lambda )\otimes \Phi
_{m-1}^{(l)}(\overset{\wedge }{\lambda }_{j})\prod_{k=1}^{j-1}\frac{%
S_{k,k+1}^{(l+1)}(\lambda _{k}-\lambda _{j})}{x_{1}^{(l+1)}(\lambda
_{k}-\lambda _{j})}\prod\limits_{\substack{ k=1  \\ k\neq j}}%
^{m}z^{(l)}(\lambda _{k}-\lambda _{j})A_{1}^{(l)}(\lambda _{j})  \notag \\
&&+B_{N_{l}-1}(\lambda )\sum_{j=2}^{m}\sum_{p=1}^{j-1}\mathbf{G}%
_{jl}^{(l)}(\lambda ,\lambda _{p},\lambda _{j})\otimes \Phi _{m-2}^{(l)}(%
\overset{\wedge }{\lambda }_{p},\overset{\wedge }{\lambda }%
_{j})\prod_{k=1}^{p-1}\frac{S_{k+1,k+2}^{(l+1)}(\lambda _{k}-\lambda _{j})}{%
x_{1}^{(l+1)}(\lambda _{k}-\lambda _{j})}\prod\limits_{k=p+1}^{j-1}\frac{%
S_{k,k+1}^{(l+1)}(\lambda _{k}-\lambda _{j})}{x_{1}^{(l+1)}(\lambda
_{k}-\lambda _{j})}  \notag \\
&&\prod\limits_{k=1}^{p-1}\frac{S_{k,k+1}^{(l+1)}(\lambda _{k}-\lambda _{p})%
}{x_{1}^{(l+1)}(\lambda _{k}-\lambda _{p})}\prod\limits_{\substack{ k=1  \\ %
k\neq p,j}}^{m}z^{(l)}(\lambda _{k}-\lambda _{j})z^{(l)}(\lambda
_{k}-\lambda _{p})A_{1}^{(l)}(\lambda _{p})A_{1}^{(l)}(\lambda _{j})+\cdots
\label{mut.2}
\end{eqnarray}%
where the indices $k,k+1$ in the matrix $S_{k,k+1}^{(l+1)}$ \ denote the
spaces where its action is not trivial. $\mathbf{G}_{jp}^{(l)}(\lambda
,\lambda _{l},\lambda _{j})$ are matrix valued functions given by%
\begin{equation}
\mathbf{G}_{jp}^{(l)}(\lambda ,\lambda _{l},\lambda _{j})=\frac{%
x_{3}^{(l)}(\lambda _{j}-\lambda )}{x_{2}^{(l)}(\lambda _{j}-\lambda )}\frac{%
\hat{Y}_{N_{l}\ 2}^{(l)}(\lambda -\lambda _{p})}{y_{N_{l}N_{l}}^{(l)}(%
\lambda -\lambda _{p})}\frac{S^{(l+1)}(\lambda _{p}-\lambda )}{%
x_{2}^{(l+1)}(\lambda _{p}-\lambda )}+\frac{y_{1N_{l}}^{(l)}(\lambda
_{p}-\lambda )}{y_{N_{l}N_{l}}^{(l)}(\lambda _{p}-\lambda )}\frac{\hat{Y}%
_{N_{l}\ 2}^{(l)}(\lambda _{p}-\lambda _{j})}{y_{N_{l}N_{l}}^{(l)}(\lambda
_{p}-\lambda _{j})}  \label{mut.3}
\end{equation}%
In (\ref{mut.2}) one can easily identify the candidate for the wanted term
in the eigenvalue problem and two groups of unwanted terms. In each group
terms differ by cyclic permutations of rapidities and consequently, by the
presence of $S$ matrices.

The action of $A_{3}^{(l)}(\lambda )$ on $\Phi _{m}^{(l)}$ has a similar
form 
\begin{eqnarray}
&&A_{3}^{(l)}(\lambda )\Phi _{m}^{(l)}(\lambda _{1},...,\lambda
_{m})=\prod_{k=1}^{m}\frac{x_{2}^{(l)}(\lambda -\lambda _{k})}{%
y_{N_{l}N_{l}}^{(l)}(\lambda -\lambda _{k})}\Phi _{m}^{(l)}(\lambda
_{1},...,\lambda _{m})A_{3}^{(l)}(\lambda )  \notag \\
&&-\sum_{j=1}^{m}\frac{\hat{Y}_{N_{l}\ 2}^{(l)}(\lambda -\lambda _{j})}{%
y_{N_{l}N_{l}}^{(l)}(\lambda -\lambda _{j})}\mathcal{B}^{\ast (l)}(\lambda
)\otimes \Phi _{m-1}^{(l)}(\overset{\wedge }{\lambda }_{j})\otimes \mathcal{D%
}^{(l)}(\lambda _{j})\prod_{k=j+1}^{m}\frac{S_{k-1,k}^{(l+1)}(\lambda
_{j}-\lambda _{k})}{x_{1}^{(l+1)}(\lambda _{j}-\lambda _{k})}\prod\limits 
_{\substack{ k=1  \\ k\neq j}}^{m}z^{(l)}(\lambda _{j}-\lambda _{k})  \notag
\\
&&+B_{N_{l}-1}(\lambda )\sum_{j=2}^{m}\sum_{p=1}^{j-1}\mathbf{H}%
_{jp}^{(l)}(\lambda ,\lambda _{p},\lambda _{j})\Phi _{m-2}^{(l)}(\overset{%
\wedge }{\lambda }_{p},\overset{\wedge }{\lambda }_{j})\otimes \mathcal{D}%
^{(l)}(\lambda _{p})\otimes \mathcal{D}^{(l)}(\lambda
_{j})\prod\limits_{k=j+1}^{m}\frac{S_{k-1,k}^{(l+1)}(\lambda _{j}-\lambda
_{k})}{x_{1}^{(l+1)}(\lambda _{j}-\lambda _{k})}  \notag \\
\times &&\prod_{k=j+1}^{m}\frac{S_{k-2,k-1}^{(l+1)}(\lambda _{l}-\lambda
_{k})}{x_{1}^{(l+1)}(\lambda _{l}-\lambda _{k})}\prod\limits_{k=p+1}^{j-1}%
\frac{S_{k-1,k}^{(l+1)}(\lambda _{p}-\lambda _{k})}{x_{1}^{(l+1)}(\lambda
_{l}-\lambda _{k})}\prod\limits_{\substack{ k=1  \\ k\neq p,j}}%
^{m}z^{(l)}(\lambda _{p}-\lambda _{k})z^{(l)}(\lambda _{j}-\lambda
_{k})+\cdots  \label{mut.4}
\end{eqnarray}%
where we have the following matrix valued functions%
\begin{equation}
\mathbf{H}_{jp}^{(l)}(\lambda ,\lambda _{l},\lambda _{j})=\frac{%
y_{N_{l}1}^{(l)}(\lambda -\lambda _{p})}{y_{N_{l}N_{l}}^{(l)}(\lambda
-\lambda _{p})}\frac{\hat{Y}_{N_{l}\ 2}^{(l)}(\lambda _{p}-\lambda _{j})}{%
y_{N_{l}\ N_{l}}^{(l)}(\lambda _{p}-\lambda _{j})}-\frac{x_{3}^{(l)}(\lambda
-\lambda _{p})}{y_{N_{l}N_{l}}^{(l)}(\lambda -\lambda _{p})}\frac{\hat{Y}%
_{N_{l}2}^{(l)}(\lambda -\lambda _{j})}{y_{N_{l}N_{l}}^{(l)}(\lambda
-\lambda _{j})}  \label{mut.5}
\end{equation}%
In (\ref{mut.4}) we have a wanted term and presence of a new group of
unwanted terms.

Acting with $\mathrm{Tr}_{a}[\mathcal{D}(\lambda )]$ \ on the vector $\Phi
_{m}^{(l)}$\ the final expression is more cumbersome 
\begin{equation*}
\mathrm{Tr}_{a}[\mathcal{D}^{(l)}(\lambda )]\Phi _{m}^{(l)}(\lambda
_{1},\ldots ,\lambda _{m})=\Phi _{m}^{(l)}(\lambda _{1},...,\lambda _{m})%
\mathrm{Tr}_{a}[\frac{\mathcal{L}_{am}^{(l+1)}(\lambda -\lambda _{m})}{%
x_{2}^{(l+1)}(\lambda -\lambda _{m)}}\cdots \frac{\mathcal{L}%
_{a1}^{(l+1)}(\lambda -\lambda _{1})}{x_{2}^{(l+1)}(\lambda -\lambda _{1)}}%
\mathcal{D}^{(l)}(\lambda )]
\end{equation*}%
\begin{equation*}
-\sum_{j=1}^{m}\frac{x_{4}^{(l)}(\lambda -\lambda _{j})}{x_{2}^{(l)}(\lambda
-\lambda _{j})}B^{(l)}(\lambda )\otimes \Phi _{m-1}^{(l)}(\overset{\wedge }{%
\lambda }_{j})\mathcal{D}^{(l)}(\lambda _{j})\otimes \mathbf{1}^{\otimes
(m-1)}\prod\limits_{k=j+1}^{m}\frac{\mathcal{R}_{k-1,k}^{(l+1)}(\lambda
_{j}-\lambda _{k})}{x_{1}^{(l+1)}(\lambda _{j}-\lambda _{k})}\prod\limits 
_{\substack{ k=1  \\ k\neq j}}^{m}z^{(l)}(\lambda _{j}-\lambda _{k})
\end{equation*}%
\begin{equation*}
+\sum_{j=1}^{m}\frac{\hat{Y}_{N_{l}\ 2}^{(l)}(\lambda -\lambda _{j})}{%
y_{N_{l}N_{l}}^{(l)}(\lambda -\lambda _{j})}B^{\ast (l)}(\lambda )\otimes
\Phi _{m-1}^{(l)}(\overset{\wedge }{\lambda }_{j})\otimes \mathbf{1}%
^{\otimes (m-2)}\prod\limits_{k=1}^{j-1}\frac{\mathcal{R}_{k,k+1}^{(l+1)}(%
\lambda _{k}-\lambda _{j})}{x_{1}^{(l+1)}(\lambda _{k}-\lambda _{j})}%
\prod\limits_{\substack{ k=1  \\ k\neq j}}^{m}z^{(l)}(\lambda _{k}-\lambda
_{j})A_{1}^{(l)}(\lambda _{j})
\end{equation*}%
\begin{eqnarray}
&&+B_{N_{l}-1}(\lambda )\sum_{j=2}^{m}\sum_{p=1}^{j-1}\mathbf{Y}%
_{jp}^{(l)}(\lambda ,\lambda _{p},\lambda _{j})\otimes \Phi _{m-2}^{(l)}(%
\overset{\wedge }{\lambda }_{p},\overset{\wedge }{\lambda }_{j})\mathcal{D}%
^{(l)}(\lambda _{j})\otimes \mathbf{1}^{\otimes (m-1)}\prod_{k=1}^{p-1}\frac{%
\mathcal{R}_{k,k+1}^{(l+1)}(\lambda _{k}-\lambda _{p})}{x_{1}^{(l+1)}(%
\lambda _{k}-\lambda _{p})}  \notag \\
&&\times \prod\limits_{k=j+1}^{m}\frac{\mathcal{R}_{k-1,k}^{(l+1)}(\lambda
_{j}-\lambda _{k})}{x_{1}^{(l+1)}(\lambda _{j}-\lambda _{k})}\prod\limits 
_{\substack{ k=1  \\ k\neq p.j}}^{m}z^{(l)}(\lambda _{k}-\lambda
_{p})z^{(l)}(\lambda _{j}-\lambda _{k})A_{1}^{(l)}(\lambda _{p})  \notag \\
&&+B_{N_{l}-1}(\lambda )\sum_{j=2}^{m}\sum_{p=1}^{j-1}\mathbf{Y}%
_{jp}^{(l)}(\lambda ,\lambda _{j},\lambda _{p})\otimes \Phi _{m-2}^{(l)}(%
\overset{\wedge }{\lambda }_{p},\overset{\wedge }{\lambda }_{j})\mathcal{D}%
^{(l)}(\lambda _{p})\otimes \mathbf{1}^{\otimes (m-1)}\prod_{k=1}^{l-1}\frac{%
\mathcal{R}^{(l+1)}(\lambda _{k}-\lambda _{j})}{x_{1}^{(l+1)}(\lambda
_{k}-\lambda _{j})}  \notag \\
&&\times \prod\limits_{k=j+1}^{m}\frac{\mathcal{R}_{k-1,k}^{(l+1)}(\lambda
_{p}-\lambda _{k})}{x_{1}^{(l+1)}(\lambda _{p}-\lambda _{k})}\prod\limits 
_{\substack{ k=1  \\ k\neq p,j}}^{m}z^{(l)}(\lambda _{k}-\lambda
_{p})z^{(l)}(\lambda _{j}-\lambda _{k})\frac{\mathcal{R}_{lj}^{(l+1)}(%
\lambda _{p}-\lambda _{j})}{x_{1}^{(l+1)}(\lambda _{p}-\lambda _{j})}%
A_{1}^{(l)}(\lambda _{j})+\cdots  \label{mut.6}
\end{eqnarray}%
where we have defined the matrix valued functions%
\begin{equation}
\mathbf{Y}_{jp}^{(l)}(\lambda ,\lambda _{p},\lambda _{j})=[z^{(l)}(\lambda
-\lambda _{p})\frac{x_{4}^{(l)}(\lambda -\lambda _{j})}{x_{2}^{(l)}(\lambda
-\lambda _{j})}-\frac{x_{4}^{(l)}(\lambda -\lambda _{p})}{%
x_{2}^{(l)}(\lambda -\lambda _{p})}\frac{x_{4}^{(l)}(\lambda _{p}-\lambda
_{j})}{x_{2}^{(l)}(\lambda _{p}-\lambda _{j})}]\frac{\hat{Y}%
_{N_{l}2}^{(l)}(\lambda -\lambda _{p})}{y_{N_{l}N_{l}}^{(l)}(\lambda
-\lambda _{p})}  \label{mut.7}
\end{equation}%
Here we observe the presence of $\mathcal{R}$ matrices instead of $S$
matrices as in the expressions for $A_{1}(\lambda )$ and $A_{3}(\lambda )$.
This difference is fundamental for the Bethe ansatz construction. For
instance, the trace (\ref{mut.6}) is to be understood as a $L+m$ site
inhomogeneous transfer matrix with $m$ inhomogeneous sites coming from the
Lax operators $\mathcal{L}^{(l+1)}(\lambda -\lambda _{k})\circeq \mathcal{R}%
^{(l+1)}(\lambda -\lambda _{k})$. The solution of this inhomogeneous
eigenvalue problem is the candidate for the wanted term in the eigenvalue
problem of the $L$ site homogeneous transfer matrix $\tau _{L}(\lambda )$.
Moreover, the remained terms of (\ref{mut.6}) have the exact form to cancel
the unwanted terms coming from (\ref{mut.2}) and (\ref{mut.4}). As before,
the ellipses used in (\ref{mut.2}), (\ref{mut.4}) and (\ref{mut.6}) denote
normally ordered terms containing annihilation operators.

In our nested Bethe ansatz language the \emph{ground} ($l=0$) for a
particular vertex model is prepared by a $L$ site homogeneous transfer
matrix $\tau _{L}(\lambda )$%
\begin{equation}
\tau _{L}(\lambda )=A_{1}(\lambda )+\sum\limits_{\alpha =1}^{N-2}D_{\alpha
\alpha }(\lambda )+A_{3}(\lambda )  \label{mut.8}
\end{equation}%
and the multi-particle Bethe state is defined by the linear combination%
\begin{equation}
\Psi _{m}(\lambda _{1},\cdots ,\lambda _{m})=\Phi _{m}(\lambda
_{1},...,\lambda _{m})\mathcal{F}_{m}\left\vert 0_{L}\right\rangle
\label{mut.9}
\end{equation}%
where $\mathcal{F}_{m}$ is a vector matrix with $(N-2)^{m}$ entries $%
f^{\alpha _{1}\cdots \alpha _{m}}$.

The corresponding eigenvalue is obtained from the first term on the right
hand side of (\ref{mut.2}), (\ref{mut.4}) and (\ref{mut.6}):%
\begin{equation}
\Lambda _{L}(\lambda |\{\lambda _{i}\})=X_{1}(\lambda
)\prod\limits_{k=1}^{m}z(\lambda _{k}-\lambda )+X_{3}(\lambda
)\prod\limits_{k=1}^{m}\frac{x_{2}(\lambda -\lambda _{k})}{y_{NN}(\lambda
-\lambda _{k})}+X_{2}(\lambda )\frac{\Lambda _{m}^{(1)}(\lambda ,\{\lambda
_{i}\}|\{\lambda _{i}^{(1)}\})}{X_{2}^{(1)}(\lambda )}  \label{mut.10}
\end{equation}%
Here we have used (\ref{eig.9}) and $\Lambda _{m}^{(1)}(\lambda ,\{\lambda
_{i}\}|\{\lambda _{i}^{(1)}\})$ is the eigenvalue of the eigenvalue problem
for a $m$-site row-to-row inhomogeneous transfer matrix with its Lax
operators identified with the matrices $\mathcal{R}^{(1)}(\lambda -\lambda
_{k})$, $k=1,...,m$ \ where $\lambda _{k}$ are the inhomogenity parameters
and $\{\lambda _{i}^{(1)}\}$ are rapidities: 
\begin{eqnarray}
\tau _{m}^{(1)}(\lambda )\mathcal{F}_{m} &=&\mathrm{Tr}_{a}\left( \mathcal{L}%
_{am}^{(1)}(\lambda -\lambda _{m})\cdots \mathcal{L}_{a1}^{(1)}(\lambda
-\lambda _{1}\right) \mathcal{F}_{m}  \notag \\
&=&\Lambda _{m}^{(1)}(\lambda ,\{\lambda _{i}\}|\{\lambda _{i}^{(1)}\})%
\mathcal{F}_{m}  \label{mut.11}
\end{eqnarray}%
where the choice 
\begin{equation}
\mathcal{F}_{m}=\Phi _{m}^{(1)}(\lambda _{1}^{(1)},...,\lambda
_{m}^{(1)})\left\vert 0_{m}^{(1)}\right\rangle  \label{mut.12}
\end{equation}%
is implicit.

The remained terms of (\ref{mut.2}), (\ref{mut.4}) and (\ref{mut.6})
multiplied by $\mathcal{F}_{m}\left\vert 0_{L}\right\rangle $ are known as
unwanted terms. There are many of these terms but they can be collected in
only three different groups. The \ first group contains $m$ terms of the
type $\mathcal{B}(\lambda )\otimes \Phi _{m-1}(\overset{\wedge }{\lambda }%
_{j})$. To see how is proceeding the cancel in this group, we start with $%
j=1 $. The corresponding term has the form 
\begin{equation}
-\frac{x_{3}(\lambda _{1}-\lambda )}{x_{2}(\lambda _{1}-\lambda )}\left(
X_{1}(\lambda _{1})\prod\limits_{k=2}^{m}z(\lambda _{k}-\lambda
_{1})-X_{2}(\lambda _{1})\prod\limits_{k=2}^{m}\frac{\mathcal{R}%
_{k-1,k}^{(1)}(\lambda _{1}-\lambda _{k})}{x_{1}^{(1)}(\lambda _{1}-\lambda
_{k})}\prod\limits_{k=2}^{m}z(\lambda _{1}-\lambda _{k})\right) \mathcal{F}%
_{m}\left\vert 0_{L}\right\rangle  \label{mut.13}
\end{equation}%
Now we take the limit of $\tau ^{(1)}(\lambda )$ at $\lambda =\lambda _{1}$
in order to identify the product of $\mathcal{R}$ matrices%
\begin{eqnarray}
\tau _{m}^{(1)}(\lambda _{1}) &=&\underset{\lambda =\lambda _{1}}{\mathrm{%
\lim }}\tau _{m}^{(1)}(\lambda )=\underset{\lambda =\lambda _{1}}{\mathrm{%
\lim }}\mathrm{Tr}_{a}\left( \mathcal{L}_{am}^{(1)}(\lambda -\lambda
_{m})\cdots \mathcal{L}_{a1}^{(1)}(\lambda -\lambda _{1}\right)  \notag \\
&=&\prod\limits_{k=1}^{m-1}\mathcal{R}_{k,k+1}^{(1)}(\lambda _{1}-\lambda
_{k})  \label{mut.14}
\end{eqnarray}%
Therefore we can use the solution of the second eigenvalue problem (\ref%
{mut.11}) at $\lambda =\lambda _{1}$ to see that this term will be cancelled
provided that%
\begin{equation}
\frac{X_{1}(\lambda _{1})}{X_{2}(\lambda _{1})}\mathcal{F}_{m}\left\vert
0_{L}\right\rangle =\prod\limits_{k=2}^{m}\frac{z(\lambda _{1}-\lambda _{k})%
}{z(\lambda _{k}-\lambda _{1})}\Lambda _{m}^{(1)}(\lambda _{1},\{\lambda
_{i}\}|\{\lambda _{i}^{(1)}\})\prod\limits_{k=2}^{m}\frac{1}{%
x_{1}^{(1)}(\lambda _{1}-\lambda _{k})}\mathcal{F}_{m}\left\vert
0_{L}\right\rangle  \label{mut.15}
\end{equation}%
The remained terms of this group can be written as cyclic permutations of (%
\ref{mut.13}). Thus, we can recall the relations (\ref{two.30})-(\ref{two.32}%
) to prove that all terms of this group are eliminated provided that%
\begin{eqnarray}
\frac{X_{1}(\lambda _{a})}{X_{2}(\lambda _{a})} &=&\left(
\prod\limits_{b\neq a}^{m}\frac{z(\lambda _{a}-\lambda _{b})}{z(\lambda
_{b}-\lambda _{a})}\right) \frac{\Lambda _{m}^{(1)}(\lambda _{a},\{\lambda
_{i}\}|\{\lambda _{i}^{(1)}\})}{X^{(1)}(\lambda _{a})}  \notag \\
a &=&1,2,...,m  \label{mut.16}
\end{eqnarray}%
where%
\begin{equation}
X^{(1)}(\lambda _{a})=\prod\limits_{b\neq a}^{m}x_{1}^{(1)}(\lambda
_{a}-\lambda _{b}).  \label{mut.16a}
\end{equation}

The second group contains $m$ terms of the type $\mathcal{B}^{\ast }(\lambda
)\otimes \Phi _{m-1}(\overset{\wedge }{\lambda }_{j})$ and they are also
cancelled by the \emph{partial} Bethe equations (\ref{mut.16}). To accept
this statement one can follow the same steps used in the first group.
Finally, the third group contains of the type $B_{N-1}(\lambda )\otimes \Phi
_{m-2}(\overset{\wedge }{\lambda }_{p},\overset{\wedge }{\lambda }_{j})$ are
also cancelled by (\ref{mut.16}). Here we would like to stress that the
technicalities involving the calculation of the third group of unwanted
terms are very laborious.

At this point we have concluded the first step in an algebraic nested Bethe
ansatz. The next step consists in taking into account the auxiliary
eigenvalue problem (\ref{mut.11}). After we have presented the results for
one and two particle states it is not frivolous to affirm that we will only
need to repeat everything once more with trivial modifications. \ It is
enough replace $L$ by $m$ and introduce the inhomogeneity parameters \{$%
\lambda _{k}^{(l-1)}$\} and the rapidities \{$\lambda _{k}^{(l)}$\} for each
label $l$. \ Indeed this is true until we arrive at the last \emph{layer} $%
l=n-1$, where the models behave differently.

The last \emph{layer}s for the B$_{n}^{(1)}$ and A$_{2n}^{(2)}$ vertex
models are building with the B$_{1}^{(1)}$ Zamolodchivok-Fateev model and
the A$_{2}^{(2)}$ Izegin-Korepin model \cite{Tarasov, ALS}, respectively. \
For D$_{n}^{(1)}$ models the last \emph{layer} is the D$_{2}^{(1)}$ vertex
model which is mapped in a direct product of two A$_{1}^{(1)}$ six-vertex
models. The C$_{n}^{(2)}$ and A$_{2n-1}^{(2)}$ models have their last \emph{%
layer} extended to $l=n$ \ where we will find the C$_{1}^{(1)}$ and A$%
_{1}^{(2)}$ six-vertex models, respectively.

From these considerations we can summarize the results in the following way:
\ the eigenvalue of the $L$ site homogeneous transfer matrix $\tau
_{L}(\lambda )$ for a m-particle Bethe state is given by

\begin{eqnarray}
\Lambda _{L}(\lambda |\{\lambda _{i}\}) &=&X_{1}(\lambda
)\prod\limits_{k=1}^{m}z(\lambda _{k}-\lambda )+X_{3}(\lambda
)\prod\limits_{k=1}^{m}\frac{x_{2}(\lambda -\lambda _{k})}{y_{NN}(\lambda
-\lambda _{k})}  \notag \\
&&+X_{2}(\lambda )\left( \sum_{l=1}^{n-2}G_{m}^{(l)}(\lambda ,\{\lambda
_{i}^{(l-1)}\}|\{\lambda _{i}^{(l)}\})+\mathcal{T}\right)  \label{mut.17a}
\end{eqnarray}%
where%
\begin{equation}
G_{m}^{(l)}(\lambda ,\{\lambda _{i}^{(l-1)}\}|\{\lambda _{i}^{(l)}\})=\frac{%
X_{1}^{(l)}(\lambda )}{X_{2}^{(l)}(\lambda )}\prod\limits_{k=1}^{m}z^{(l)}(%
\lambda _{k}^{(l)}-\lambda )+\frac{X_{3}^{(l)}(\lambda )}{%
X_{2}^{(l)}(\lambda )}\prod\limits_{k=1}^{m}\frac{x_{2}^{(l)}(\lambda
-\lambda _{k}^{(l)})}{y_{N_{l}N_{l}}^{(l)}(\lambda -\lambda _{k}^{(l)})}
\label{mut.17}
\end{equation}%
and we are again using the shorthand notation presented in (\ref{two.26})
and (\ref{two.27}). i.e.,%
\begin{eqnarray}
X_{a}(\lambda ) &=&[x_{a}(\lambda )]^{L},\quad \qquad X_{a}^{(l)}(\lambda
)=\prod\limits_{k=1}^{m}x_{a}^{(l)}(\lambda -\lambda _{k}^{(l-1)})\qquad
a=1,2.\qquad  \notag \\
X_{3}(\lambda ) &=&[y_{NN}(\lambda )]^{L},\qquad \ \ X_{3}^{(l)}(\lambda
)=\prod\limits_{k=1}^{m}y_{N_{l}N_{l}}^{(l)}(\lambda -\lambda _{k}^{(l-1)})
\label{mut.18}
\end{eqnarray}%
The $\mathcal{T}$ \ term in (\ref{mut.17a}) makes the result difference for
different models:%
\begin{equation}
\mathcal{T}=G_{m}^{(n-1)}(\lambda ,\{\lambda _{i}^{(n-2)}\}|\{\lambda
_{i}^{(n-1)}\})+\prod\limits_{k=1}^{m}\frac{z^{(n-1)}(\lambda -\lambda
_{k}^{(n-1)})}{\omega (\lambda -\lambda _{k}^{(n-1)})}  \label{mut.19}
\end{equation}%
for B$_{n}^{(1)}$ and A$_{2n}^{(2)}$ vertex models,%
\begin{equation}
\mathcal{T}=G_{m}^{(n-1)}(\lambda ,\{\lambda _{i}^{(n-2)}\}|\{\lambda
_{i}^{(n-1)}\})+G_{m}^{(n)}(\lambda ,\{\lambda _{i}^{(n-1)}\}|\{\lambda
_{i}^{(n)}\})  \label{mut.20}
\end{equation}%
for C$_{n}^{(1)}$ and A$_{2n-1}^{(2)}$ vertex models. For D$_{n}^{(1)}$
vertex models the $\mathcal{T}$ term is a little bit different due to the
direct product of two A$_{1}^{(1)}$ six-vertex models: 
\begin{eqnarray}
\mathcal{T} &=&[X_{1}^{+}(\lambda )\prod\limits_{k=1}^{m}z(\lambda
_{k}^{+}-\lambda )+X_{2}^{+}(\lambda )\prod\limits_{k=1}^{m}z(\lambda
-\lambda _{k}^{+})]  \notag \\
&&\times \lbrack X_{1}^{-}(\lambda )\prod\limits_{k=1}^{m}z(\lambda
_{k}^{-}-\lambda )+X_{2}^{-}(\lambda )\prod\limits_{k=1}^{m}z(\lambda
-\lambda _{k}^{-})]  \label{mut.21}
\end{eqnarray}%
where%
\begin{eqnarray}
z(\lambda ^{\pm }) &=&\frac{x_{1}(\lambda ^{\pm })}{x_{2}(\lambda ^{\pm })}%
,\qquad X_{a}^{\pm }(\lambda )=\prod\limits_{k=1}^{m}x_{a}(\lambda -\lambda
_{k}^{n-3}),\qquad (a=1,2)  \notag \\
\{x_{1},x_{2}\} &\in &\mathrm{A}_{1}^{(1)}  \label{mut.21a}
\end{eqnarray}

\bigskip The \ corresponding sequences of Bethe equations also have a common
part 
\begin{equation*}
\frac{X_{1}(\lambda _{a})}{X_{2}(\lambda _{a})}=\prod\limits_{b\neq a}^{m}%
\frac{z(\lambda _{a}-\lambda _{b})}{z(\lambda _{b}-\lambda _{a})}%
x_{1}^{(1)}(0)\prod\limits_{k=1}^{m}z^{(1)}(\lambda _{k}^{(1)}-\lambda
_{a})\quad
\end{equation*}%
\begin{eqnarray}
\frac{X_{1}^{(l)}(\lambda _{a}^{(l)})}{X_{2}^{(l)}(\lambda _{a}^{(l)})}
&=&\prod\limits_{b\neq a}^{m}\frac{z^{(l)}(\lambda _{a}^{(l)}-\lambda
_{b}^{(l)})}{z^{(l)}(\lambda _{b}^{(l)}-\lambda _{a}^{(l)})}%
x_{1}^{(l+1)}(0)\prod\limits_{k=1}^{m}z^{(l+1)}(\lambda _{k}^{(l+1)}-\lambda
_{a}^{(l)}),  \notag \\
l &=&1,2,...,n-2.\quad a\neq b=1,2,...,m  \label{mut.22}
\end{eqnarray}%
where we are substituting the residues%
\begin{eqnarray}
\Lambda _{m}^{(l+1)}(\lambda _{a}^{(l)},\{\lambda _{i}^{(l)}\}|\{\lambda
_{i}^{(l+1)}\}) &=&\underset{\lambda =\lambda _{a}^{(l)}}{\mathrm{res}}%
\Lambda _{m}^{(l+1)}(\lambda ,\{\lambda _{i}^{(l)}\}|\{\lambda
_{i}^{(l+1)}\})  \notag \\
&=&x_{1}^{(l+1)}(0)X_{1}^{(l+1)}(\lambda
_{a}^{(l)})\prod\limits_{k=1}^{2}z^{(l+1)}(\lambda _{k}^{(l+1)}-\lambda
_{a}^{(l)})  \notag \\
l &=&0,1,...,n-2,\quad a\neq b=1,2,...,m.  \label{mut.22a}
\end{eqnarray}

As happened with the eigenvalue sequence (\ref{mut.17a}) this sequence will
end in different ways depending on the model:%
\begin{equation}
\frac{X_{1}^{(n-1)}(\lambda _{a}^{(n-1)})}{X_{2}^{(n-1)}(\lambda
_{a}^{(n-1)})}=\prod\limits_{b\neq a}^{m}\frac{z^{(n-1)}(\lambda
_{a}^{(n-1)}-\lambda _{b}^{(n-1)})}{z^{(n-1)}(\lambda _{b}^{(n-1)}-\lambda
_{a}^{(n-1)})}\omega (\lambda _{b}^{(n-1)}-\lambda _{a}^{(n-1)})
\label{mut.23}
\end{equation}%
for B$_{n}^{(1)}$ and A$_{2n}^{(2)}$ vertex models,%
\begin{eqnarray}
\prod\limits_{k=1}^{m}\frac{x_{1}(\lambda _{a}^{\pm }-\lambda _{k}^{(n-2)})}{%
x_{2}(\lambda _{a}^{\pm }-\lambda _{k}^{(n-2)})} &=&\prod\limits_{b\neq
a}^{m}\frac{z(\lambda _{a}^{\pm }-\lambda _{b}^{\pm })}{z(\lambda _{b}^{\pm
}-\lambda _{a}^{\pm })},  \notag \\
\{x_{1},x_{2}\} &\in &\mathrm{A}_{1}^{(1)}  \label{mut.24}
\end{eqnarray}%
for D$_{n}^{(1)}$ vertex models. For C$_{n}^{(1)}$ and A$_{2n-1}^{(2)}$%
models we have more two equations: one equation for the \emph{layer} $l=n-1$ 
\begin{equation*}
\frac{X_{1}^{(n-1)}(\lambda _{a}^{(n-1)})}{X_{2}^{(n-1)}(\lambda
_{a}^{(n-1)})}=\prod\limits_{b\neq a}^{m}\frac{z^{(n-1)}(\lambda
_{a}^{(n-1)}-\lambda _{b}^{(n-1)})}{z^{(n-1)}(\lambda _{b}^{(n-1)}-\lambda
_{a}^{(n-1)})}x_{1}^{(n)}(0)\prod\limits_{k=1}^{m}z^{(n)}(\lambda
_{k}^{(n)}-\lambda _{a}^{(n-1)}),\quad
\end{equation*}%
and other equation for the \emph{layer} $l=n$%
\begin{equation}
\frac{X_{1}^{(n)}(\lambda _{a}^{(n)})}{X_{3}^{(n)}(\lambda _{a}^{(n)})}%
=\prod\limits_{b\neq a}^{m}\frac{x_{1}^{(n)}(\lambda _{a}^{(n)}-\lambda
_{b}^{(n)})}{x_{1}^{(n)}(\lambda _{b}^{(n)}-\lambda _{a}^{(n)})}\frac{%
y_{22}^{(n)}(\lambda _{b}^{(n)}-\lambda _{a}^{(n)})}{y_{22}^{(n)}(\lambda
_{a}^{(n)}-\lambda _{b}^{(n)})}  \label{mut.25}
\end{equation}%
Here we note that the function $\omega (\lambda )$ used in the B$_{n}^{(1)}$
and A$_{2n}^{(2)}$ was defined in (\ref{two.38}).

There is a technical point which we already touched in the one-particle
case. To solve the eigenvalue problem for a $L$ site homogeneous transfer
matrix $\tau _{L}(\lambda )$ with eigenstate $\Psi _{m}(\lambda _{1},\lambda
_{2},\cdots ,\lambda _{m})$ we are left with a second eigenvalue problem for
a $m$ site inhomogeneous transfer matrix $\tau _{m}^{(1)}(\lambda ,\{\lambda
_{i}\})$ for which the vector $\mathcal{F}_{m}$ must be an eigenstate.
However, $\mathcal{F}_{m}$ defines $\Psi _{m}$ as a linear combinations of
the components of the vector $\Phi _{m}$. The dimensions of $\Phi _{m}$ , $%
\left\vert 0_{m}\right\rangle $ and $\mathcal{F}_{m}$ \ are suggesting the
choice of $\mathcal{F}_{m}$ \ as a $m$-particle state in the second
eigenvalue problem. \ The choice of $\mathcal{F}_{m}$ as a $r$-particle
state could give particular value for $\Psi _{m}$ if $r<m$ and, increasing
considerably the number of parameters \{$\lambda _{i}^{(l)}$\} in each \emph{%
layer} if $r>m$ . However, these different choices are not necessary to fix
the rapidities of the states via the Bethe equations.

\section{Conclusion}

In this paper the nested Bethe ansatz \ formulation is used to solve exactly
a series of trigonometric vertex models based on the non-exceptional Lie
algebras. Here a detailed account of this method was described in order to
complement the results in the literature \cite{Martins1}-\cite{Martins4}.

There are several issues for which this paper could useful:

a) The off-shell Bethe ansatz - Gaudin theory - Solution of the
Kniznik-Zamolodchikov equation. The explicit expressions for the eigenvalue
problem \ as presented in this paper define the off-shell Bethe ansatz
equation. Now, if one can extend the Babujian and Flume formalism \cite{BF}
for nested Bethe ansatz such a issue could be possible.

b) Graded matrix Bethe ansatz. A recent work \cite{Martins3} about the
vertex models based on superalgebras assures the possibility to extend our
result for graded vertex models.

c) A Bethe ansatz with open boundary conditions. The analytical Bethe ansatz
for quantum-algebra-invariant spin chains \cite{Artz1, Artz2} gives us the
eigenvalues for the models considered in this paper with quantum open
boundary. The nested Bethe ansatz with diagonal reflection K-matrices was
used by de Vega and Gonz\'{a}lez-Ruiz \cite{de Vega}\ to find eigenvectors
and eigenvalues of the A$_{n}^{(1)}$ vertex models. More recently, the
nested Bethe ansatz \ with diagonal K-matrices boundary conditions for the B$%
_{n}^{(1)}$ vertex model \cite{Li1} and the A$_{2n}^{(2)}$ vertex models 
\cite{Li2} have been considered. Looking at the difficulties of obtaining
these results we can see how could be useful an unified Bethe ansatz
formulation with open boundary conditions.

\vspace{1cm}

\textbf{Acknowledgment:} This work was supported in part by Funda\c{c}\~{a}o
de Amparo \`{a} Pesquisa do Estado de S\~{a}o Paulo-\textsl{FAPESP}-Brasil
and Conselho Nacional de Desenvolvimento-\textsl{CNPq}-Brasil. The author
would like to thank M.J. Martins by his interest in this work. He also thank
to W. Galleas by the useful discussions.

\newpage

\section*{Appendix: A$_{n}^{(1)}$ vertex models}

\setcounter{equation}{0} \renewcommand{\theequation}{A.\arabic{equation}}

For sake of completeness we will describe in this appendix how the A$%
_{n}^{(1)}$ result is obtained by reduction from the our previous results

The A$_{n}^{(1)}$ matrix $\mathcal{R}$ has the form%
\begin{eqnarray}
\mathcal{R}^{(l)} &=&x_{1}\sum_{\alpha \neq \alpha ^{\prime
}}^{N_{l}}E_{\alpha \alpha }\otimes E_{\alpha \alpha }+x_{2}\sum_{\alpha
\neq \beta }^{N_{l}}E_{\alpha \alpha }\otimes E_{\beta \beta
}+x_{3}\sum_{\alpha <\beta }^{N_{l}}E_{\alpha \beta }\otimes E_{\beta \alpha
}  \notag \\
&&+x_{4}\sum_{\alpha >\beta }^{N_{l}}E_{\alpha \beta }\otimes E_{\beta
\alpha }  \label{An0}
\end{eqnarray}%
where the Boltzmann weights are given by 
\begin{eqnarray}
x_{1}(\lambda ) &=&\mathrm{e}^{\lambda }-q^{2},\qquad \ x_{2}(\lambda )=q(%
\mathrm{e}^{\lambda }-1)  \label{An0a} \\
x_{3}(\lambda ) &=&(q^{2}-1),\qquad x_{4}(\lambda )=\mathrm{e}^{\lambda
}x_{3}(\lambda )  \notag
\end{eqnarray}%
Here $N_{l}=n-l+1.$ \ Note that these weights are not labeled by $l$ because
they are the same ones for all A$_{n-l}^{(1)}$ models.

After we have identified the Lax operator with this $\mathcal{R}$ matrix the
corresponding monodromy matrix can be written as a $N_{l}$ by $N_{l}$ matrix 
\begin{equation}
T^{(l)}=\left( 
\begin{array}{ccccc}
A_{1}^{(l)} & B_{1}^{(l)} & B_{2}^{(l)} & \cdots & B_{N_{l}-1}^{(l)} \\ 
C_{1}^{(l)} & D_{11}^{(l)} & D_{12}^{(l)} & \cdots & D_{1,N_{l}-1}^{(l)} \\ 
C_{2}^{(l)} & D_{21}^{(l)} & D_{22}^{(l)} & \cdots & D_{2,N_{l}-1}^{(l)} \\ 
\vdots & \vdots & \vdots & \ddots & \vdots \\ 
C_{N_{l}-1}^{(l)} & D_{N_{l}-1,1}^{(l)} & D_{N_{l}-1,2}^{(l)} & \cdots & 
D_{N_{l}-1,N_{l}-1}^{(l)}%
\end{array}%
\right)  \label{An.1}
\end{equation}%
The usual reference state in a $L$ site homogeneous lattice is the highest
vector of $T^{(l)}$ 
\begin{eqnarray}
A_{1}^{(l)}(\lambda )\left\vert 0_{L}\right\rangle ^{(l)}
&=&X_{1}^{(l)}(\lambda )\left\vert 0_{L}\right\rangle ^{(l)},\quad D_{\alpha
\alpha }^{(l)}(\lambda )\left\vert 0_{L}\right\rangle
^{(l)}=X_{2}^{(l)}(\lambda )\left\vert 0_{L}\right\rangle ^{(l)}  \notag \\
C^{(l)}(\lambda )\left\vert 0_{L}\right\rangle ^{(l)} &=&0,\quad \qquad
\qquad \quad D_{\alpha \beta }^{(l)}(\lambda )\left\vert 0_{L}\right\rangle
^{(l)}=0  \notag \\
\quad B_{\alpha }^{(l)}(\lambda )\left\vert 0_{L}\right\rangle ^{(l)} &\neq
&\{0,\left\vert 0_{L}\right\rangle ^{(l)}\},\quad \quad \alpha \neq \beta
=1,2,...,N_{l}-1  \label{An.2}
\end{eqnarray}%
where%
\begin{equation}
X_{1}^{(l)}(\lambda )=[x_{1}(\lambda )]^{L}\qquad \mathrm{and}\qquad
X_{2}^{(l)}(\lambda )=[x_{2}(\lambda )]^{L}  \label{An.3}
\end{equation}%
Therefore we can write the monodromy matrix as a $2$ by $2$ matrix%
\begin{equation}
T^{(l)}(\lambda )=\left( 
\begin{array}{cc}
A_{1}^{(l)}(\lambda ) & \mathcal{B}^{(l)}\mathcal{(\lambda )} \\ 
\mathcal{C}^{(l)}\mathcal{(\lambda )} & \mathcal{D}^{(l)}\mathcal{(\lambda )}%
\end{array}%
\right)  \label{An.4}
\end{equation}%
where we identify a scalar $A^{(1)}(\lambda )$, two vector $\mathcal{B}%
^{(l)}(\lambda )$ and $\mathcal{C}^{(l)}(\lambda )$ with $N_{l}-1$ entries
and a $N_{l}-1$ by $N_{l}-1$ matrix $\mathcal{D}^{(l)}(\lambda ).$The
commutation relations among the matrix elements of (\ref{An.4}) can be
obtained using the intertwining relation with 
\begin{equation}
\lbrack S^{(l)}]=\left( 
\begin{array}{cccc}
x_{1} & 0 & 0 & 0 \\ 
0 & x_{4} & x_{2} & 0 \\ 
0 & x_{2} & x_{3} & 0 \\ 
0 & 0 & 0 & S^{(l+1)}%
\end{array}%
\right)  \label{An.5}
\end{equation}%
where $S^{(l+1)}$ is exactly the permutation of (\ref{mod.8}) with $l$
replaced by $l+1$.\ In (\ref{An.5}) $x_{1}$ is scalar and \{$%
x_{2},x_{3},x_{4}$\} are proportional to the identity matrix.

Now it is easy to see how our previous general results can be reduced in
order to obtain the well-known results of the A$_{n}^{(1)}$ vertex models 
\cite{Babelon}. \ Removing the third row and the third column of (\ref%
{eig.11}) we have (\ref{An.4}). It means that the entries $\left\{
B_{N_{l}-1},\mathcal{B}^{\ast },A_{3},\mathcal{C}^{\ast
},C_{N_{l}-1}\right\} $ are vanishing in the A$_{n}^{(1)}$ cases. Moreover,
removing from (\ref{fcr.1}) all row and column with entries $\{y_{\alpha
\beta }\}$ we will get (\ref{An.5}). Indeed these reductions already
expected since that the structure of $\mathcal{R}$ matrices (\ref{mod.1})
and (\ref{An0}) differ (up to normalization) by the $y_{\alpha \beta
}(\lambda )$ terms (\ref{mod.3}).\ 

Using these reductions in the previous results we are working with the A$%
_{n}^{(1)}$ vertex models. For instance, one can use the intertwining
relation (\ref{fcr.5}) with (\ref{An.4}) and (\ref{An.5}) in order to derive
the following commutation relations 
\begin{eqnarray}
A_{1}^{(l)}(\lambda )\mathcal{B}^{(l)}(\mu ) &=&z(\mu -\lambda )\mathcal{B}%
^{(l)}(\mu )A_{1}^{(l)}(\lambda )-\frac{x_{3}(\mu -\lambda )}{x_{2}(\mu
-\lambda )}\mathcal{B}^{(l)}(\lambda )A_{1}^{(l)}(\mu )  \notag \\
\mathrm{Tr}_{a}[\mathcal{D}^{(l)}(\lambda )]B^{(l)}(\mu ) &=&\mathcal{B}%
^{(l)}(\mu )\mathrm{Tr}_{a}\left[ \frac{\mathcal{L}^{(l+1)}(\lambda -\mu )}{%
x_{2}(\lambda -\mu )}\mathcal{D}^{(l)}(\lambda )\right] -\frac{x_{4}(\lambda
-\mu )}{x_{2}(\lambda -\mu )}\mathcal{B}^{(l)}(\lambda )\mathcal{D}%
^{(l)}(\mu )  \notag \\
C^{(l)}(\lambda )\otimes B^{(l)}(\mu ) &=&\mathcal{B}^{(l)}(\mu )\otimes 
\mathcal{C}^{(l)}(\lambda )+\frac{x_{4}(\lambda -\mu )}{x_{2}(\lambda -\mu )}%
\left[ A_{1}^{(l)}(\mu )\mathcal{D}^{(l)}(\lambda )-A_{1}^{(l)}(\lambda )%
\mathcal{D}^{(l)}(\mu )\right]  \notag \\
\mathcal{B}^{(l)}(\lambda )\otimes \mathcal{B}^{(l)}(\mu ) &=&\mathcal{B}%
^{(l)}(\mu )\otimes \mathcal{B}^{(l)}(\lambda )\frac{S^{(l+1)}(\lambda -\mu )%
}{x_{1}(\lambda -\mu )}  \label{An.6}
\end{eqnarray}%
which are the reductions of (\ref{fcr.8}), (\ref{one.2}), (\ref{two.9}) and (%
\ref{two.2}), respectively

Now we define the normal ordered $m$-particle vector by the reduction of (%
\ref{mut.1}) to%
\begin{equation}
\Phi _{m}^{(l)}(\lambda _{1},\lambda _{2},\ldots ,\lambda _{m})=\mathcal{B}%
^{(l)}(\lambda _{1})\otimes \mathcal{B}^{(l)}(\lambda _{2})\otimes \cdots
\otimes \mathcal{B}^{(l)}(\lambda _{m})  \label{An.7}
\end{equation}

Using (\ref{An.6}) one can compute the action of the diagonal elements of (%
\ref{An.4}) on this vector%
\begin{eqnarray}
&&A_{1}^{(l)}(\lambda )\Phi _{m}^{(l)}(\lambda _{1},\lambda _{2},\ldots
,\lambda _{m})  \notag \\
&=&\prod\limits_{k=1}^{m}z(\lambda _{k}-\lambda )\Phi _{m}^{(l)}(\lambda
_{1},\lambda _{2},\ldots ,\lambda _{m})A_{1}^{(l)}(\lambda )  \notag \\
&&-\sum_{j=1}^{m}\frac{x_{3}(\lambda _{j}-\lambda )}{x_{2}(\lambda
_{j}-\lambda )}\mathcal{B}^{(l)}(\lambda )\otimes \Phi _{m-1}^{(l)}(\overset{%
\wedge }{\lambda }_{j})\prod_{k=1}^{j-1}\frac{S_{k,k+1}^{(l+1)}(\lambda
_{k}-\lambda _{j})}{x_{1}(\lambda _{k}-\lambda _{j})}\prod\limits 
_{\substack{ k=1  \\ k\neq j}}^{m}z(\lambda _{k}-\lambda
_{j})A_{1}^{(l)}(\lambda _{j})  \label{An.8}
\end{eqnarray}%
and%
\begin{eqnarray}
&&\mathrm{Tr}_{a}[\mathcal{D}^{(l)}(\lambda )]\Phi _{m}^{(l)}(\lambda
_{1},\ldots ,\lambda _{m})  \notag \\
&=&\Phi _{m}^{(l)}(\lambda _{1},...,\lambda _{m})\mathrm{Tr}_{a}[\frac{%
\mathcal{L}_{am}^{(l+1)}(\lambda -\lambda _{m})}{x_{2}(\lambda -\lambda _{m})%
}\cdots \frac{\mathcal{L}_{a1}^{(l+1)}(\lambda -\lambda _{1})}{x_{2}(\lambda
-\lambda _{1})}\mathcal{D}^{(l)}(\lambda )]  \notag \\
&&-\sum_{j=1}^{m}\frac{x_{4}(\lambda -\lambda _{j})}{x_{2}(\lambda -\lambda
_{j})}B^{(l)}(\lambda )\otimes \Phi _{m-1}^{(l)}(\overset{\wedge }{\lambda }%
_{j})\mathcal{D}^{(l)}(\lambda _{j})\otimes \mathbf{1}^{(l)\otimes
(m-1)}\prod\limits_{k=j+1}^{m}\frac{\mathcal{R}_{k-1,k}^{(l+1)}(\lambda
_{j}-\lambda _{k})}{x_{1}(\lambda _{j}-\lambda _{k})}\prod\limits 
_{\substack{ k=1  \\ k\neq j}}^{m}z(\lambda _{j}-\lambda _{k})  \notag \\
&&  \label{An.9}
\end{eqnarray}%
or we can recall (\ref{mut.2}) and (\ref{mut.6}) in order to get these
expressions.

The eigenvalue problem on the \emph{ground} ($l=0$) is 
\begin{eqnarray}
\tau _{L}(\lambda )\Psi (\lambda _{1},\ldots ,\lambda _{m}) &=&\left(
A(\lambda )+\mathrm{Tr}_{a}[\mathcal{D}(\lambda )]\right) \Psi (\lambda
_{1},\ldots ,\lambda _{m})  \notag \\
&=&\Lambda _{L}(\lambda |\{\lambda _{i}\})\Psi (\lambda _{1},\ldots ,\lambda
_{m})  \label{An.10}
\end{eqnarray}%
where the $m$-particle Bethe state is defined by the linear combination%
\begin{equation*}
\Psi (\lambda _{1},\ldots ,\lambda _{m})=\Phi _{m}(\lambda _{1},\ldots
,\lambda _{m})\mathcal{F}_{m}\left\vert 0_{L}\right\rangle .
\end{equation*}%
Here, $\mathcal{F}_{m}$ is a vector with entries $f^{\alpha _{1}\cdots
\alpha _{m}},\ \alpha _{i}=1,2,...,N-1.$

Substituting (\ref{An.8}) and (\ref{An.9}) into (\ref{An.10}) the eigenvalue
problem gets the form%
\begin{eqnarray}
\tau _{L}(\lambda )\Psi _{m}(\lambda _{1},\ldots ,\lambda _{m})
&=&X_{1}(\lambda )\prod\limits_{k=1}^{m}z(\lambda _{k}-\lambda )\Psi
_{m}(\lambda _{1},\lambda _{2},\ldots ,\lambda _{m})  \notag \\
&&+X_{2}(\lambda )\frac{\Lambda _{m}^{(1)}(\lambda ,\{\lambda
_{i}\}|\{\lambda _{i}^{(1)}\})}{X_{2}^{(1)}(\lambda )}\Psi _{m}(\lambda
_{1},\lambda _{2},\ldots ,\lambda _{m})  \notag \\
&&-\sum_{j=1}^{m}\frac{x_{3}(\lambda _{j}-\lambda )}{x_{2}(\lambda
_{j}-\lambda )}\mathcal{B}(\lambda )\otimes \Phi _{m-1}(\overset{\wedge }{%
\lambda }_{j})  \notag \\
&&\times \lbrack M_{m}^{(1)}(\lambda _{j},\{\lambda _{i}\})]^{j-1}\left\{
X_{1}(\lambda _{j})\prod\limits_{k=1,k\neq j}^{m}z(\lambda _{k}-\lambda
_{j})\right.  \notag \\
&&\left. -X_{2}(\lambda _{j})\prod\limits_{k=1,k\neq j}^{m}z(\lambda
_{j}-\lambda _{k})\frac{\Lambda _{m}^{(1)}(\lambda _{j},\{\lambda
_{i}\}|\{\lambda _{i}^{(1)}\})}{X_{1}^{(1)}(\lambda _{j})}\right\} \mathcal{F%
}_{m}\left\vert 0_{L}\right\rangle  \label{An.11}
\end{eqnarray}%
where we have made the choice%
\begin{equation}
\mathcal{F}_{m}=\Phi _{m}^{(1)}(\lambda _{1}^{(1)},\ldots ,\lambda
_{m}^{(1)})\left\vert 0_{m}\right\rangle ^{(1)}  \label{An.12}
\end{equation}%
and we are collected the unwanted terms taking into account the cyclic
permutation property (\ref{two.32}).

Therefore, the eigenvalue is%
\begin{equation}
\Lambda _{L}(\lambda |\{\lambda _{i}\})=X_{1}(\lambda
)\prod\limits_{k=1}^{m}z(\lambda _{k}-\lambda )+X_{2}(\lambda )\frac{\Lambda
_{m}^{(1)}(\lambda ,\{\lambda _{i}\}|\{\lambda _{i}^{(1)}\})}{%
X_{2}^{(1)}(\lambda )}  \label{An.13}
\end{equation}%
provided%
\begin{eqnarray}
\frac{X_{1}(\lambda _{j})}{X_{2}(\lambda _{j})} &=&\left(
\prod\limits_{k=1,k\neq j}^{m}\frac{z(\lambda _{j}-\lambda _{k})}{z(\lambda
_{k}-\lambda _{j})}\right) \frac{\Lambda _{m}^{(1)}(\lambda _{j},\{\lambda
_{i}\}|\{\lambda _{i}^{(1)}\})}{X_{1}^{(1)}(\lambda _{j})}  \notag \\
&=&\prod\limits_{k=1,k\neq j}^{m}\frac{z(\lambda _{j}-\lambda _{k})}{%
z(\lambda _{k}-\lambda _{j})}x_{1}(0)\prod\limits_{k\neq j}^{m}z(\lambda
_{k}^{(1)}-\lambda _{j})  \label{An.14}
\end{eqnarray}%
where we have substitute $\Lambda _{m}^{(1)}(\lambda _{j},\{\lambda
_{i}\}|\{\lambda _{i}^{(1)}\})$ by the residue of $\Lambda
_{m}^{(1)}(\lambda ,\{\lambda _{i}\}|\{\lambda _{i}^{(1)}\})$ at $\lambda
=\lambda _{j}$ which is find by the second eigenvalue problem. The second
eigenvalue problem gives us 
\begin{equation}
\frac{\Lambda _{m}^{(1)}(\lambda ,\{\lambda _{i}\}|\{\lambda _{i}^{(1)}\})}{%
X_{2}^{(1)}(\lambda )}=\frac{X_{1}^{(1)}(\lambda )}{X_{2}^{(1)}(\lambda )}%
\prod\limits_{k=1}^{m}z(\lambda _{k}^{(1)}-\lambda )+\frac{\Lambda
_{m}^{(2)}(\lambda ,\{\lambda _{i}^{(1)}\}|\{\lambda _{i}^{(2)}\})}{%
X_{2}^{(2)}(\lambda )}  \label{An.15}
\end{equation}%
provided that%
\begin{eqnarray*}
\frac{X_{1}^{(1)}(\lambda _{j}^{(1)})}{X_{2}^{(1)}(\lambda _{j}^{(1)})}
&=&\left( \prod\limits_{k=1,k\neq j}^{m}\frac{z(\lambda _{j}^{(1)}-\lambda
_{k}^{(1)})}{z(\lambda _{k}^{(1)}-\lambda _{j}^{(1)})}\right) \frac{\Lambda
_{m}^{(2)}(\lambda _{j}^{(1)},\{\lambda _{i}^{(1)}\}|\{\lambda _{i}^{(2)}\})%
}{X_{1}^{(2)}(\lambda _{j}^{(1)})} \\
&=&\prod\limits_{k=1,k\neq j}^{m}\frac{z(\lambda _{j}^{(1)}-\lambda
_{k}^{(1)})}{z(\lambda _{k}^{(1)}-\lambda _{j}^{(1)})}x_{1}(0)\prod%
\limits_{k\neq j}^{m}z(\lambda _{k}^{(2)}-\lambda _{j}^{(1)})
\end{eqnarray*}%
where we have substitute $\Lambda _{m}^{(2)}(\lambda _{j}^{(1)},\{\lambda
_{i}^{(1)}\}|\{\lambda _{i}^{(2)}\})$ by the residue of $\Lambda
_{m}^{(2)}(\lambda ,\{\lambda _{i}^{(1)}\}|\{\lambda _{i}^{(2)}\})$ at $%
\lambda =\lambda _{j}^{(1)}$ which is given by the third eigenvalue problem,
and so on. We follow this procedure till we reach the last layer which
consists of the six-vertex model whose transfer-matrix diagonalization is
well known in the literature. Therefore, the eigenvalue of the transfer
matrix for the A$_{n}^{(1)}$ vertex models is given by 
\begin{equation}
\Lambda _{L}(\lambda |\{\lambda _{i}\})=X_{1}(\lambda
)\prod\limits_{k=1}^{m}z(\lambda _{k}-\lambda )+X_{2}(\lambda )\left(
\sum_{l=1}^{n-1}G_{m}^{(l)}(\lambda ,\{\lambda _{i}^{(l-1)}\}|\{\lambda
_{i}^{(l)}\})+\prod\limits_{k=1}^{m}z(\lambda -\lambda _{k}^{(n-1)})\right)
\label{An.16}
\end{equation}%
where 
\begin{equation}
G_{m}^{(l)}(\lambda ,\{\lambda _{i}^{(l-1)}\}|\{\lambda _{i}^{(l)}\})=\frac{%
X_{1}^{(l)}(\lambda )}{X_{2}^{(l)}(\lambda )}\prod\limits_{k=1}^{m}z(\lambda
_{k}^{(l)}-\lambda )  \label{An.17}
\end{equation}%
The Bethe equations are%
\begin{eqnarray}
\frac{X_{1}(\lambda _{j})}{X_{2}(\lambda _{j})} &\mathcal{=}%
&=\prod\limits_{k=1,k\neq j}^{m}\frac{z(\lambda _{j}-\lambda _{k})}{%
z(\lambda _{k}-\lambda _{j})}x_{1}(0)\prod\limits_{k\neq j}^{m}z(\lambda
_{k}^{(1)}-\lambda _{j})  \notag \\
\frac{X_{1}^{(l)}(\lambda _{j}^{(l)})}{X_{2}^{(l)}(\lambda _{j}^{(l)})}
&=&\prod\limits_{k=1,k\neq j}^{m}\frac{z(\lambda _{j}^{(l)}-\lambda
_{k}^{(l)})}{z(\lambda _{k}^{(l)}-\lambda _{j}^{(l)})}x_{1}(0)\prod%
\limits_{k\neq j}^{m}z(\lambda _{k}^{(l+1)}-\lambda _{j}^{(l)})  \notag \\
l &=&1,2,...,n-1  \label{An.18}
\end{eqnarray}%
These results complete our study about the nested Bethe ansatz for vertex
models based in non-exceptional Lie algebras.

\end{document}